\documentclass[acmtog]{acmart}

\AtBeginDocument{%
  }

\usepackage{multirow}
\usepackage{subcaption}
\usepackage{pifont}
\usepackage{colortbl}
\usepackage{cleveref}    
\newcommand{\alphabox}[1]{%
    \begingroup
    \setlength{\fboxsep}{1.2pt}%
    \colorbox{black!60}{
        \scalebox{.7}{\color{white}\hspace{1pt}#1\hspace{1pt}}%
    }%
    \endgroup
}

\crefname{figure}{Fig.}{Figs.}
\crefname{table}{Tab.}{Tabs.}
\crefname{equation}{Eq.}{Eqs.}
\copyrightyear{2026}
\acmYear{2026}
\setcopyright{cc}
\setcctype{by-nc-nd}
\acmConference[SIGGRAPH Conference Papers '26]{Special Interest Group on Computer Graphics and Interactive Techniques Conference Conference Papers}{July 19--23, 2026}{Los Angeles, CA, USA}
\acmBooktitle{Special Interest Group on Computer Graphics and Interactive Techniques Conference Conference Papers (SIGGRAPH Conference Papers '26), July 19--23, 2026, Los Angeles, CA, USA}
\acmDOI{10.1145/3799902.3811064}
\acmISBN{979-8-4007-2554-8/2026/07}

\acmSubmissionID{332}

\begin{document}

\title{Learning View-Dependent Splatting Kernels}

\author{Huakeng Ding}
\authornote{Equal contributions. Corresponding authors: \{kunzhou,hwu\}@acm.org}
\orcid{0009-0000-5632-6489}
\author{Zhanpeng Liu}
\authornotemark[1]
\affiliation{%
  \institution{State Key Lab of CAD and CG, Zhejiang University}
  \city{Hangzhou}
  \state{Zhejiang}
  \country{China}
}

\author{Fan Pei}
\affiliation{%
  \institution{State Key Lab of CAD and CG, Zhejiang University}
  \country{China}}

\author{Kun Zhou}
\affiliation{%
  \institution{State Key Lab of CAD and CG, Zhejiang University}
  \country{China}}
\affiliation{%
  \institution{Hangzhou Research Institute of Holographic and AI Technology}
  \country{China}}

\author{Hongzhi Wu}
\affiliation{%
  \institution{State Key Lab of CAD and CG, Zhejiang University}
  \country{China}}









\renewcommand{\shortauthors}{Ding et al.}



\begin{abstract}
We present a differentiable framework to automatically learn view-dependent 2D kernels in a splatting-based pipeline to improve reconstruction quality and representation efficiency for novel 3D view synthesis. Our volumetric primitive is defined as a bounding ellipsoid and a 3D-kernel latent vector. We first learn a projection network to output a 2D-kernel latent, taking the attributes of the ellipsoid and the 3D-kernel latent as input. Next, the result is sent to a decoder to produce a radially symmetric 2D kernel in terms of Mahalanobis distance, bounded by the projected ellipsoid. The neural networks along with per-primitive attributes are jointly optimized. The effectiveness of our approach is demonstrated on standard benchmarks, comparing favorably against state-of-the-art techniques on both analytical and learned kernels. Finally, we extend the idea to learn general 2D kernels for 2D splatting as well as image representation.

\end{abstract}


\begin{teaserfigure}
\centering
\includegraphics[width=\linewidth]{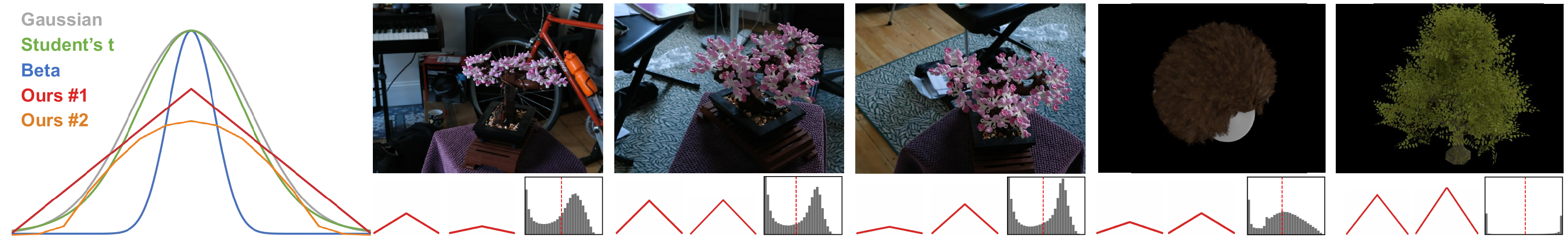}
\hbox to \linewidth{%
  \makebox[0.24\linewidth][c]{(a)}%
  \makebox[0.15\linewidth][c]{(b)}%
  \makebox[0.15\linewidth][c]{(c)}%
  \makebox[0.15\linewidth][c]{(d)}%
  \makebox[0.15\linewidth][c]{(e)}%
  \makebox[0.15\linewidth][c]{(f)}%
}
\caption{We present a differentiable framework to automatically learn view-dependent 2D kernels in a splatting-based pipeline to improve reconstruction quality and representation efficiency for novel 3D view synthesis. We show 1D profile comparisons between our learned kernels and existing ones in (a). For each set of 4 images from (b) to (f), the top one is a view in a scene; the left 2 images in the bottom are the 1D profiles of two of our learned kernels; and the right image in the bottom shows a histogram of parameter $d_1$ (\cref{{eq:sample_d}}) of our kernels corresponding to the view above it, with the red dotted line indicating the mean (the range of the horizontal axis is [0,1]). Our learned primitives vary across views from (b) to (d), and are adaptive to different scenes (b-d), (e) and (f).}
\label{fig:teaser}
\end{teaserfigure}

\maketitle

\section{Introduction}
Since its debut in~\cite{kerbl3Dgaussians}, 3D Gaussian Splatting (3DGS) has achieved tremendous success in image-based representation and rendering. First, a standard 3D Gaussian \emph{kernel} is scaled, rotated, and translated to produce a semi-transparent anisotropic \emph{shape}. This forms a basic primitive, a collection of which are used to model an input scene. 
Next, this 3D shape is approximately projected to the 2D screen (i.e., splatting), and blended according to a sorted order via efficient tile-based differentiable rendering.
All related parameters are optimized with respect to multi-view input images.

The splatting kernel is a crucial factor in the reconstruction quality and representation efficiency of any related pipeline. Consider a simple 2D example: it is far more efficient to represent a rectangle with a set of squares than Gaussians. Researchers devote considerable efforts to this topic, by \emph{manually} replacing the Gaussians in~\cite{kerbl3Dgaussians} 
with various analytical kernels~\cite{chen20243DLS,GES_hamdi_2024_CVPR,liu2025deformablebetasplatting,zhu2025sss}. This series of work naturally leads to two fundamental questions: (1) \emph{what are optimal splatting kernels?} and (2) \emph{how can we automatically discover them?}

Recent work explores \emph{automatically} learning alternative volumetric primitives from data~\cite{Held20243DConvex}. However, closed-form equations do not exist for computing line integrals over general 3D primitives, a step in splatting. So a ``slicing'' approximation is adopted to avoid the expensive numerical integration. On the other hand, exact integral can be avoided/computed, by restricting the learned primitive to be a planar shape~\cite{huang2024drk}, or represented by a specific type of network~\cite{zhou2025splatnet}. In both cases, the expressiveness of the representation is limited. It is yet unclear how to systematically learn general splatting kernels in a data-driven fashion.

In this paper, we present a differentiable framework to automatically learn view-dependent 2D splatting kernels, as an implicit representation of volumetric primitives. Our approach avoids the computation of line integral, by explicitly modeling the integral results. We also introduce learned view-dependency to our kernels to improve reconstruction quality. Specifically, we start with a bounding ellipsoid as in vanilla 3DGS. Next, a neural 3D-to-2D projection is performed by a bounding-ellipsoid-aware multi-layer perceptron (MLP), which takes as input a latent vector that represents a 3D kernel, and outputs another one that represents a 2D kernel after the projection. Afterward, a decoder transforms the resulting latent into a radially symmetric 2D kernel in terms of Mahalanobis distance, as the final splatting shape. This shape is passed on to the rest of a standard splatting pipeline for further processing (e.g., coloring, alpha-blending).

The effectiveness of our framework is demonstrated on 4 standard benchmarks. In both 3D/2D splatting, our approach is the best or second-best in all reconstruction quality metrics across all benchmarks, compared with state-of-the-art techniques. Moreover, our volumetric primitives also demonstrate superior representation efficiency, as well as good scalability with respect to the number of primitives. We also apply our planar primitives to learn to efficiently represent 2D images. Our code is publicly available at https://optkernel.github.io.

\section{Related Work}
\label{sec:related_work}
This section focuses on research on improving the shapes of splatting primitives in the context of novel view synthesis. We divide existing work based on whether its primitives are volumetric or planar.
We do not cover other aspects of improvements, including but not limited to density control \cite{kheradmand20243dgsmcmc,park2025dropgaussianstructuralregularizationsparseview}, appearance~\cite{yang2024spec-gs,LatentSpecGaussian,bi2024rgs}, and anti-aliasing~\cite{Yu2024MipSplatting}, as they are orthogonal to our focus. Interested readers are directed to excellent surveys~\cite{survey0,survey1,survey2} for a broader view of the topic.



\subsection{Volumetric Primitives}

\subsubsection{Analytical Kernels} 
The majority of related work falls into this category, which replaces 3D Gaussians in vanilla 3DGS with alternative, explicitly defined volumetric primitives or implicitly defined ones as view-independent 2D kernels (after projection).


For the former, the 3D density distribution of a primitive is analytically defined. Notable examples include generalized exponential~\cite{GES_hamdi_2024_CVPR}, Student's t~\cite{zhu2025sss}, and deformable Beta kernels~\cite{liu2025deformablebetasplatting}. In addition, these kernels come with analytical formulas for computing splatting results,
leading to high rendering performance. While reconstruction quality and representation efficiency are constantly improving, related approaches are limited in terms of expressiveness: their shape variations are entirely determined by hand-crafted, analytical equations, and thus cannot fully adapt to input scenes in a data-driven fashion.


On the other hand, existing work implicitly models volumetric primitives as view-independent 2D splatting kernels, such as linear~\cite{chen20243DLS}, overlapping Gaussians~\cite{liu2025TNT}, B-spline~\cite{Thomas2025SplineSplatRR}, or even Gaussian multiplied with a 2D alpha texture~\cite{chao2025texturedgaussians}. However, no 3D consistency among 2D kernels across different views is imposed, resulting in suboptimal reconstructions. In comparison, our learned view-dependency of kernels serves as a form of data-driven consistency, leading to improved reconstructions.


It is worth mentioning that most of the aforementioned kernels are \emph{radially symmetric}, the reason of which will be elaborated in~Sec.~\ref{sec:justify}. Also note that while the \emph{shape} of a popular primitive~\cite{GES_hamdi_2024_CVPR,zhu2025sss,liu2025deformablebetasplatting,kerbl3Dgaussians} is view-dependent, its \emph{kernel} is view-independent as a function of Mahalanobis distance (e.g., according to~\cref{eq:3dgs} for Gaussians). In comparison, our kernel is view dependent, which brings in extra degrees of freedom, as illustrated in~\cref{fig:kernel_comparison}.

\subsubsection{Learned Representations} 
Few papers explore learning optimal splatting primitives directly from data. Held et al.~\shortcite{Held20243DConvex} optimize the vertices of a convex shape as a primitive, and simplify the line integral for splatting as a constant value, which gets further softened near the boundary. Zhou et al.~\shortcite{zhou2025splatnet} essentially splat neural fields as primitives. They are represented as a \emph{specific type} of shallow MLPs, so that their exact line integrals can be derived as closed-form equations.


Unlike the above work, we do not explicitly model the density distribution of a 3D primitive. Instead, the view-dependent projection of a primitive is modeled, which avoids the challenging line integral computation over a general 3D density field. Moreover, our approach is not tied to any specific type of neural networks.


\subsection{Planar Primitives}


\subsubsection{Analytical Kernels}
Starting with the pioneering work of 2DGS \cite{Huang2DGS2024}, researchers improve the original 2D surfels with alternative analytical 2D kernels, such as Hermite polynomials~\cite{yu20242dgh2dgaussianhermitesplatting}, Gabor~\cite{watanabe3Dgabors}, and even triangles~\cite{Held2025Triangle,held2025meshsplattingdifferentiablerenderingopaque}. Similar to related work on volumetric primitives, existing approaches here are limited in expressiveness: the splatting shape variations are determined by analytical equations and cannot fully adapt to input scenes.



%
\subsubsection{Learned Representations.}
Svitov et al.~\shortcite{svitov2024billboard} multiply a learnable 2D RGBA texture with a 2D Gaussian primitive to introduce spatial variations with a shared uv paramterization. This can be viewed as a refined splatting shape with the details from the alpha channel of the texture. Note that research which augments a primitive with an RGB texture does not change its shape~\cite{rong2024gstex,zhang2025nest,xu2024supergaussians,Papantonakis2025ContentAwareTF}, and thus is orthogonal to our work. Recently, Huang et al.~\shortcite{huang2024drk} propose a learnable 2D radial kernel with a number of control points corresponding to different angles.

In comparison, our work is a unified approach for both volumetric and planar primitives, while the methods above are limited to handle the latter. Moreover, the view-dependency of our kernels achieves higher quality with existing work, where the 2D kernels (prior to affine transformations) are view independent. Note that our planar primitives (Sec.~\ref{sec:neural_planar}) are asymmetric and can be viewed as a generalized version of~\cite{huang2024drk}, as~\cref{eq:planar_dec} is an MLP (i.e., we do not assume parametric curves as in their paper).

\begin{figure*}
    \centering
    \includegraphics[width=\linewidth]{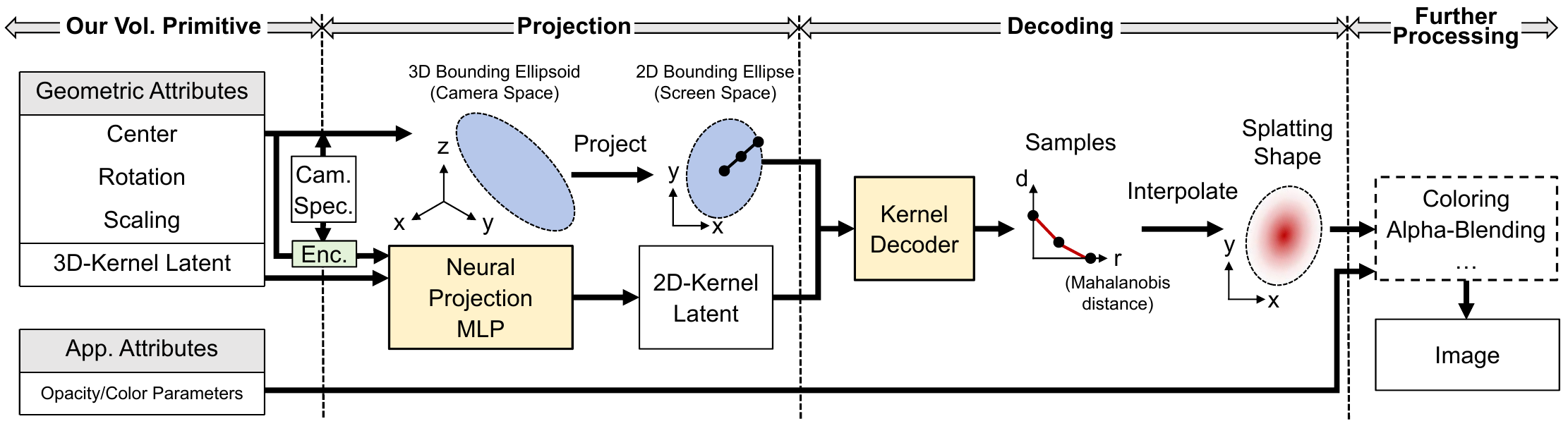}
    \caption{Our pipeline for splatting volumetric primitives. For each primitive, we first project the bounding ellipsoid to the image plane as a 2D bounding ellipse. Next, we project our 3D kernel to 2D, by transforming the 3D-kernel latent into one that represents a 2D kernel, via a projection MLP that is also aware of attributes of the ellipsoid. The result is then sent to a decoder to output a radially symmetric 2D kernel in terms of Mahalanobis distance, bounded by the 2D ellipse, as the final splatting shape. To accelerate the process, we pre-sample the 1D profile of the 2D kernel and draw the splat by linearly interpolating the samples. Finally, the shape is passed on to the rest of a standard splatting-based pipeline for further processing (e.g., coloring, alpha-blending). Vol. = volumetric, Enc. = encoding, and Cam. Spec. = camera specification.}
    
    \label{fig:pipeline}
\end{figure*}
\section{Preliminaries}
\label{sec:preliminaries}

This section briefly describes the key terms in 3D/2D Gaussian splatting that are most related to the definitions of our kernels in the subsequent text. 

For 3DGS~\cite{kerbl3Dgaussians}, the attributes of a 3D Gaussian primitive include its mean $\boldsymbol{\mu}_{3D}$ and covariance $\boldsymbol{\Sigma}_{3D}=\mathbf{R}\mathbf{S}\mathbf{S^T}\mathbf{R^T}$, where $\mathbf{S}$ is a scaling matrix and $\mathbf{R}$ is a rotation one. The projection of a Gaussian onto an image plane is approximated via EWA splatting~\cite{zwicker2001ewa}. The resulting 2D covariance matrix $\boldsymbol{\Sigma}_{2D}$ is computed as:
\begin{equation}
\boldsymbol{\Sigma}_{2D} = \mathbf{J} \mathbf{W} \boldsymbol{\Sigma}_{3D} \mathbf{W}^T \mathbf{J}^T,
    \label{eq:gsproj}
\end{equation}
where $\mathbf{W}$ is the viewing transformation, and $\mathbf{J}$ is the Jacobian of affine approximation of projection. We denote the projected mean as $\boldsymbol{\mu}_{2D}$. The screen-space shape of a projected primitive can be defined as:
\begin{align}
    G(\mathbf{x}) &= \exp\left(-\frac{1}{2} (\mathbf{x} - \boldsymbol{\mu}_{2D})^T \boldsymbol{\Sigma}_{2D}^{-1} (\mathbf{x} - \boldsymbol{\mu}_{2D}) \right) = \exp(-\frac{1}{2} r^2),
    \label{eq:3dgs}
\end{align}
Here $\mathbf{x}$ are the coordinates of a pixel, and $r^2$ is the squared Mahalanobis distance. The latter can be viewed as a transform from an anisotropic Gaussian distribution to a standard isotropic normal distribution, which \emph{solely depends} on $r^2$.

For 2DGS~\cite{Huang2DGS2024}, its primitives are oriented disks. The screen-space shape of a projected primitive is defined as:
\begin{equation}
    G'(\mathbf{x}) = \exp\left(-\frac{1}{2} (u^2 + v^2) \right),
\end{equation}
where $(u, v)$ are the local 2D coordinates of the intersection of a camera ray corresponding to $\mathbf{x}$ against a primitive on its tangent plane. This native $(u, v)$ parameterization in 2DGS makes it convenient to define spatial variations on a primitive (e.g., general 2D kernel or 2D texture).



\section{Overview}
The geometric attributes of our primitive consist of a 3D bounding ellipsoid, similar to the truncated Gaussian in vanilla 3DGS, and a latent vector that represents a 3D kernel (Sec.~\ref{sec:neural_primitive}). For each primitive, our pipeline first projects the ellipsoid to the image plane as a 2D bounding ellipse. Next, we project our 3D kernel to 2D, by transforming the 3D-kernel latent into one that represents a 2D kernel, via a projection MLP that is also aware of attributes of the ellipsoid (Sec.~\ref{sec:projector}). The result is then sent to a decoder to output a radially symmetric 2D kernel in terms of Mahalanobis distance, bounded by the 2D ellipse, as the final splatting shape (Sec.~\ref{sec:decoder}). This shape is passed on to the rest of any standard splatting-based pipeline for further processing (e.g., coloring, alpha-blending). Please refer to~\cref{fig:pipeline} for an illustration.

\section{Our Approach}
\label{sec:method}
\subsection{Volumetric Primitives}
\label{sec:neural_primitive}

To define its geometry, each of our volumetric primitives includes a 3D bounding ellipsoid and a 3D-kernel latent: the attributes for the ellipsoid consist of its center $\boldsymbol{\mu}_{3D}$, a 3D vector of scaling factors $\mathbf{s}$ (to represent the scaling matrix $\boldsymbol{S}$), and a 4D quaternion $q$ (to represent the rotation matrix $\boldsymbol{R}$); the latent $\mathbf{z}_{3D}$ is a vector (whose dimension is 5 in most of our experiments), which stores the parameters of an implicit 3D kernel.




We employ a bounding ellipsoid for the following reasons. First, it directly models scaling/rotation/translation transformations to the kernel; our MLPs can focus on learning ``normalized" shapes, prior to applying affine transformations. We believe this is a good balance between quality and performance. A more aggressive approach that directly learns shapes would probably require larger (slower) MLPs to handle the extra complexity. Second, the ellipsoid serves as an efficient geometric proxy for sorting and culling. Its similarity with truncated Gaussians in vanilla 3DGS also allows us to reuse much of that pipeline (Sec.~\ref{sec:preliminaries}).


The learnable appearance of each primitive is similarly defined as in 3DGS, which includes an overall opacity and the parameters of a color model~\cite{kerbl3Dgaussians,liu2025deformablebetasplatting}.

\subsection{Neural Projection}
\label{sec:projector}
Below we describe how to project our volumetric primitive to an image plane, for a given camera specification. First, for the 3D bounding ellipsoid, we directly project it with~\cref{eq:gsproj}  to obtain a 2D bounding ellipse. 


Next, similar to how~\cref{eq:gsproj} projects a 3D ellipsoid to a 2D ellipse, we learn a global projection MLP $\Phi_{\operatorname{proj}}$ to transform the implicit 3D kernel (i.e., $\mathbf{z}_{3D}$) to a latent vector $\mathbf{z}_{2D}$ (whose dimension is 5 in most experiments), which represents a 2D kernel after the projection:

\begin{equation}
    \mathbf{z}_{2D} = \Phi_{\operatorname{proj}}\big(\mathbf{z}_{3D}, \boldsymbol{\mu}^{\operatorname{cam}}_{3D}, \mathbf{s}, \mathbf{R}^{\operatorname{cam}}\big).
\label{eq:mlp_project}
\end{equation}
A straightforward solution might take the 3D geometry (represented as $\mathbf{z}_{3D}$ in our case) and a camera specification only as input, as in the majority of related work. However, we discover through experiments (\cref{tab:ablation_inputs}) that adding other factors as input to $\Phi_{\operatorname{proj}}$ brings quality improvement while maintaining generalization ability. 

Specifically, we add the camera-space coordinates of the ellipsoidal center, $\boldsymbol{\mu}^{\operatorname{cam}}_{3D}$, to provide spatial awareness for the MLP. Also, the scaling factors, $\mathbf{s}$, add scale awareness. Finally, we send in $\mathbf{R}^{\operatorname{cam}}$, the rotation matrix of our primitive in the camera space, because it is a compact representation of the orientation of the bounding ellipsoid/primitive in the camera space, efficiently encoding the 3D view information. By default, $\Phi_{\operatorname{proj}}$ is implemented as a lightweight, 4-layer MLP, with 64 neurons per hidden layer and leaky-ReLU activation after each layer except for the last. The global MLP is jointly trained with the kernel decoder, which will be introduced in the next subsection, with respect to all input images.



\subsection{Kernel Decoder}
\label{sec:decoder}
\subsubsection{Definition.}
Our kernel decoder, $\Phi_{\operatorname{dec}}$, decodes a radially symmetric 2D kernel from a latent $\mathbf{z}_{2D}$. Specifically, for a given pixel, we first compute its squared Mahalanobis distance $r^2$ with respect to the projected primitive center $\boldsymbol{\mu}_{2D}$ according to~\cref{eq:3dgs}, based on the attributes from the corresponding 2D bounding ellipse. Next, we replace the analytical shape function in related work (e.g., Gaussian~\cite{kerbl3Dgaussians} or Beta~\cite{liu2025deformablebetasplatting}) with a global MLP $\Phi_{\operatorname{dec}}$. It takes $r^2$ and $\mathbf{z}_{2D}$ as input, and outputs an opacity value $d$ as the result:
\begin{equation}
    d = \Phi_{\operatorname{dec}}(r^2, \mathbf{z}_{2D}).
\end{equation}
This MLP essentially describes a 1D profile of a radially symmetric 2D splatting kernel, for $r^2$ in the range of [0,1]. Our default implementation of $\Phi_{\operatorname{dec}}$ is a lightweight, 3-layer MLP, with 4 neurons per hidden layer and leaky-ReLU activation after each layer except for the last. The final layer is a sigmoid function to bound the output.

\subsubsection{Justification of Radial Symmetry.}
\label{sec:justify}
The reason we choose to output a radially symmetric 2D kernel, rather than a general one, is due to the theoretical discontinuity in mapping an arbitrary 3D view condition to a 2D frame (on which a general 2D kernel can be defined). Mathematically, this is known as the hairy ball theorem~\cite{hatcher2002algebraic}. In our pilot study, we apply a smoothing trick to alleviate this discontinuity. But it results in a rapid rotation of the 2D kernel around certain view directions, which is disturbing during animation. On the other hand, by restricting our output to a radially symmetric 2D kernel, we avoid the aforementioned mapping in the first place, as the 2D frame is no longer needed. Note that this challenge is also mentioned in existing work on learnable kernels (cf. Sec.~3.4 of~\cite{huang2024drk}).

\subsubsection{Acceleration.}
\label{sec:acc}


Directly executing $\Phi_{\text{dec}}$ for every pixel within our bounding ellipse could be computationally expensive. To improve rendering performance, we pre-sample the 1D profile 
that $\Phi_{\text{dec}}$ represents, and perform interpolation during rendering, substantially reducing the runtime cost of calling $\Phi_{\text{dec}}$. This essentially decouples the number of pixels within our primitive from the number of calls to $\Phi_{\text{dec}}$.

Specifically, for each primitive, we place $k$ uniform samples of $r$ to cover the range of [0,1], which are denoted as $\{r_i\}_{1 \leq i \leq k}$. Next, we run $\Phi_{\text{dec}}$ only at these samples, and store the results as $\{d_i\}_{1 \leq i \leq k}$:
\begin{equation}
    d_i = \Phi_{\text{dec}}(r^2_i, \mathbf{z}_{2D}).
    \label{eq:sample_d}
\end{equation}
Finally, during rasterization, for any $r$ computed from a particular pixel, we find its closet samples such that $r_c \le r \le r_{c+1}$, and efficiently compute the final result via linear interpolation:
\begin{align}
    d &= (1-\lambda) d_c + \lambda d_{c+1},
    \text{\, with \,} \lambda = \frac{r - r_c}{r_{c+1} - r_c}.
\end{align}
Note that our approach is not tied to any type of interpolation. More sophisticated interpolation can also be plugged in, to bring higher-order smoothness at the cost of more computational budget. We use $k=2$ in most experiments, as it suffices to produce results with high quality, according to Sec.~\ref{sec:ablation}.




\subsection{Extension to Planar Primitives}
\label{sec:neural_planar}
In addition to volumetric primitives, our idea can also be applied to learn general 2D kernels for planar primitives in the context of 2D splatting~\cite{Huang2DGS2024}. To reduce repetitive text, below we primarily describe the differences in defining/processing our planar primitives versus volumetric ones.

First, to define the geometry, each of our planar primitives includes a bounding 
ellipse, and a 10D latent code $\mathbf{z}^{\operatorname{prior}}_{2D}$ that represents a general 2D kernel \emph{prior} to projection. Next, similar to Sec.~\ref{sec:projector}, we learn a projection MLP with the same architecture as $\Phi_{\operatorname{proj}}$ in Sec.~\ref{sec:projector}, to transform $\mathbf{z}^{\operatorname{prior}}_{2D}$ to another 10D latent $\mathbf{z}^{\operatorname{after}}_{2D}$, which implicitly represents a general 2D kernel after projection. Finally, for each pixel within the projected bounding ellipse, a kernel decoder takes $\mathbf{z}^{\operatorname{after}}_{2D}$ and local 2D coordinates as input, and outputs an opacity value $d$:
\begin{equation}
    d = \Phi^{\operatorname{planar}}_{\operatorname{dec}}(u, v, \mathbf{z}^{\operatorname{after}}_{2D}).
    \label{eq:planar_dec}
\end{equation}
Here ($u$, $v$) are the 2D coordinates of the intersection of a camera ray corresponding to a pixel against our planar primitive on its tangent plane. The architecture of $\Phi^{\operatorname{planar}}_{\operatorname{dec}}$ is almost the same as its counterpart $\Phi_{\operatorname{dec}}$ for volumetric primitives, with the number of neurons per hidden layer doubled to 8, to cope with the extra complexity of a general 2D kernel.

\subsection{Implementation Details}
\label{sec:imp}

We pre-train neural projection and kernel decoder prior to 3D reconstruction. To start, the 5D latent code $\mathbf{z}_{3D}$ is initialized to all zeros, while all weights in $\Phi_{\operatorname{proj}}$ and $\Phi_{\operatorname{dec}}$ are initialized with~\cite{He_2015_ICCV}. We then jointly optimize $\Phi_{\operatorname{proj}}$ and $\Phi_{\operatorname{dec}}$. To synthesize training data, we randomly sample $\boldsymbol{\mu}^{\operatorname{cam}}_{3D}$, $\mathbf{s}$, and $\mathbf{R}^{\operatorname{cam}}$, and then randomly sample $r$ to obtain target profile sample according to~\cref{eq:sample_d}. In the volumetric case, we employ a 1D cosine function as the target profile (i.e., $d = \cos(\frac{\pi}{2}r^2)$); for the planar case, 2D Gaussian surfels are used. We find that pre-training results in higher reconstruction quality than without it. Please see~\cref{tab:ablation_init} for a detailed comparison.


For training, we employ a two-stage strategy to improve stability, as our primitives have higher degrees of freedom compared with standard 3D/2DGS. After pre-training, in the first stage of training, we freeze neural projection and kernel decoder for the first 2000 iterations for a scene whose initial point cloud comes from SfM/random points, respectively. In the second stage, all parameters are optimized jointly with respect to the loss function defined in~\cite{liu2025deformablebetasplatting}, which is an image loss plus the regularization terms for opacity and scale. Note that we also adopt the kernel-agnostic MCMC density control strategy from the same paper, as it is effective and not tied to specific types of kernels. The learning rate for neural projection/kernel decoder decays exponentially from $1.6\times 10^{-4}$ to $1.6\times 10^{-6}$. For other parameters, the learning rates are set according to~\cite{kheradmand20243dgsmcmc}. We use the Adam optimizer to train 30K iterations in each experiment.




\section{Results and Discussions}

\begin{table*}[t]
\centering
\caption{Quantitative comparisons between our approach and state-of-the-art techniques on 4 standard datasets: Mip-NeRF360~\cite{barron2022mipnerf360}, Tanks~\&~Temples~\cite{Tank&Temple}, Deep Blending~\cite{DeepBlending2018} and NeRF Synthetic~\cite{mildenhall2020nerf}. We highlight the best, second-best, and third-best results in \textcolor{red}{red}, \textcolor{orange}{orange}, and \textcolor{yellow}{yellow}, respectively. Baseline results are obtained from the original papers whenever available; otherwise, we run the official implementation and list the results in \textit{italics}. SH = spherical harmonics, SB = spherical Beta.}
\resizebox{\textwidth}{!}{
\begin{tabular}{ll|ccc|ccc|ccc|ccc}
\toprule
\multicolumn{2}{c|}{} & \multicolumn{3}{c|}{Mip-NeRF360} & \multicolumn{3}{c|}{Tanks~\&~Temples} & \multicolumn{3}{c|}{Deep Blending} & \multicolumn{3}{c}{NeRF Synthetic} \\
\multicolumn{2}{c|}{} & PSNR$\uparrow$ & SSIM$\uparrow$ & LPIPS$\downarrow$ & PSNR$\uparrow$ & SSIM$\uparrow$ & LPIPS$\downarrow$ & PSNR$\uparrow$ & SSIM$\uparrow$ & LPIPS$\downarrow$ & PSNR$\uparrow$ & SSIM$\uparrow$ & LPIPS$\downarrow$ \\
\midrule
\multirow{5}{*}{\rotatebox{90}{2D Splatting}} 
& 2DGS~\cite{Huang2DGS2024}
& \cellcolor{yellow!25}27.04 & \cellcolor{yellow!25}0.805 & \cellcolor{orange!25}0.223
& \textit{23.13} & \textit{0.831} & \textit{0.212} 
& \cellcolor{orange!25}\textit{29.50} & \textit{0.902} & \textit{0.257}
& 33.07 & \textit{0.967} & \cellcolor{yellow!25}\textit{0.035} \\
& BBSplat~\cite{svitov2024billboard}
& \textit{26.88} & \textit{0.794} & \textit{0.253}
& \cellcolor{yellow!25}\textit{23.59 }& \textit{0.851} & \cellcolor{red!25}\textit{0.150} 
& \textit{29.28} & \textit{0.902} & \cellcolor{red!25}\textit{0.245} 
& \textit{32.00} & \textit{0.956} & \textit{0.044}  \\
& DRK~\cite{huang2024drk}
& 26.76 & 0.787 & 0.236 
& \textit{22.49} & \cellcolor{red!25}\textit{0.867} & \textit{0.243}
& \textit{29.42} & \cellcolor{red!25}\textit{0.922} & \textit{0.322}
& \cellcolor{yellow!25}\textit{33.22} & \cellcolor{yellow!25}\textit{0.968} & \textit{0.043} \\
\cmidrule(lr){2-14}

& Ours~(2D+SH) 
& \cellcolor{orange!25}28.42 & \cellcolor{red!25}0.824 & \cellcolor{red!25}0.219 
& \cellcolor{orange!25}24.50 & \cellcolor{orange!25}0.857 & \cellcolor{orange!25}0.161
& \cellcolor{yellow!25}30.18 & \cellcolor{orange!25}0.911 & \cellcolor{orange!25}0.248 
& \cellcolor{red!25}34.57 & \cellcolor{red!25}0.973 & \cellcolor{red!25}0.026 \\
& Ours~(2D+SB)
& \cellcolor{red!25}28.44 & \cellcolor{orange!25}0.821 & \cellcolor{yellow!25}0.225 
& \cellcolor{red!25}24.59 & \cellcolor{yellow!25}0.857 & \cellcolor{yellow!25}0.163 
& \cellcolor{red!25}30.26 & \cellcolor{yellow!25}0.910 & \cellcolor{yellow!25}0.250 
& \cellcolor{orange!25}34.49 & \cellcolor{orange!25}0.972 & \cellcolor{orange!25}0.027 \\
\midrule
\multirow{8}{*}{\rotatebox{90}{3D Splatting}} 
& 3DGS~\cite{kerbl3Dgaussians}
& 27.20 & 0.815 & 0.214 
& 23.15 & 0.840 & 0.183 
& 29.41 & 0.903 & 0.243 
& 33.31 & 0.969 & 0.037 \\
& 3DGS-MCMC~\cite{kheradmand20243dgsmcmc}
& 28.29 & 0.840 & 0.210 
& 24.29 & 0.860 & 0.190 
& 29.67 & 0.895 & 0.320 
& 33.80 & 0.970 & 0.040 \\
& 3DCS~\cite{Held20243DConvex}
& 27.29 & 0.802 & 0.207 
& 23.95 & 0.851 & 0.157 
& 29.81 & 0.902 & 0.237 
& \textit{31.68} &\textit{0.958} & \textit{0.048} \\
& SplatNet~\cite{zhou2025splatnet}
& 27.21 & 0.791 & 0.216 
& 23.59 & 0.846 & 0.162 
& 29.20 & 0.892 & 0.264 
& 33.34 & 0.967 & 0.032 \\
& SSS~\cite{zhu2025sss}
& \textit{28.25} & \textit{0.838} & \textit{0.171} 
& 24.87 & \cellcolor{yellow!25}0.873 & \cellcolor{red!25}0.138 
& 30.07 & 0.907 & 0.247 
& \textit{34.29} & \textit{0.971} & \textit{0.029} \\
& DBS~\cite{liu2025deformablebetasplatting}
& \cellcolor{yellow!25}28.60 & \cellcolor{red!25}0.844 & \cellcolor{red!25}0.182 
& 24.79 & 0.868 & 0.148 
& 30.10 & 0.910 & 0.240 
& \cellcolor{orange!25}34.64 & \cellcolor{yellow!25}0.973 & \cellcolor{yellow!25}0.028 \\


& Ours~(3D,w/o SH) for ablation
& 28.53 & 0.835 &  0.192 
&  \cellcolor{yellow!25} 25.24 &  0.872 &  0.148 
&  \cellcolor{yellow!25} 30.48 & \cellcolor{yellow!25} 0.912 & \cellcolor{orange!25} 0.234 
&  34.37 &  0.972 &  0.028 \\

\cmidrule(lr){2-14}

& Ours~(3D+SH) 
& \cellcolor{orange!25}28.73 & \cellcolor{orange!25}0.842 & \cellcolor{orange!25}0.185 
& \cellcolor{red!25}25.42 & \cellcolor{red!25}0.877 & \cellcolor{orange!25}0.141 
& \cellcolor{orange!25}30.52 & \cellcolor{orange!25}0.913 & \cellcolor{red!25}0.233 
& \cellcolor{red!25}34.72 & \cellcolor{red!25}0.973 & \cellcolor{red!25}0.027 \\
& Ours~(3D+SB) 
& \cellcolor{red!25}28.82 & \cellcolor{yellow!25}0.840 & \cellcolor{yellow!25}0.189 
& \cellcolor{orange!25}25.35 & \cellcolor{orange!25}0.874 & \cellcolor{yellow!25}0.145 
& \cellcolor{red!25}30.54 & \cellcolor{red!25}0.913 & \cellcolor{yellow!25}0.237 
& \cellcolor{yellow!25}34.58 & \cellcolor{orange!25}0.973 & \cellcolor{orange!25}0.027 \\
\bottomrule
\end{tabular}
}
\label{tab:quantitative_eval}
\end{table*}

We conduct most experiments on a workstation with dual AMD EPYC 7763 CPUs, 768GB RAM, and an RTX 4090 GPU. The only exception is the reconstruction experiments with planar primitives on {\sc{Drjohnson}} and {\sc{Treehill}} scenes, where an RTX PRO 6000 is used due to the memory requirement.

We test on all 21 scenes from 4 standard benchmarks, including 9 scenes from Mip-NeRF 360~\cite{barron2022mipnerf360}, 2 from Tanks~\&~Temples \cite{Tank&Temple}, 2 from Deep Blending~\cite{DeepBlending2018}, and 8 from NeRF Synthetic~\cite{mildenhall2020nerf}. All scenes are physically captured images, with the exception of NeRF Synthetic. For quantitative assessments of reconstruction quality, we compute
PSNR, SSIM, and LPIPS averaged over all test images. Our average training time is 53 minutes.

\subsection{Comparisons}
\label{sec:comparisons}

\subsubsection{Reconstruction Quality.}
We compare our volumetric primitives with state-of-the-art 3D splatting techniques, including (1) analytical kernels: 3DGS~\cite{kerbl3Dgaussians}, 3DGS-MCMC~\cite{kheradmand20243dgsmcmc}, SSS~\cite{zhu2025sss} and DBS~\cite{liu2025deformablebetasplatting}, and (2) learned representations:  SplatNet~\cite{zhou2025splatnet} and 3DCS~\cite{Held20243DConvex}. For 2D splatting, we compare our planar primitives with 2DGS~\cite{Huang2DGS2024} as well as learned representations: BBSplat~\cite{svitov2024billboard} and DRK~\cite{huang2024drk}.

\cref{tab:quantitative_eval} shows quantitative comparison results. Note that for our volumetric/planar primitives, two variants with different color models, spherical harmonics (SH)~\cite{kerbl3Dgaussians} or spherical Beta (SB)~\cite{liu2025deformablebetasplatting}, are included. In both 3D/2D splatting, our approach is the \emph{best or second-best} in all quality metrics across all benchmarks. Please also see the supplemental material for a detailed, per-scene breakdown. For qualitative comparisons between some top performing approaches, please refer to~\cref{fig:main_compact} as well as the accompanying video.



\subsubsection{Visualization.} 
\begin{figure}[htbp]
\centering
\setlength{\tabcolsep}{0pt} 
\newcommand{\myrow}[7]{
    \begin{minipage}{\linewidth}
        \centering
        \begin{minipage}{.05\linewidth}
            \rotatebox{90}{\scriptsize #1}
        \end{minipage}
        \begin{minipage}{.15\linewidth} \includegraphics[width=\linewidth]{#2} \end{minipage} \hfill
        \begin{minipage}{.15\linewidth} \includegraphics[width=\linewidth]{#3} \end{minipage} \hfill
        \begin{minipage}{.15\linewidth} \includegraphics[width=\linewidth]{#4} \end{minipage} \hfill
        \begin{minipage}{.15\linewidth} \includegraphics[width=\linewidth]{#5} \end{minipage} \hfill
        \begin{minipage}{.15\linewidth} \includegraphics[width=\linewidth]{#6} 
        \end{minipage} \hfill
        \begin{minipage}{.15\linewidth} \includegraphics[width=\linewidth]{#7} \end{minipage}
    \end{minipage}  
}

\myrow{Ours~($k=2$)}{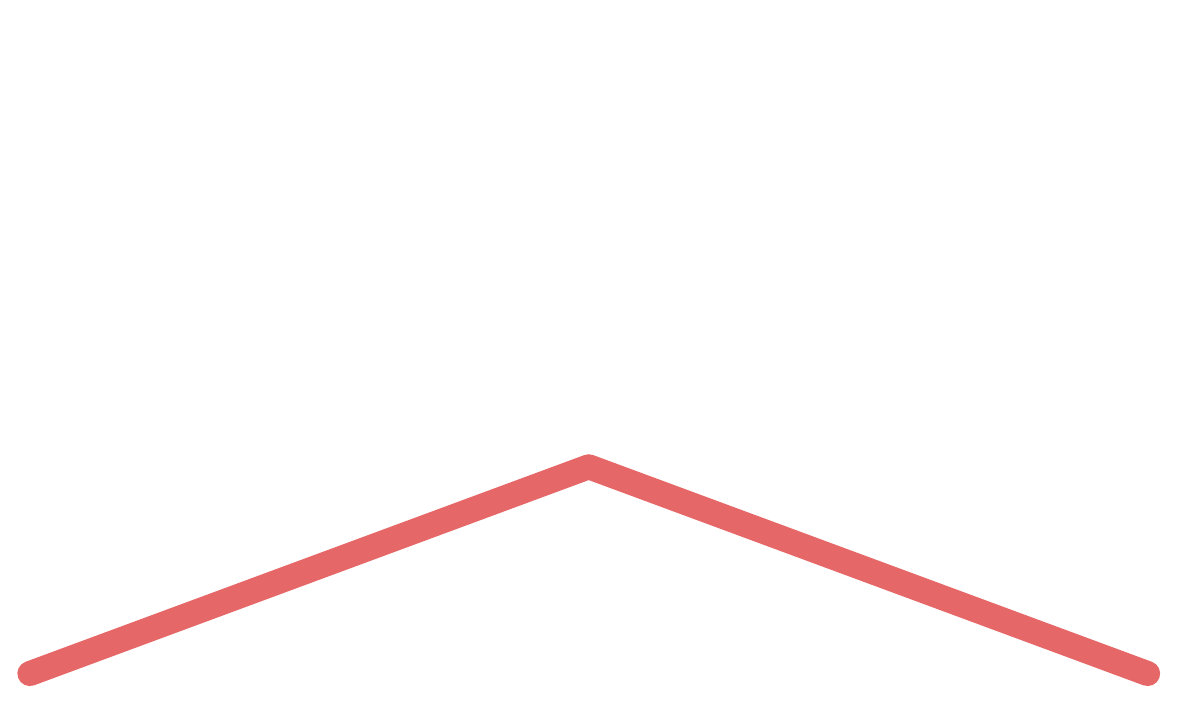}{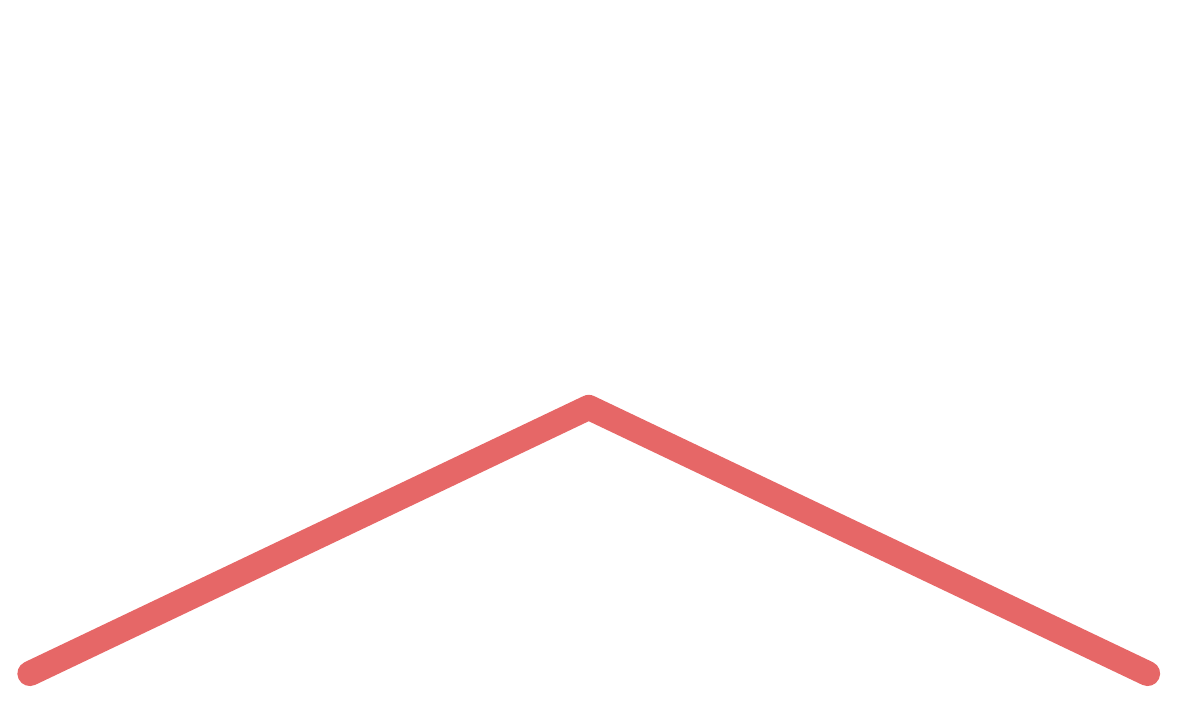}{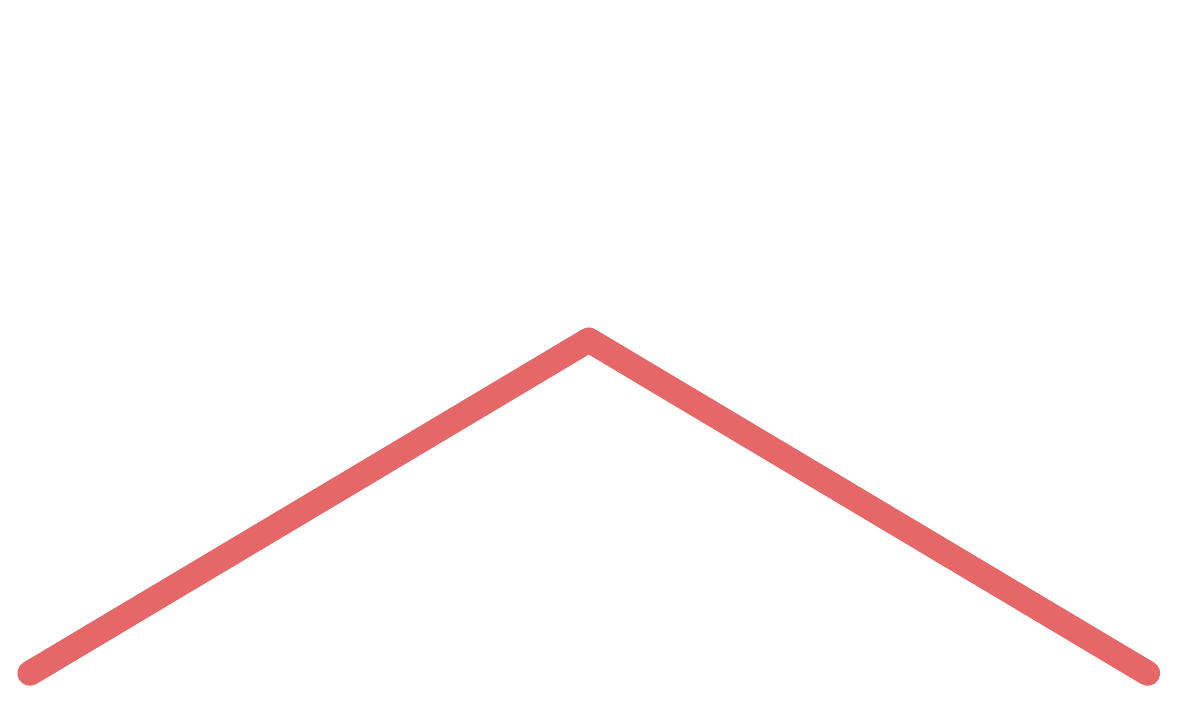}{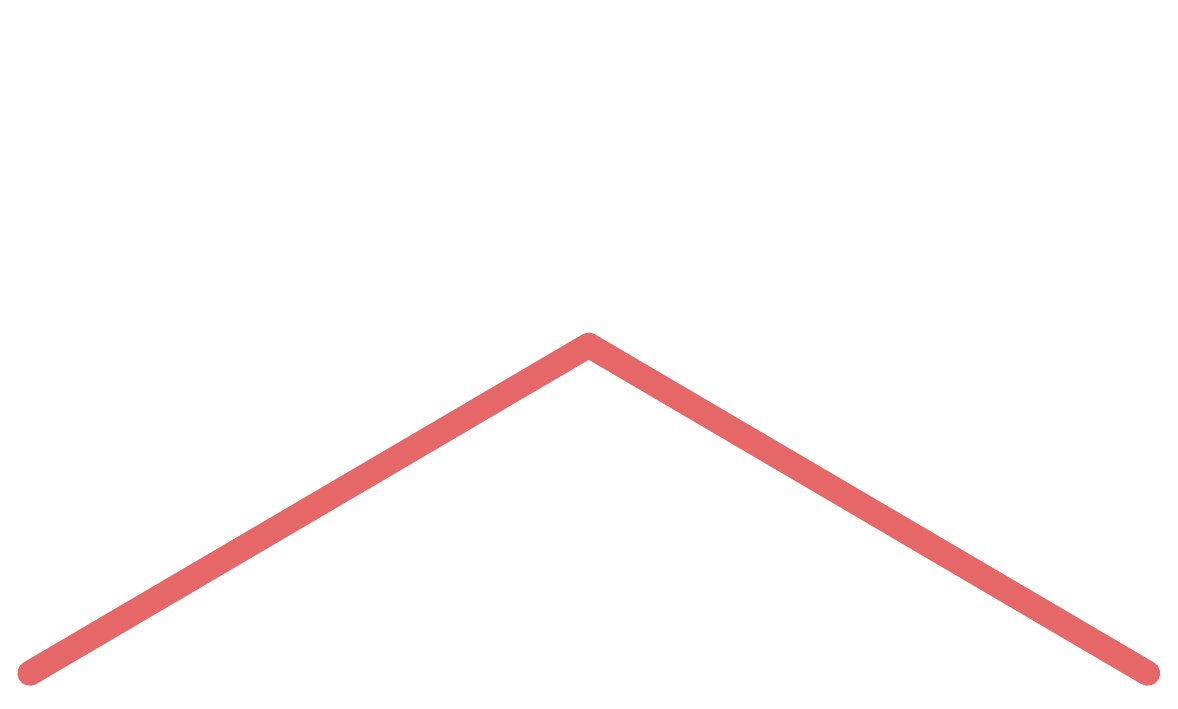}{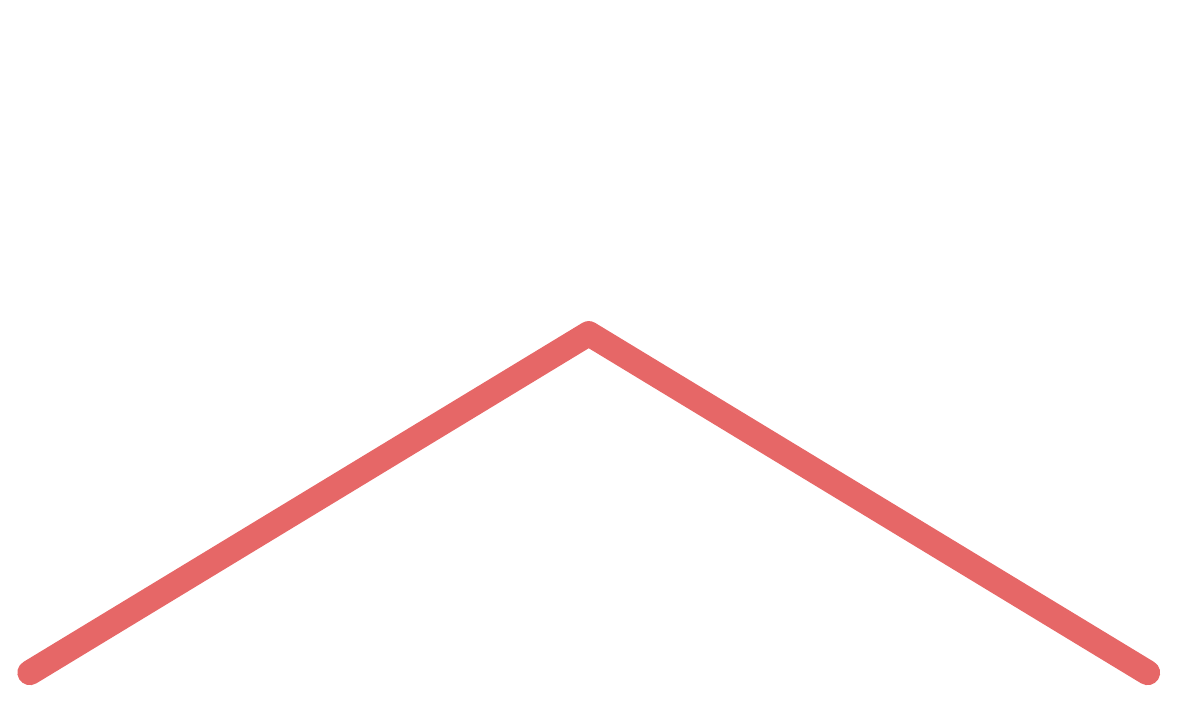}{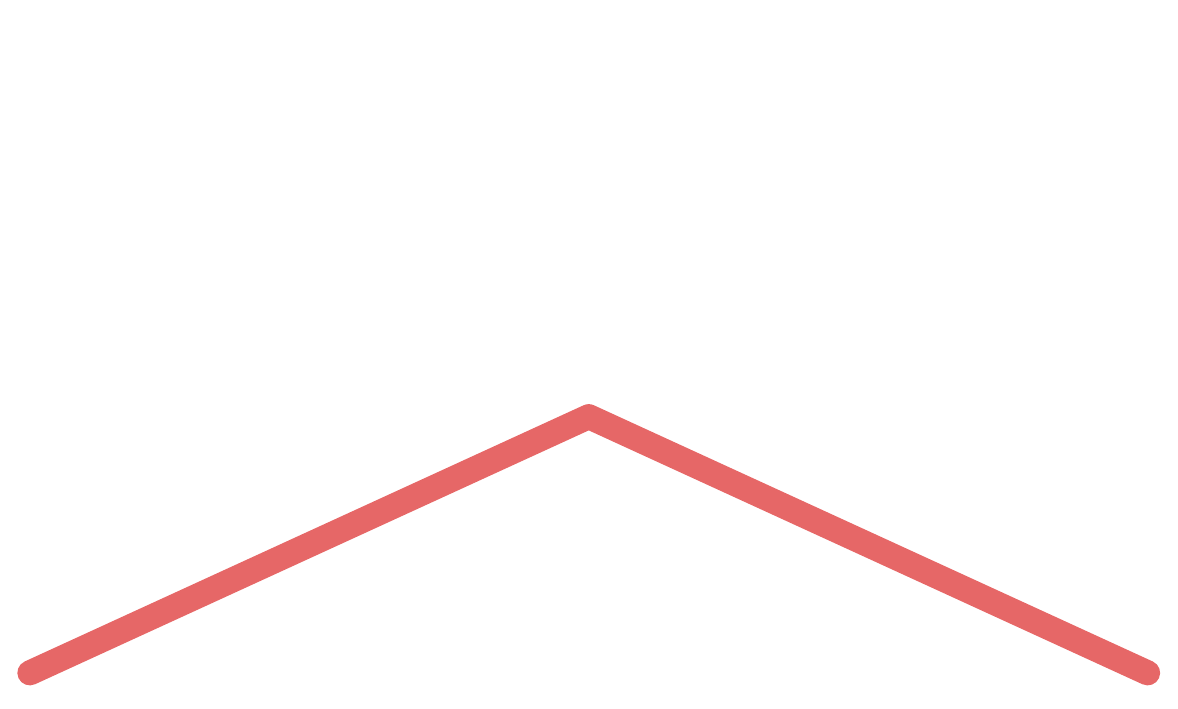}
\myrow{Ours~($k=4$)}{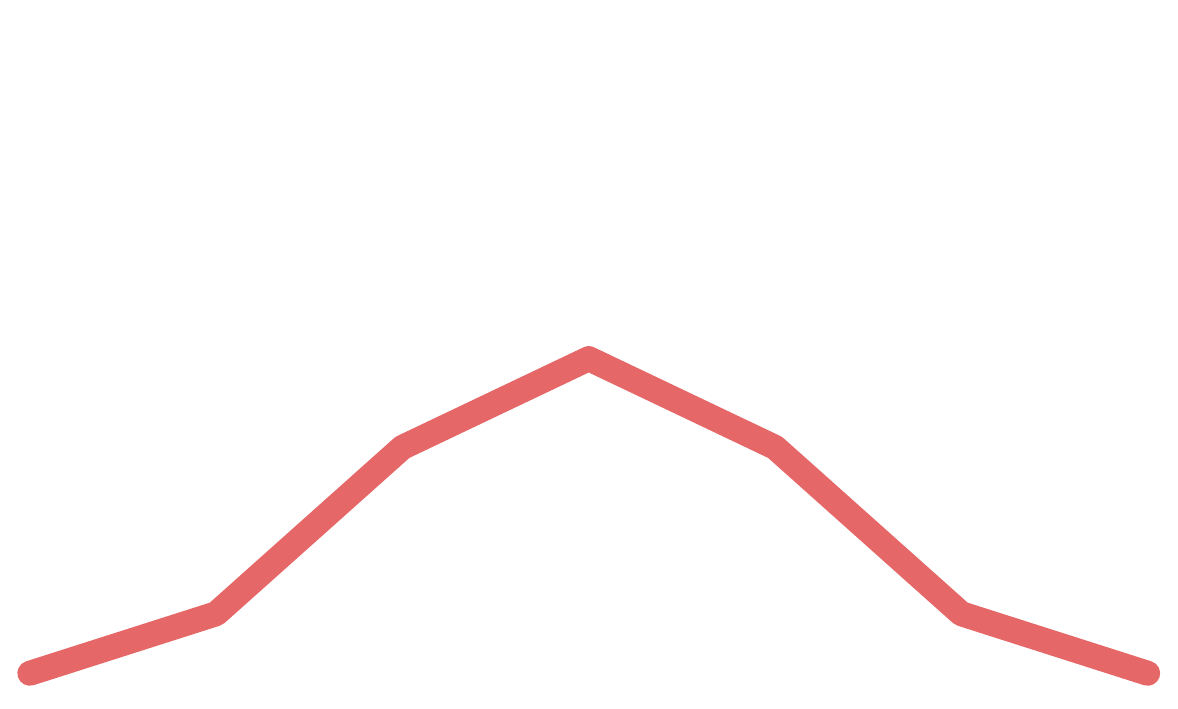}{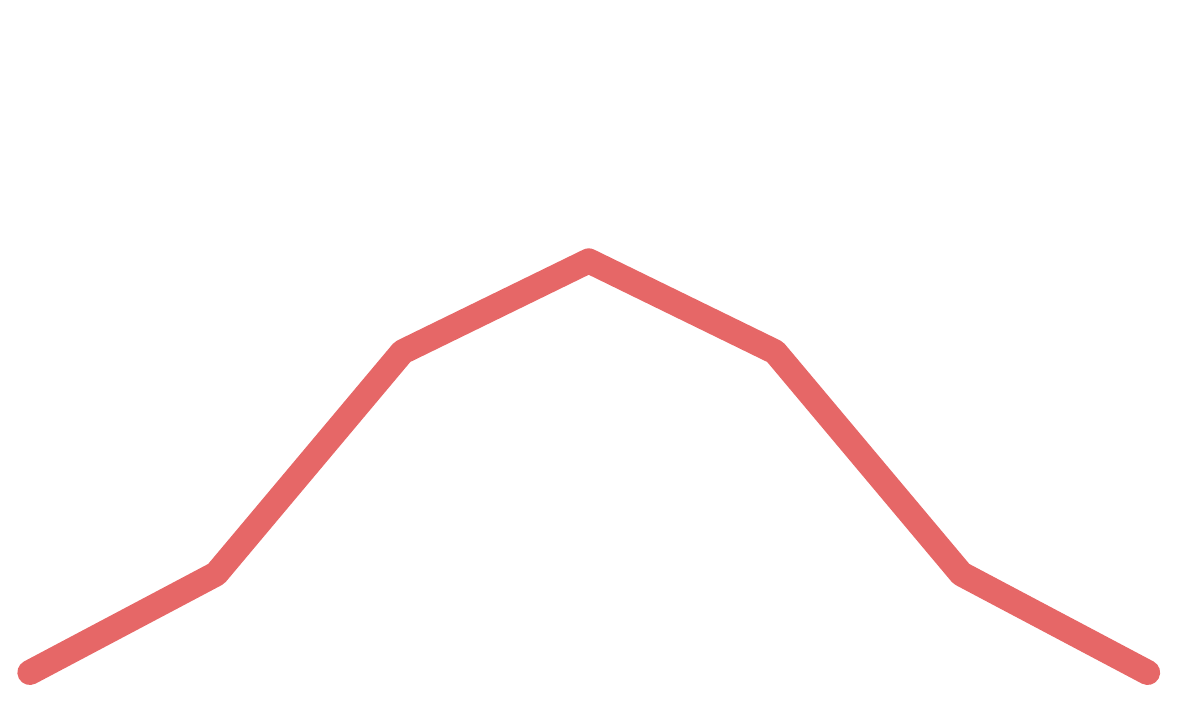}{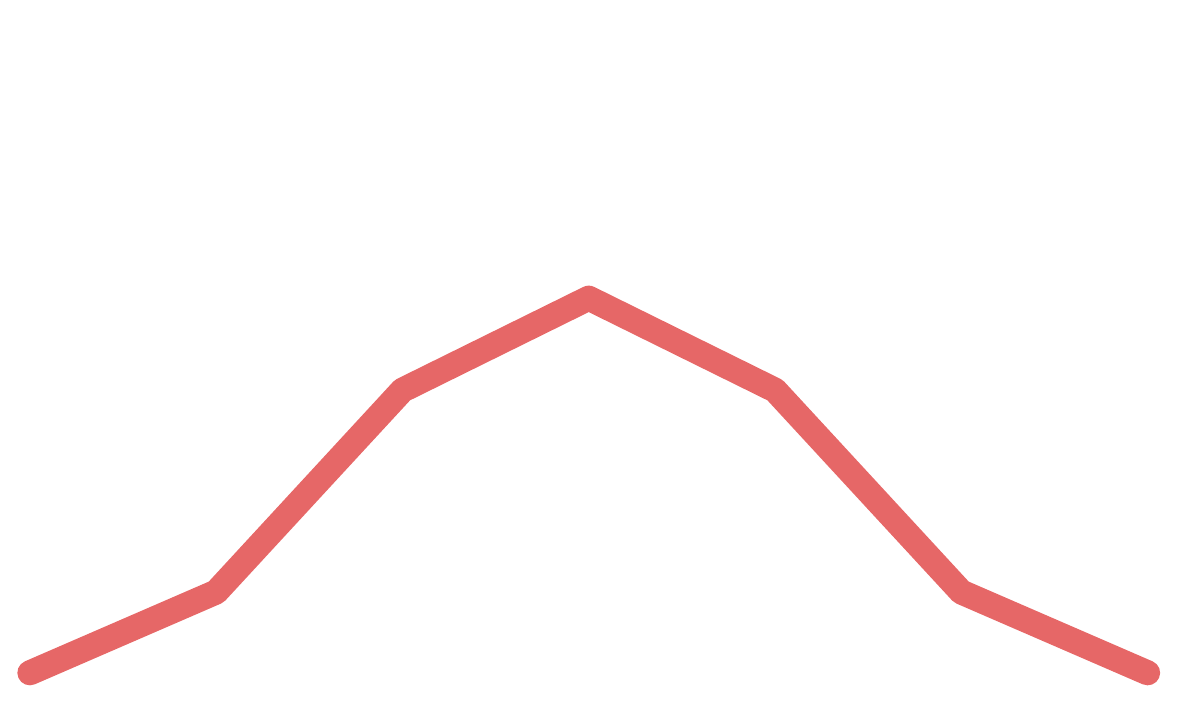}{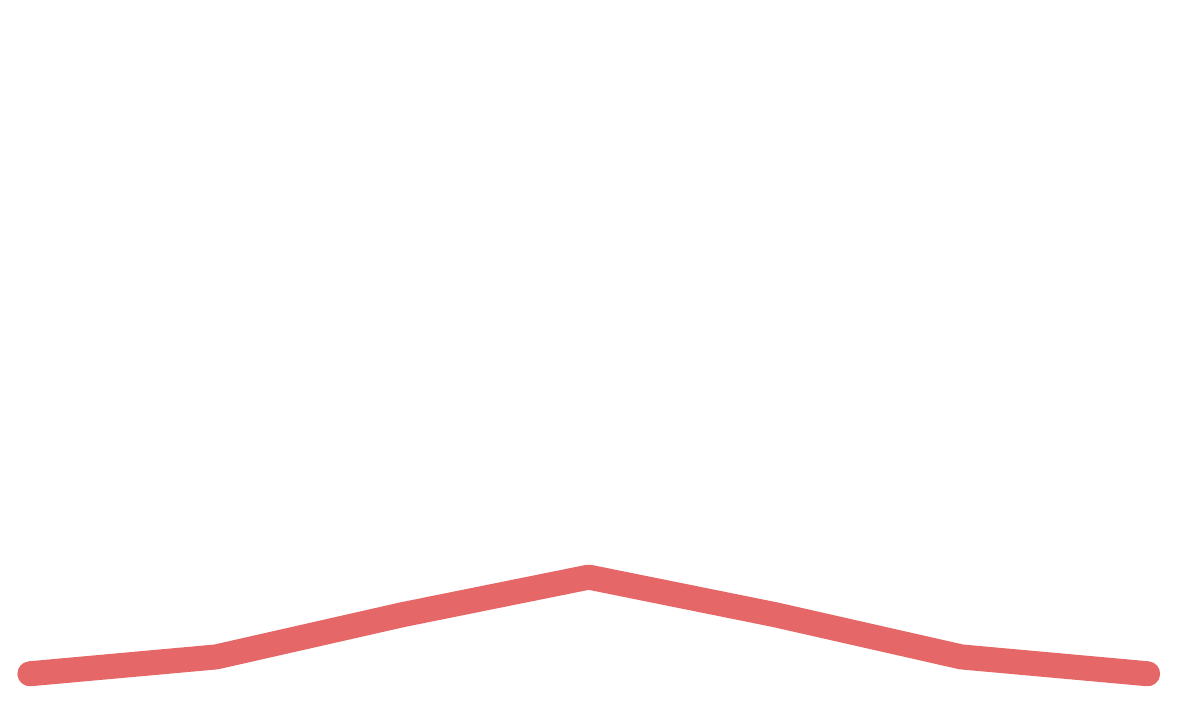}{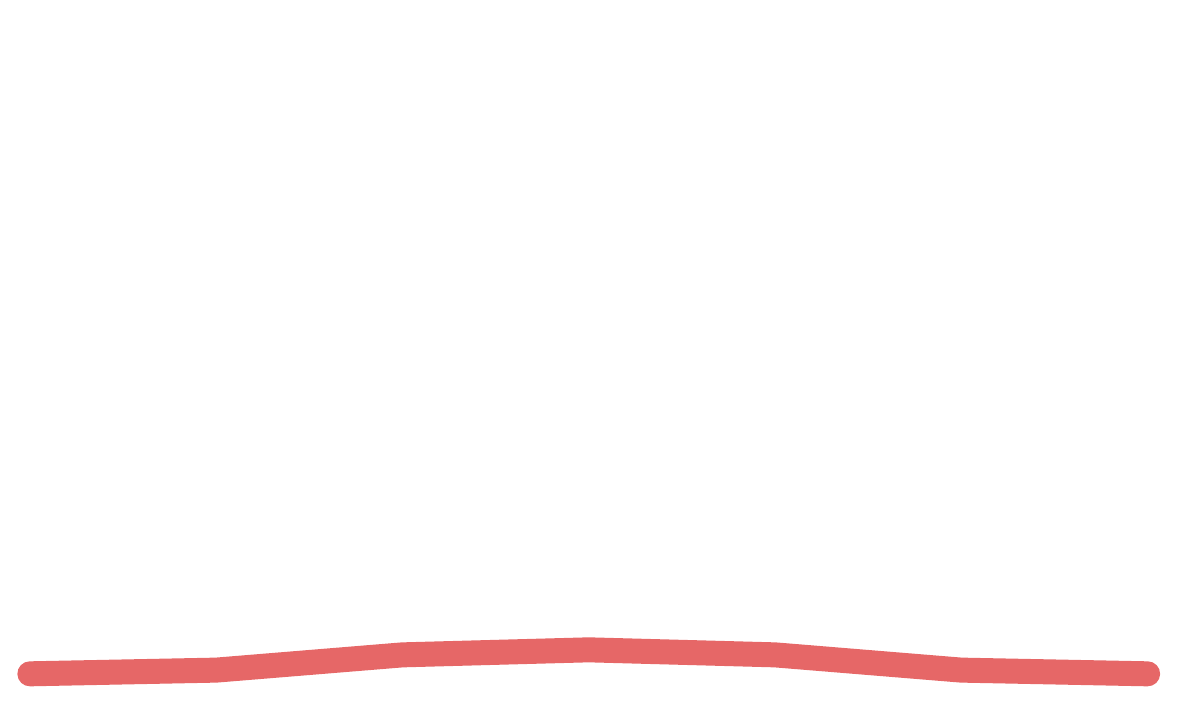}{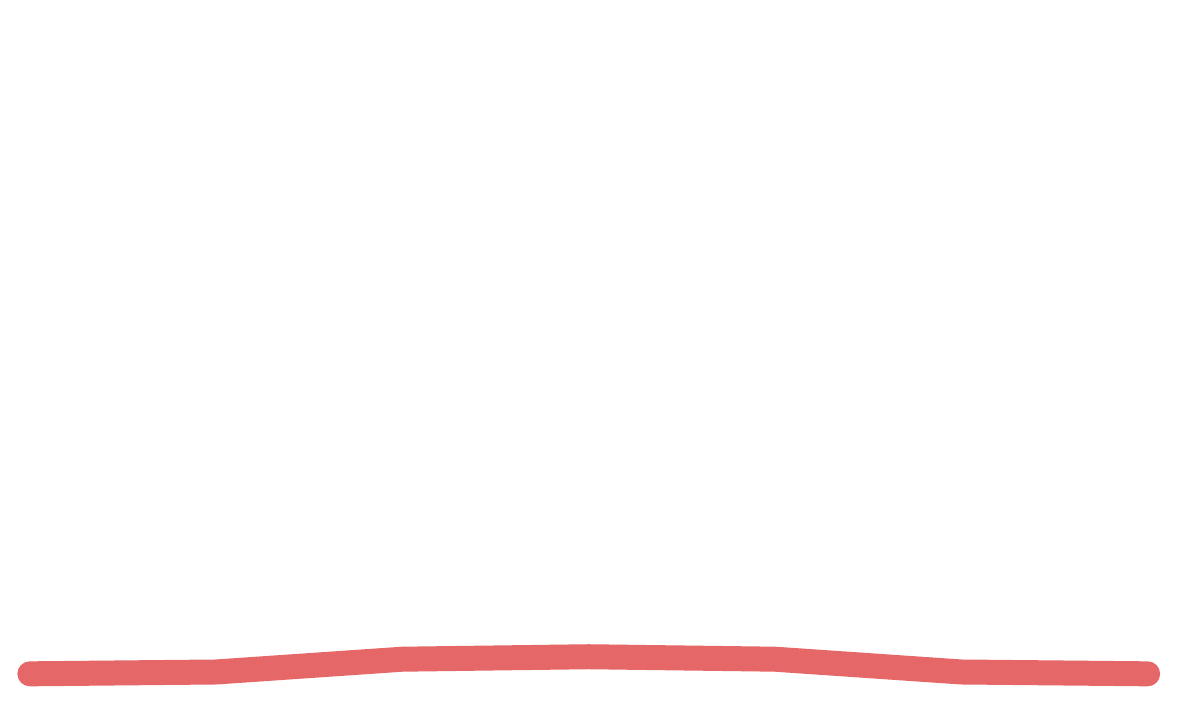}
\myrow{Ours~($k=8$)}{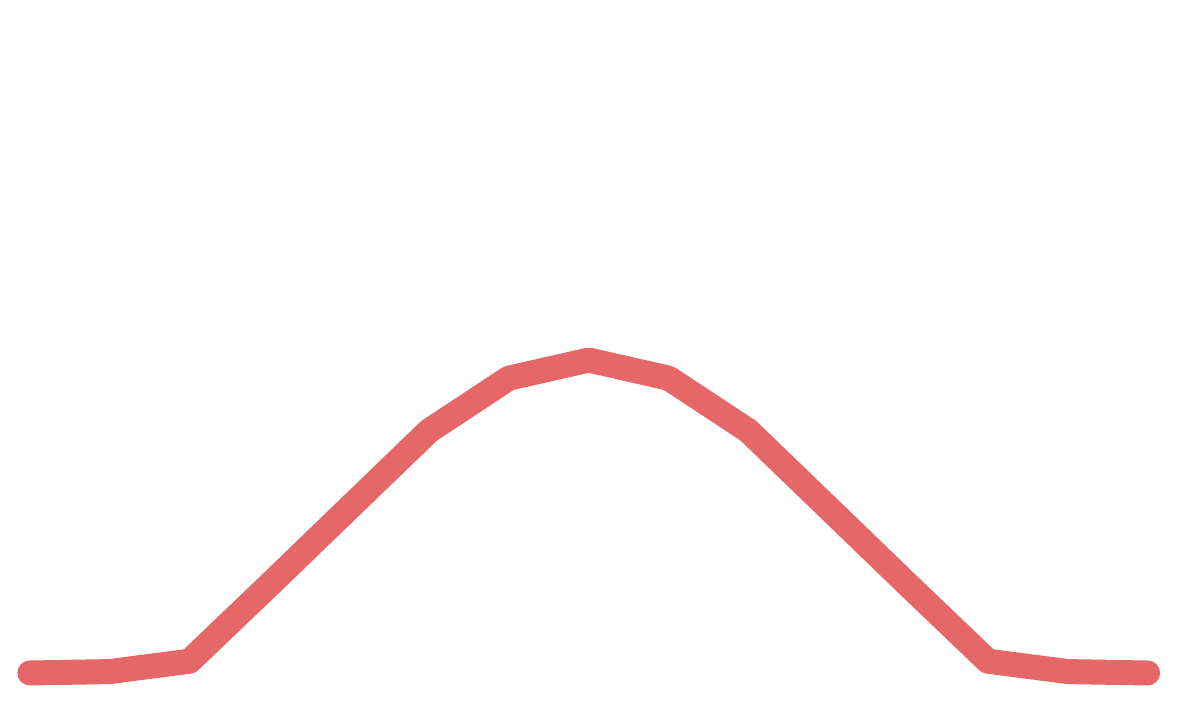}{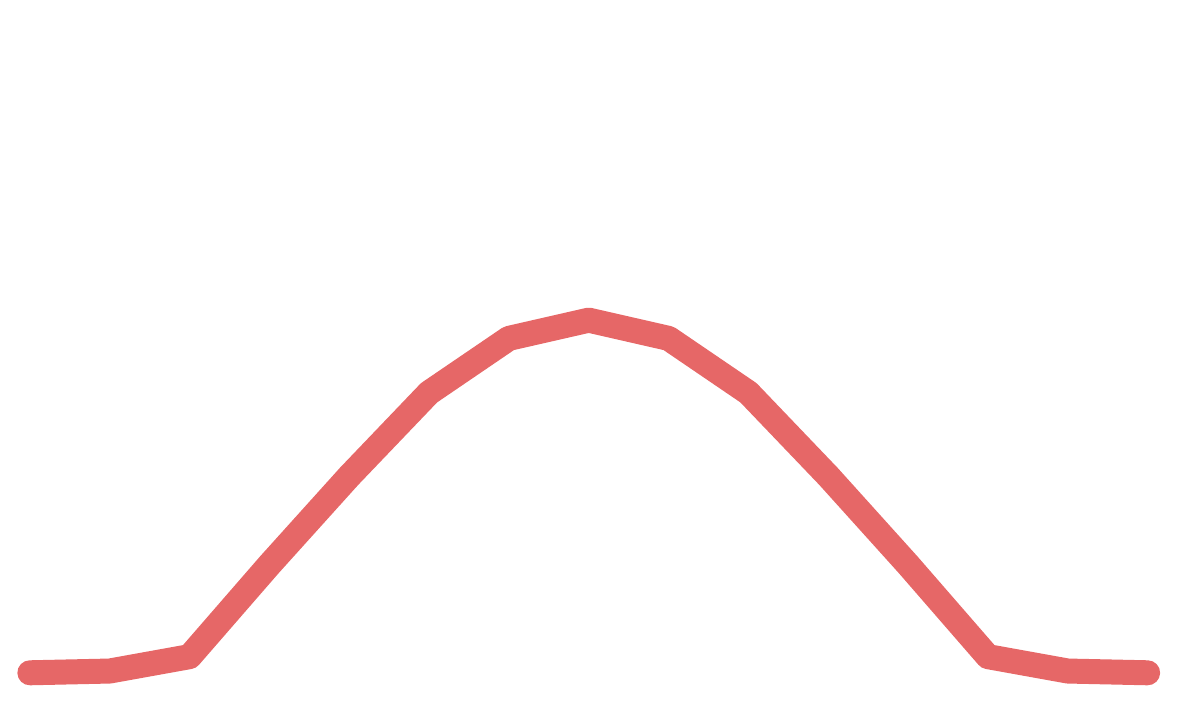}{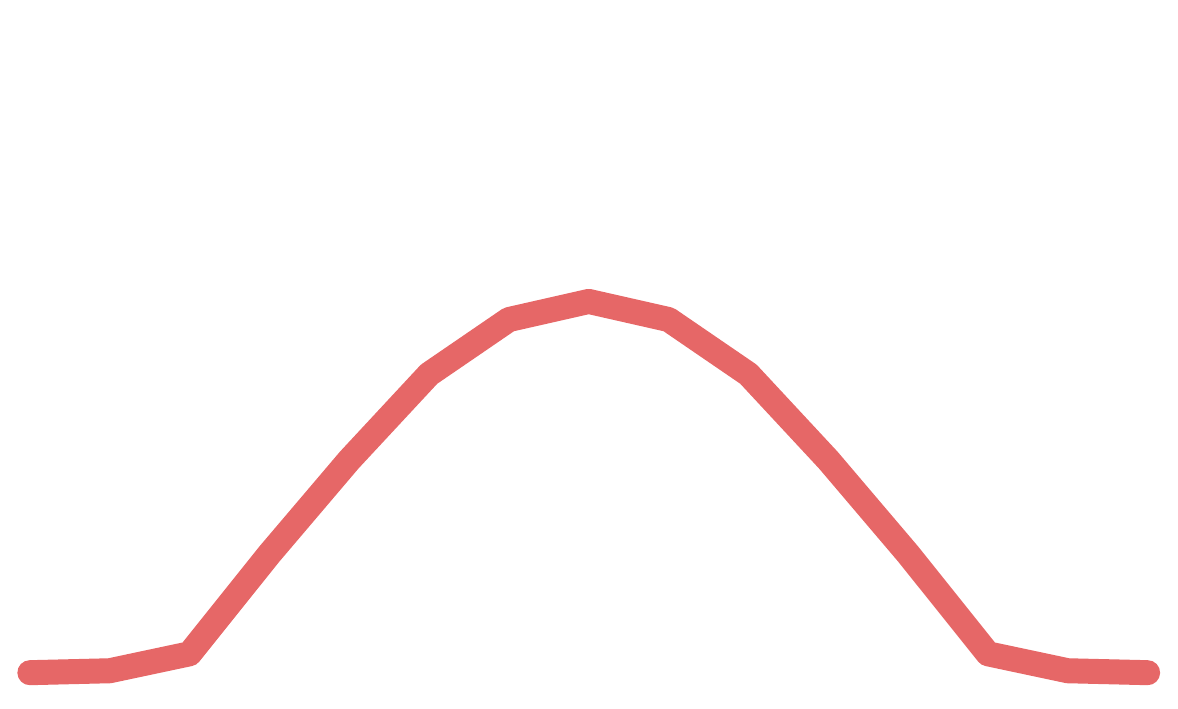}{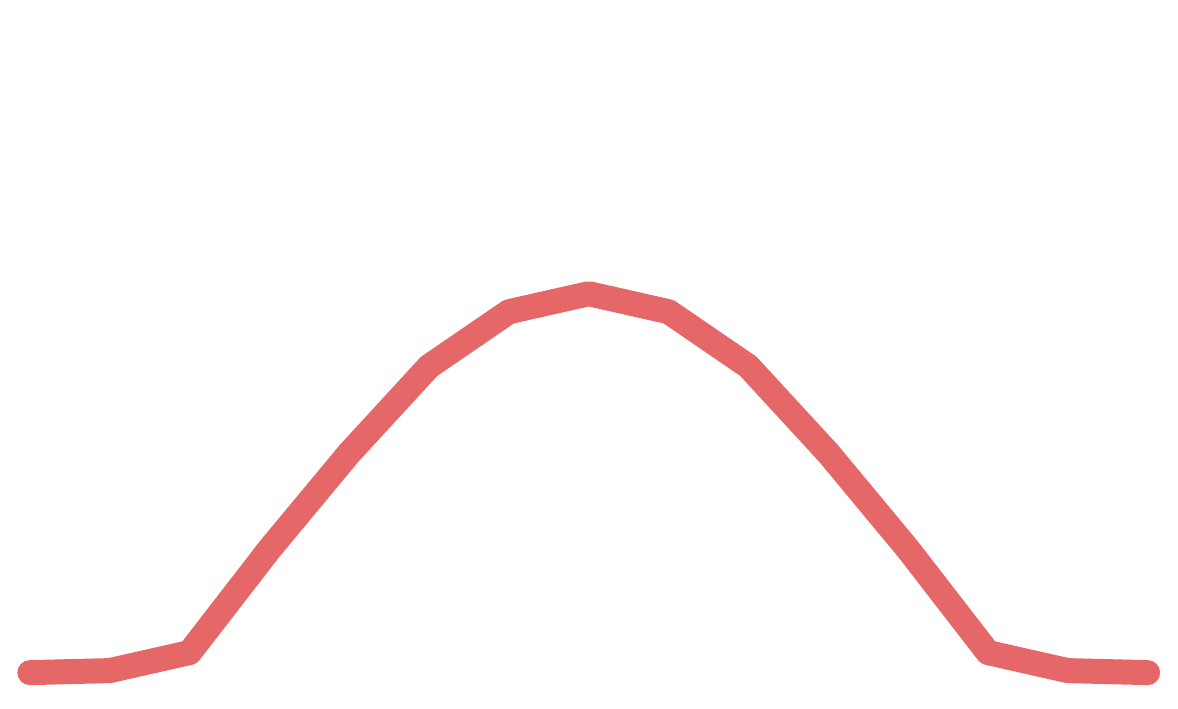}{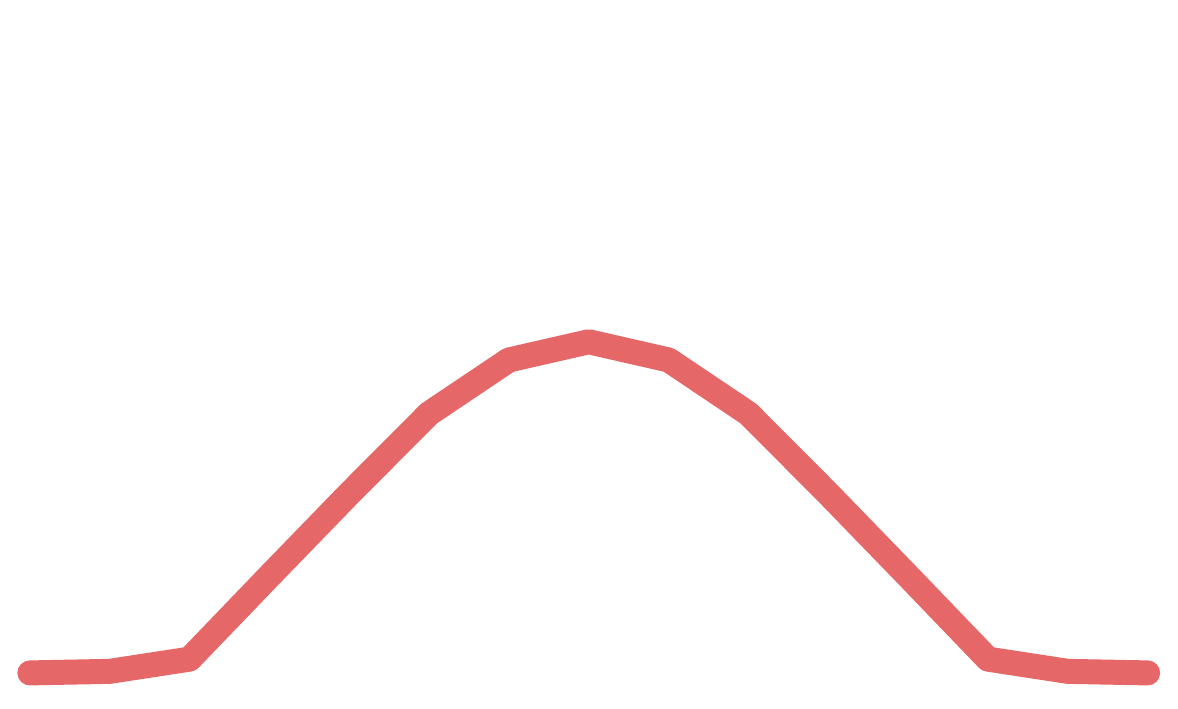}{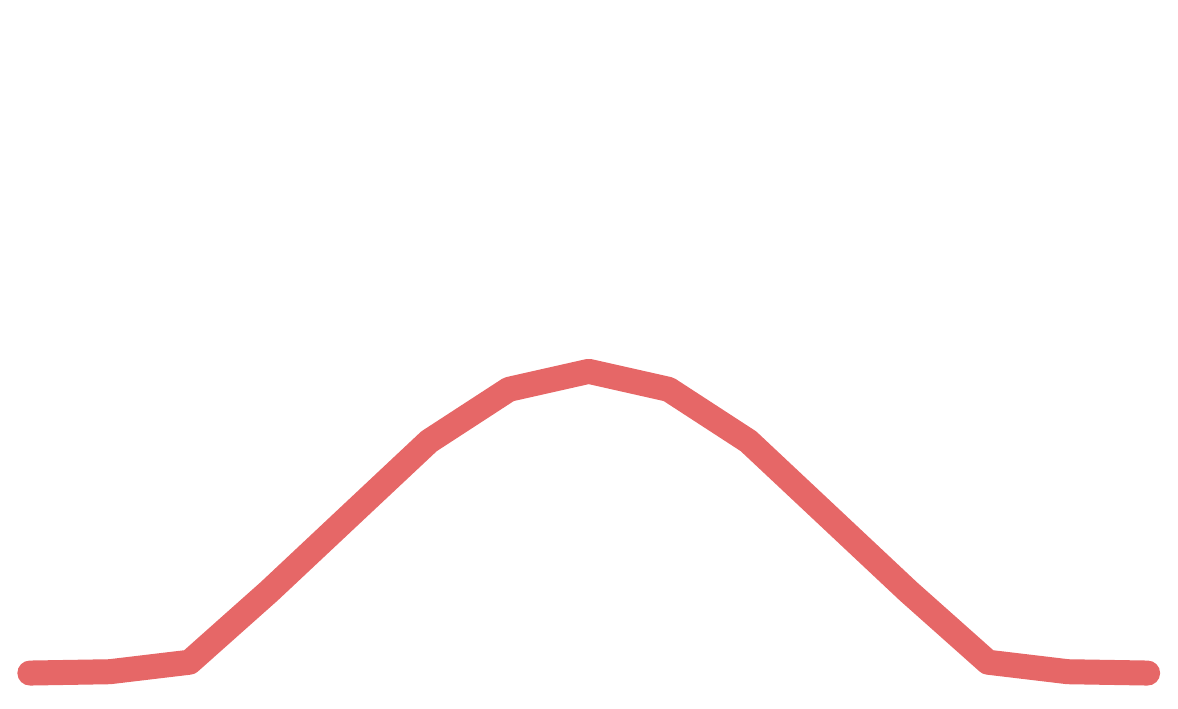}
\myrow{Gaussian}{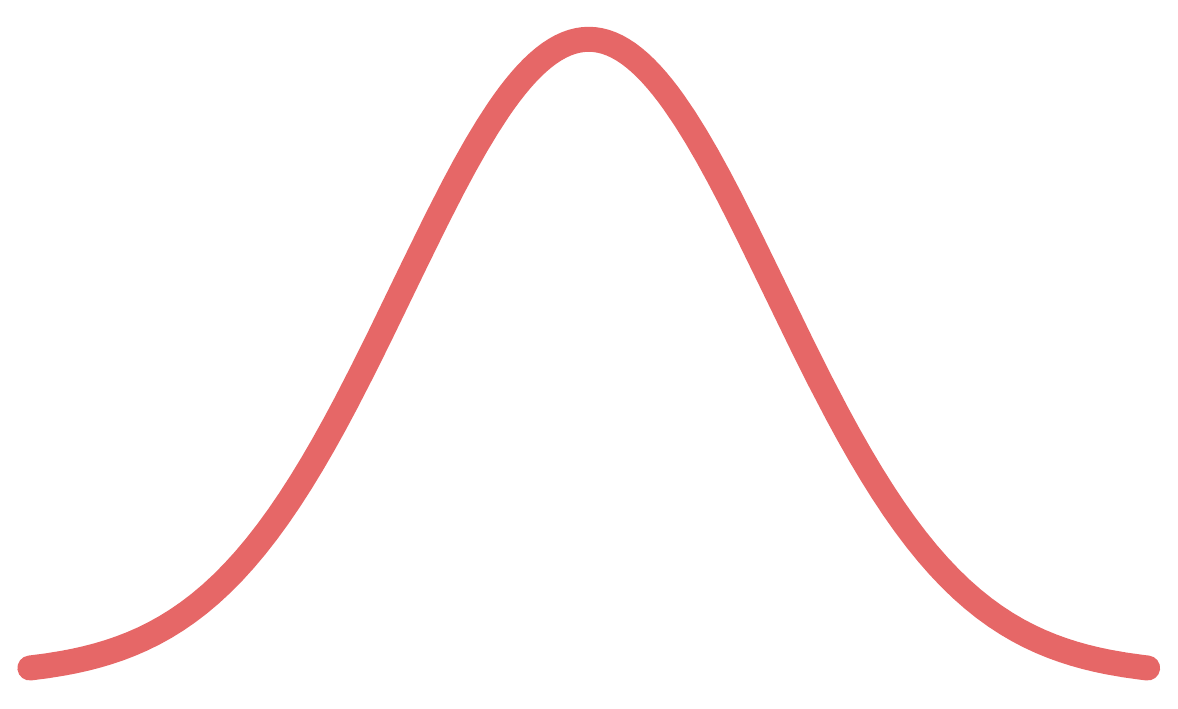}{imgs/comparison/room/gs_curve_9845.pdf}{imgs/comparison/room/gs_curve_9845.pdf}{imgs/comparison/room/gs_curve_9845.pdf}{imgs/comparison/room/gs_curve_9845.pdf}{imgs/comparison/room/gs_curve_9845.pdf}
\myrow{Student's t}{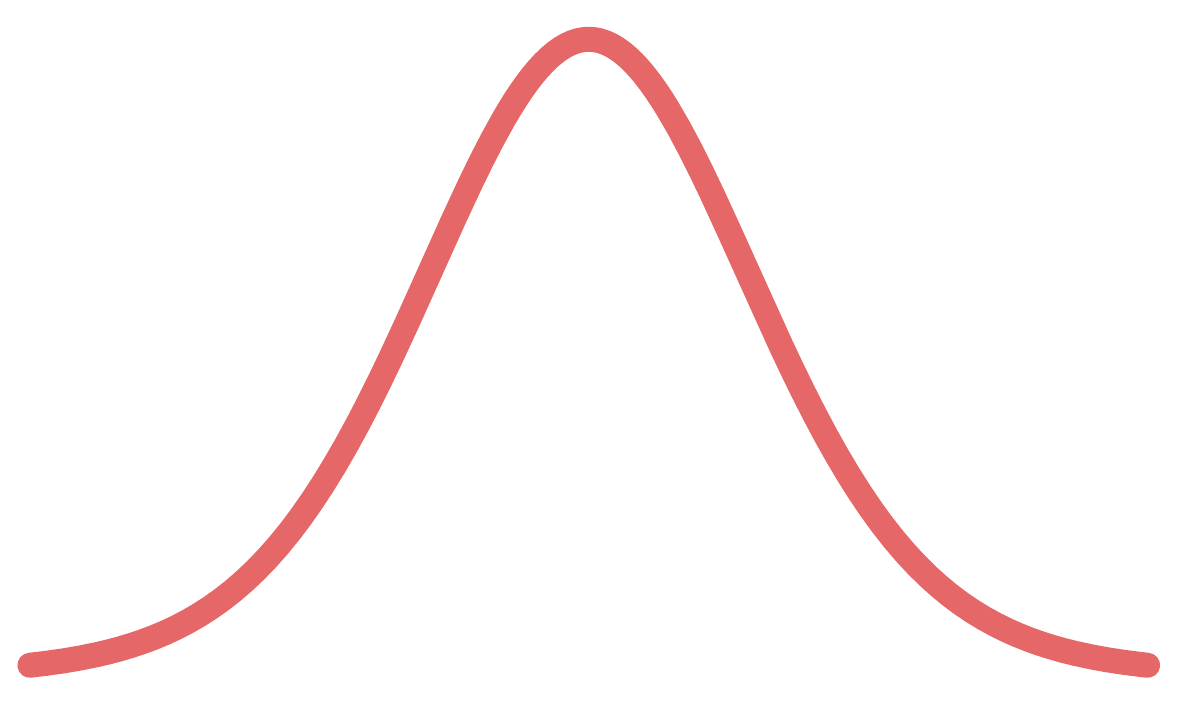}{imgs/comparison/room/sss_curve_9845.pdf}{imgs/comparison/room/sss_curve_9845.pdf}{imgs/comparison/room/sss_curve_9845.pdf}{imgs/comparison/room/sss_curve_9845.pdf}{imgs/comparison/room/sss_curve_9845.pdf}
\myrow{Beta}{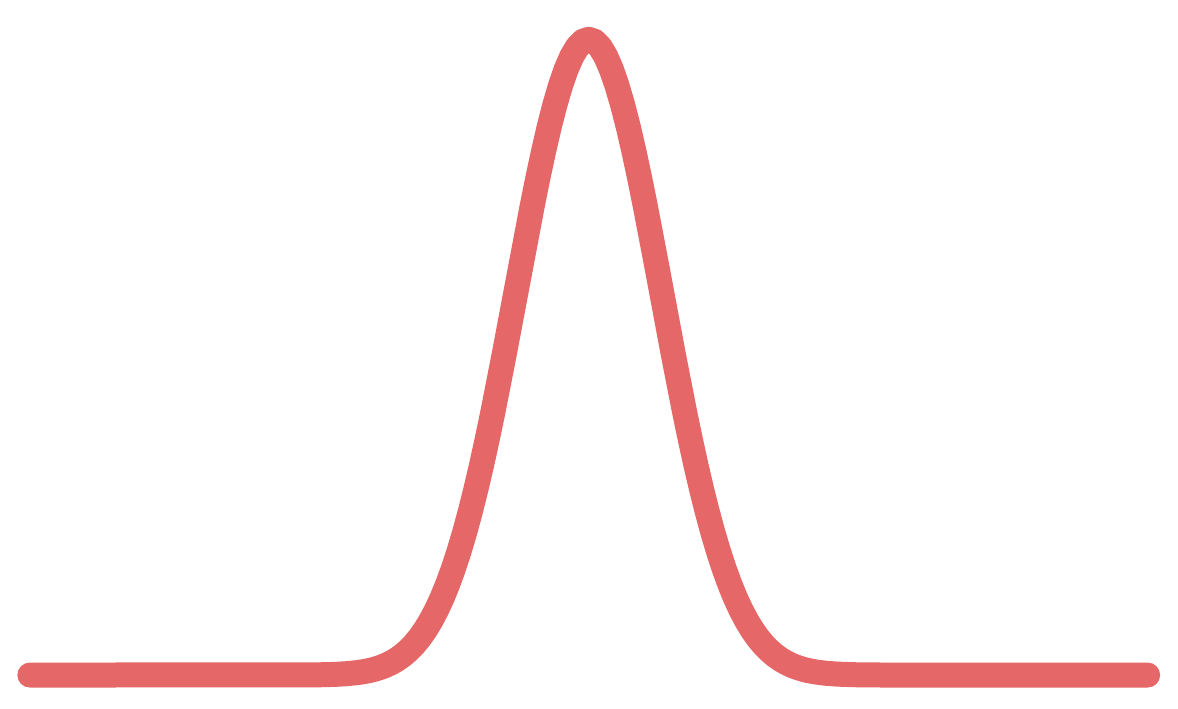}{imgs/comparison/room/beta_curve_9845.pdf}{imgs/comparison/room/beta_curve_9845.pdf}{imgs/comparison/room/beta_curve_9845.pdf}{imgs/comparison/room/beta_curve_9845.pdf}{imgs/comparison/room/beta_curve_9845.pdf}
\myrow{Corr. View}{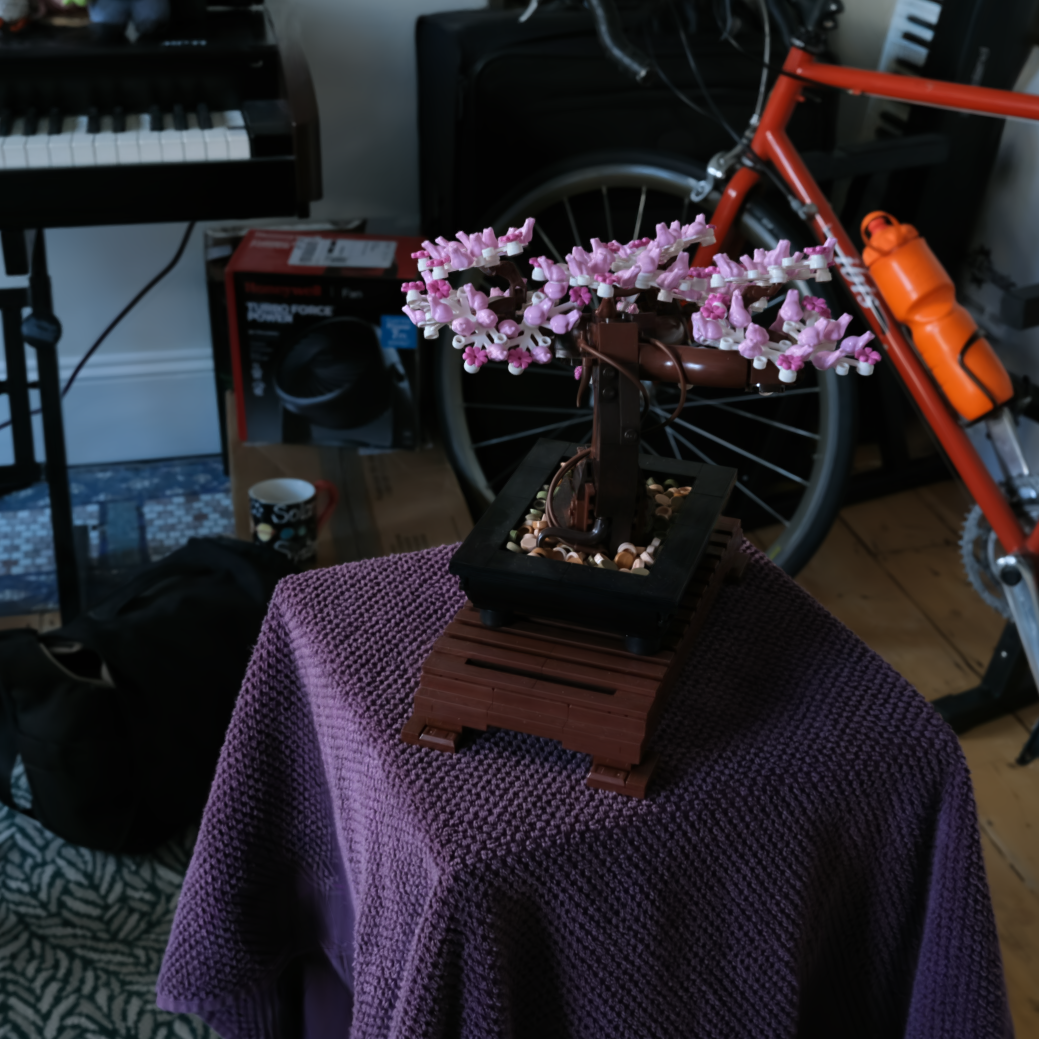}{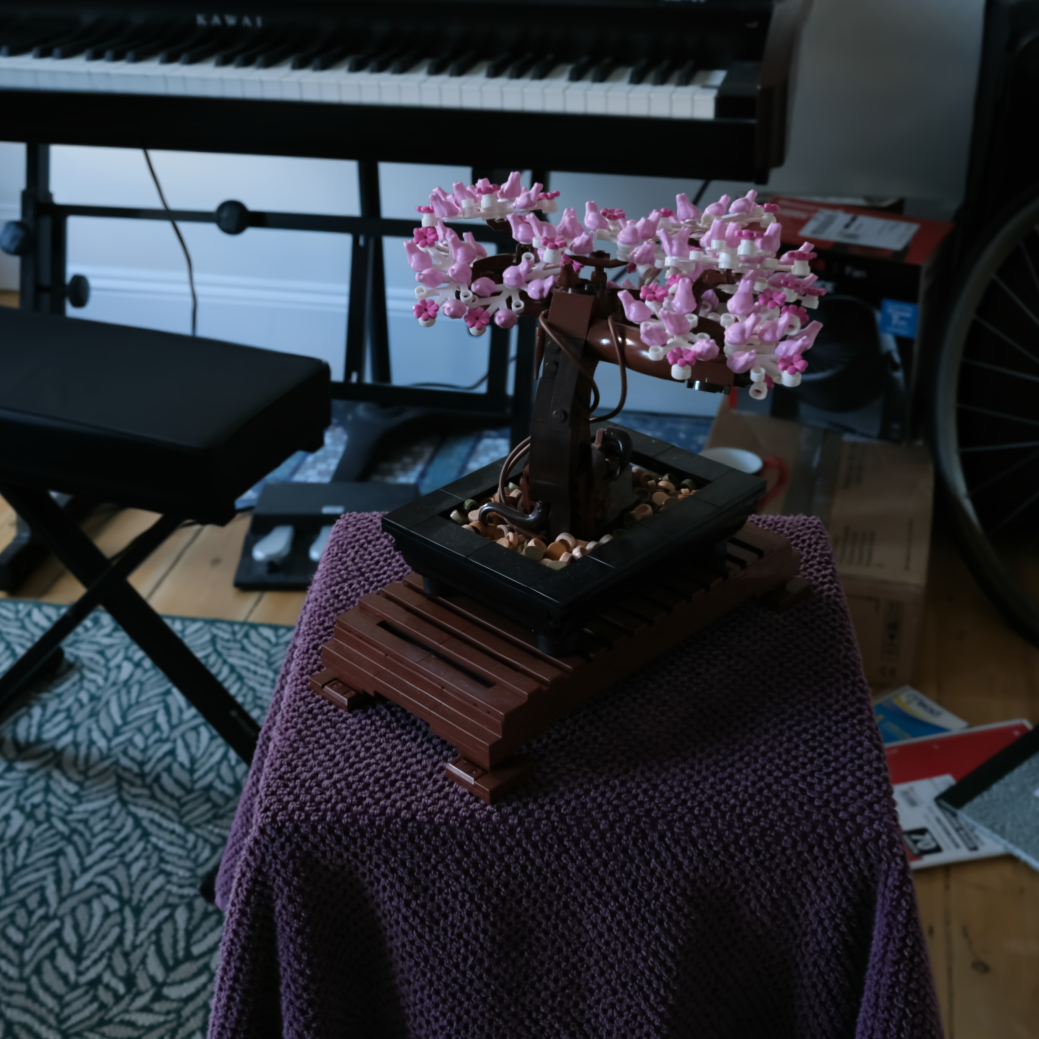}{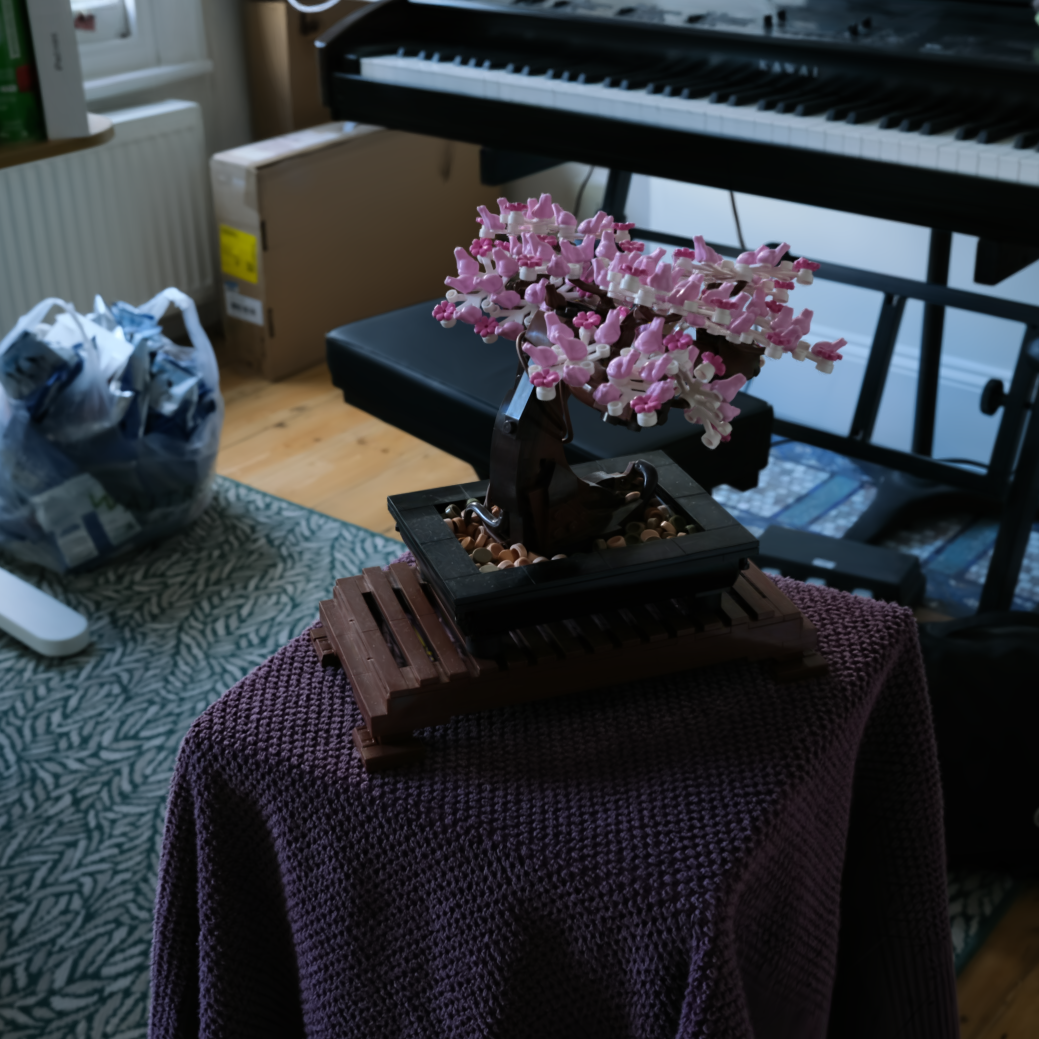}{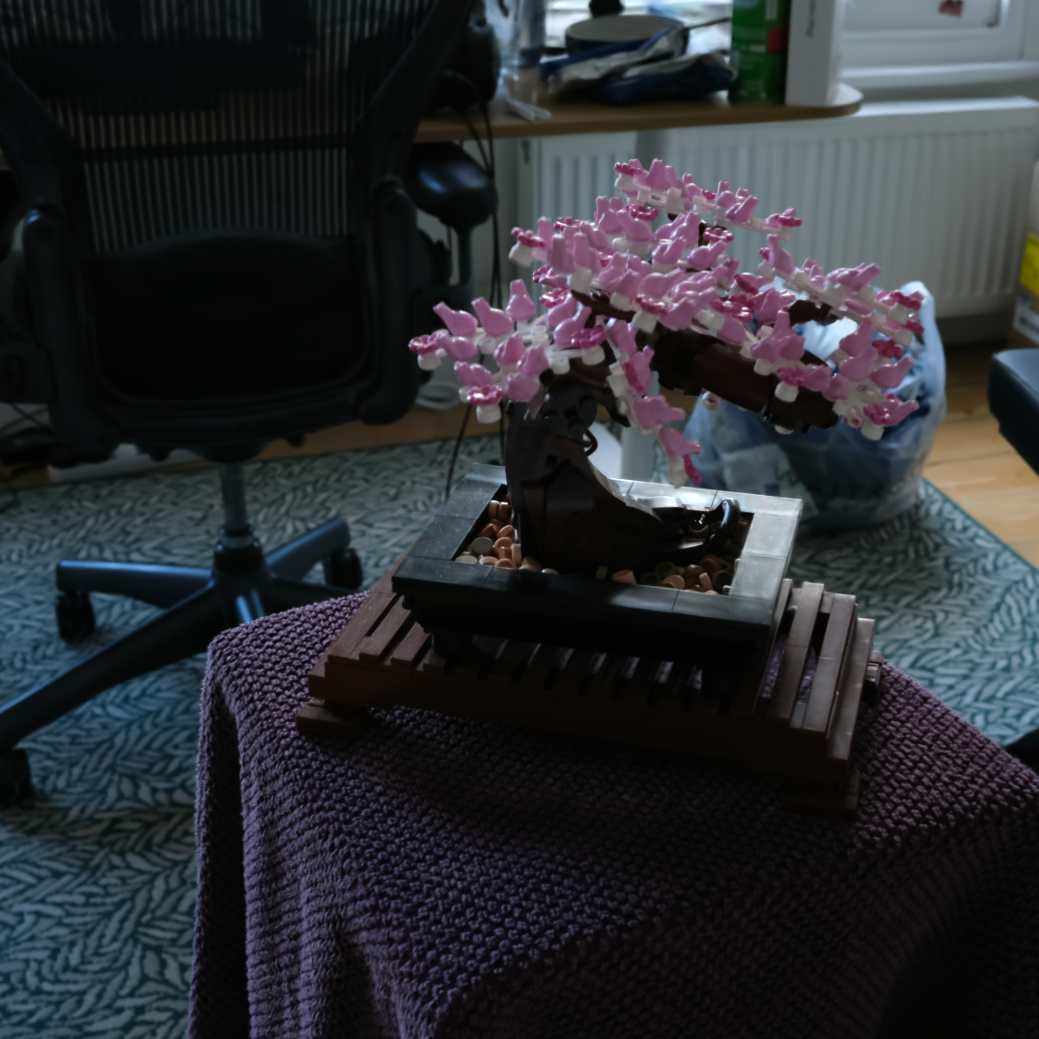}{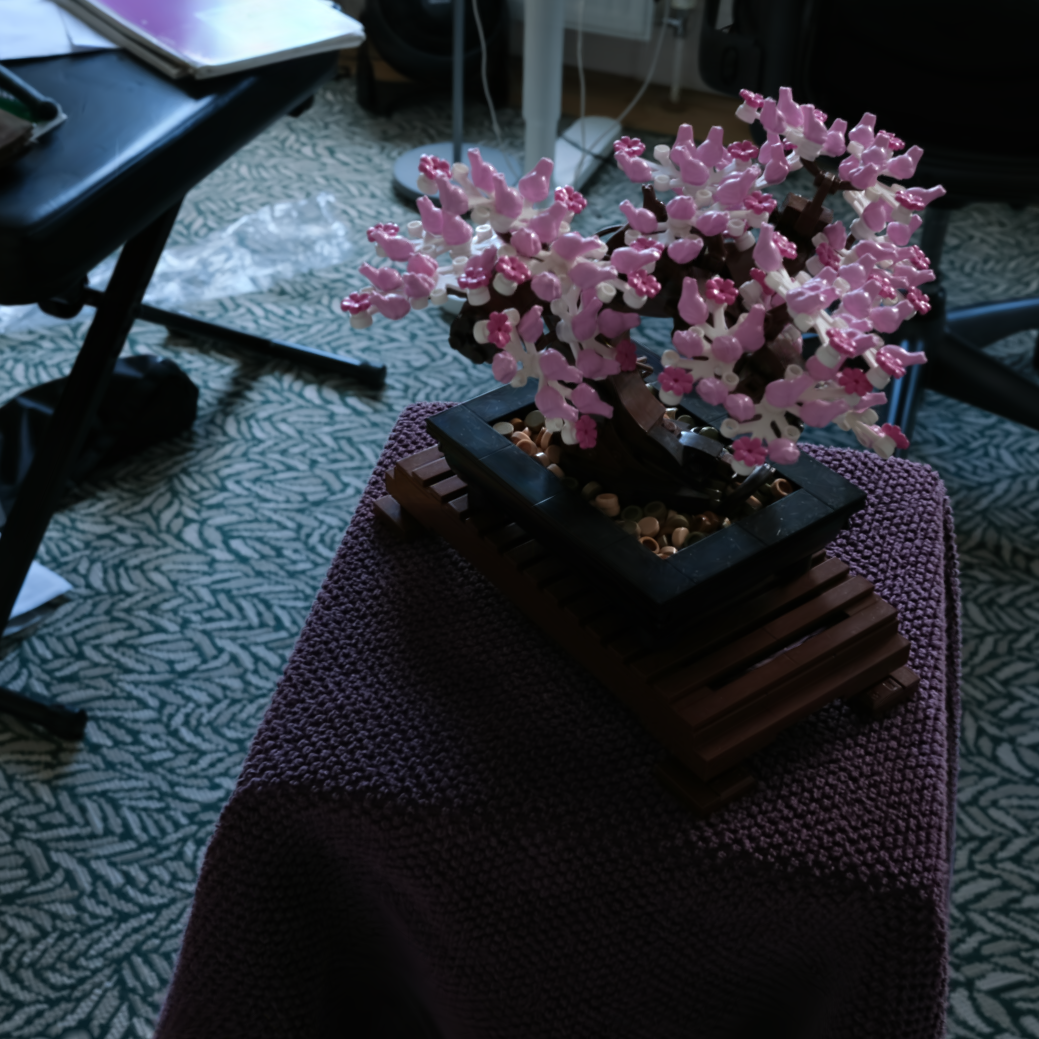}{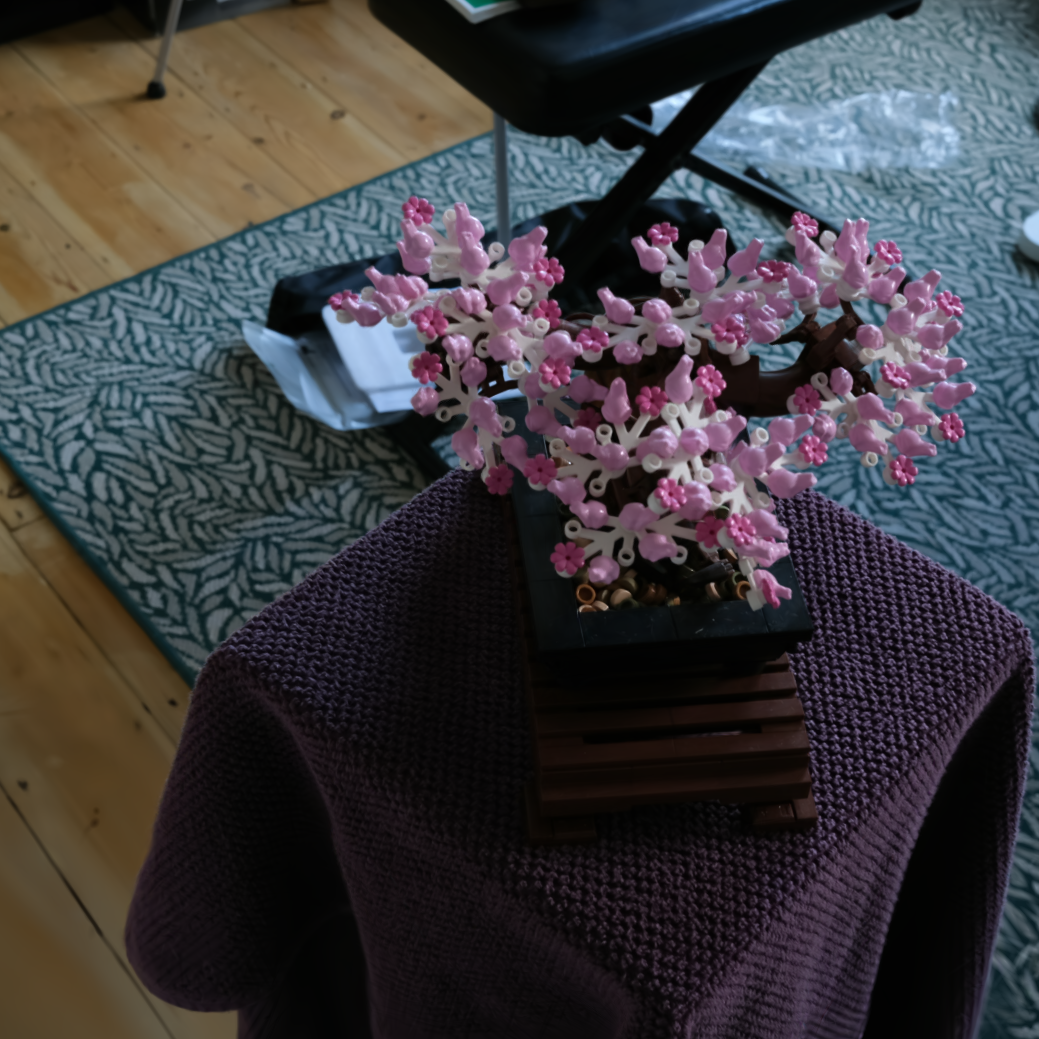}

    \caption{Visualization of 1D profiles of different radially symmetric kernels. From the top row to bottom, our kernels ($k$=2, 4, 8), Gaussian~\cite{kerbl3Dgaussians}, Student's t~\cite{zhu2025sss} and Beta kernel~\cite{liu2025deformablebetasplatting}, and the corresponding views. All profiles in a column correspond to the same view shown at the bottom row. Note that our kernels vary with the view, while the kernels in existing work are view-independent.}
    
    \label{fig:kernel_comparison}
\end{figure}
We first visualize the 1D profiles of individual radially symmetric kernels in different approaches, including variants of ours with different $k$, vanilla 3DGS, SSS and DBS in~\cref{fig:kernel_comparison}. Here one key difference is that our kernels are \emph{view-dependent} by learning, while the ones (prior to any affine transformations described in Sec.~\ref{sec:preliminaries}) in existing work are view-independent.

\begin{figure}[htbp]
    \centering
    \setlength{\tabcolsep}{0pt}
    
    \newcommand{\myrowSix}[7]{
        \begin{minipage}{\linewidth}
            \centering
            \begin{minipage}{.03\linewidth}
                \centering 
                \rotatebox{90}{\scriptsize #1} 
            \end{minipage}%
            \hfill 
            %
            \begin{minipage}{.155\linewidth} \includegraphics[width=\linewidth]{#2} \end{minipage}\hfill
            \begin{minipage}{.155\linewidth} \includegraphics[width=\linewidth]{#3} \end{minipage}\hfill
            \begin{minipage}{.155\linewidth} \includegraphics[width=\linewidth]{#4} \end{minipage}\hfill
            \begin{minipage}{.155\linewidth} \includegraphics[width=\linewidth]{#5} \end{minipage}\hfill
            \begin{minipage}{.155\linewidth} \includegraphics[width=\linewidth]{#6} \end{minipage}\hfill
            \begin{minipage}{.155\linewidth} \includegraphics[width=\linewidth]{#7} \end{minipage}
        \end{minipage} 
    }

    \myrowSix{\textsc{Bonsai}}{imgs/diffview/bonsai/crop_00002.png}{imgs/diffview/bonsai/crop_00005.png}{imgs/diffview/bonsai/crop_00008.png}{imgs/diffview/bonsai/crop_00011.png}{imgs/diffview/bonsai/crop_00014.png}{imgs/diffview/bonsai/crop_00017.png}
    
    \myrowSix{Histogram}{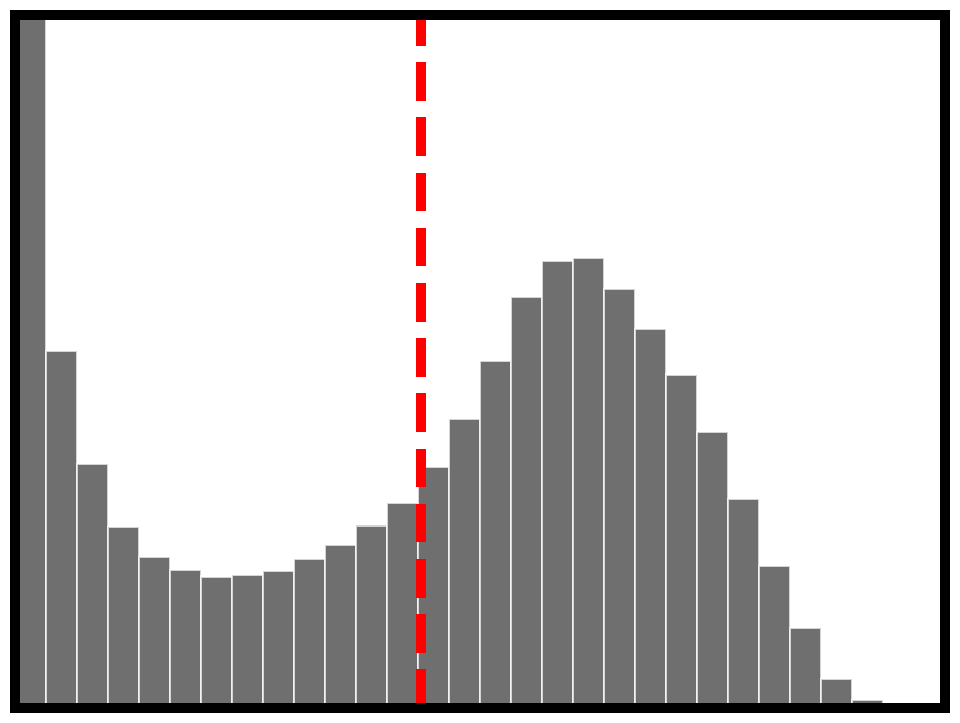}{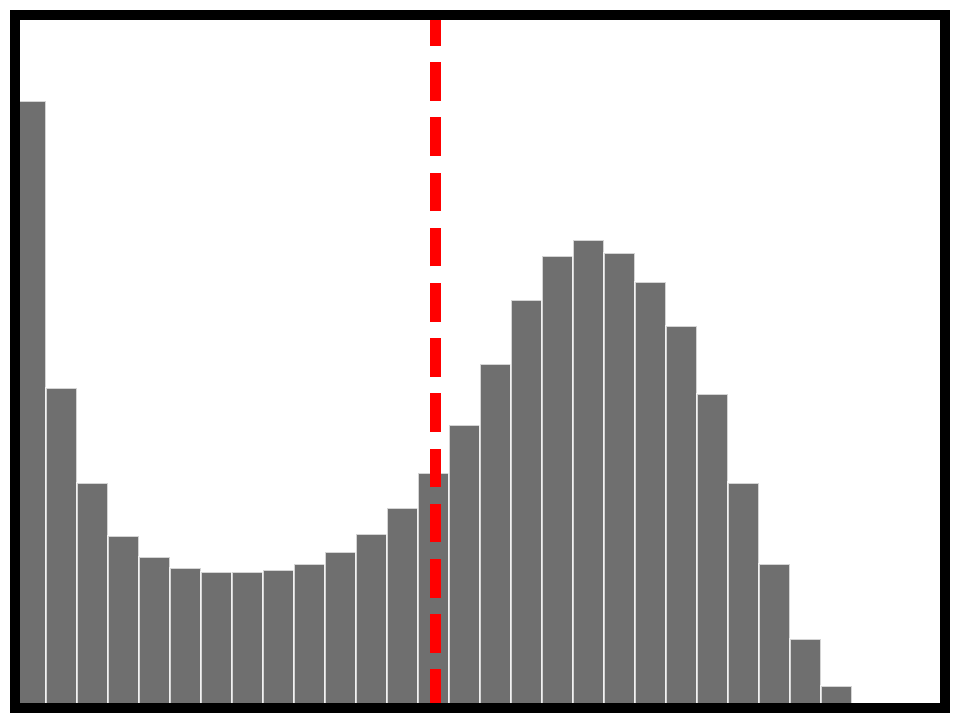}{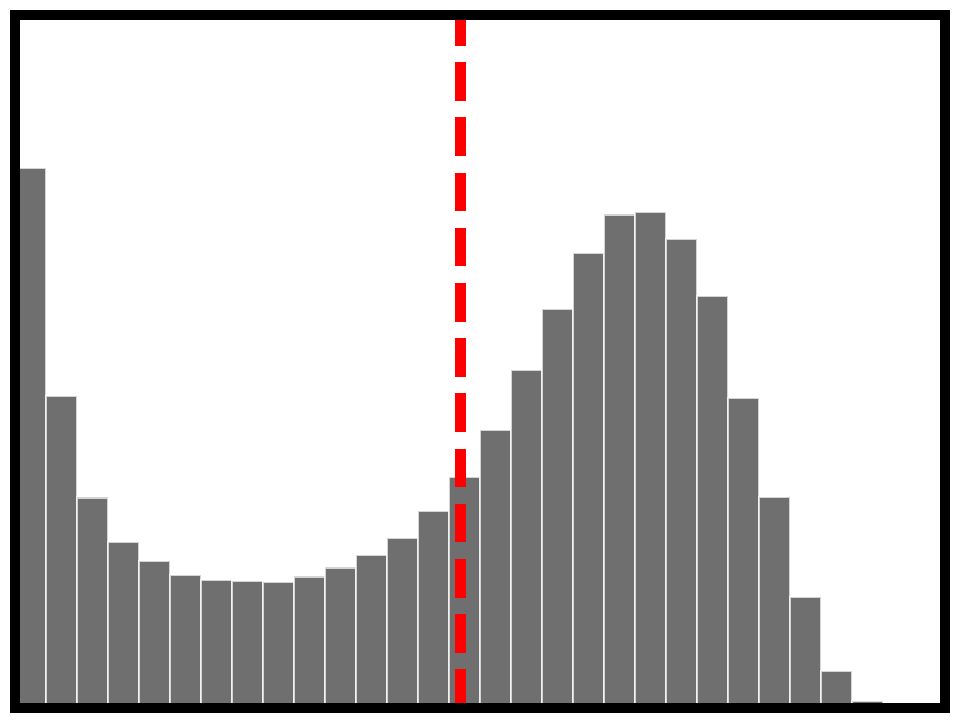}{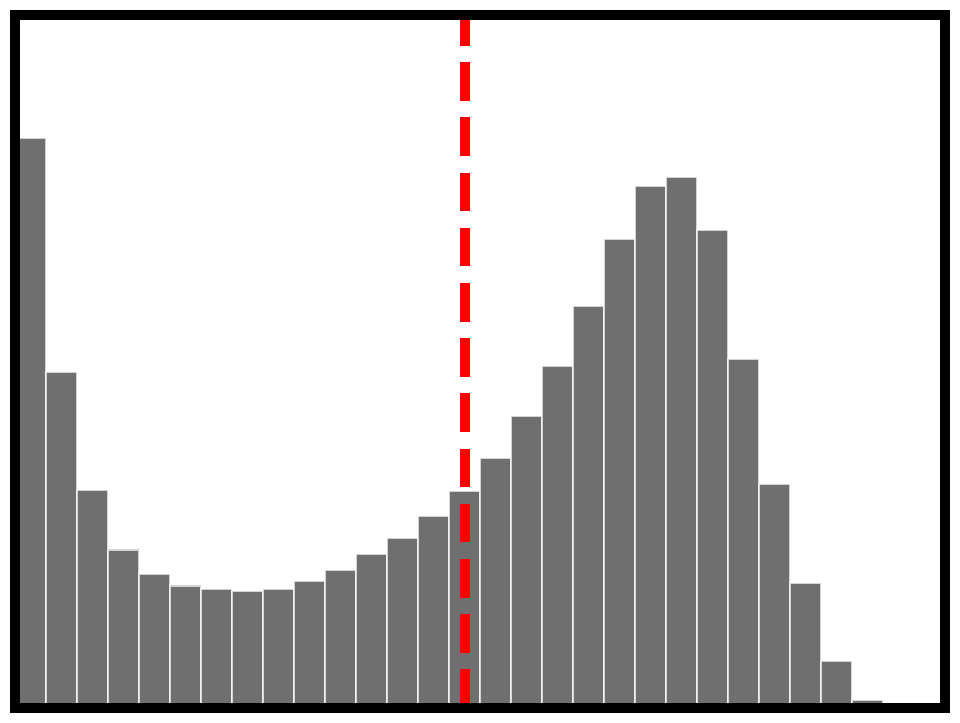}{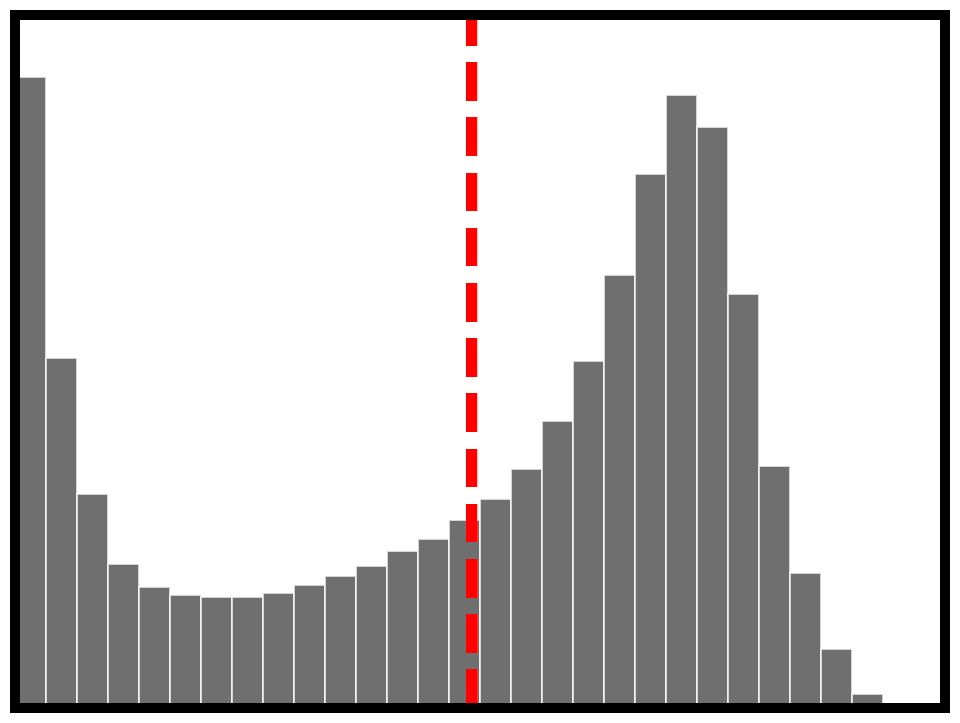}{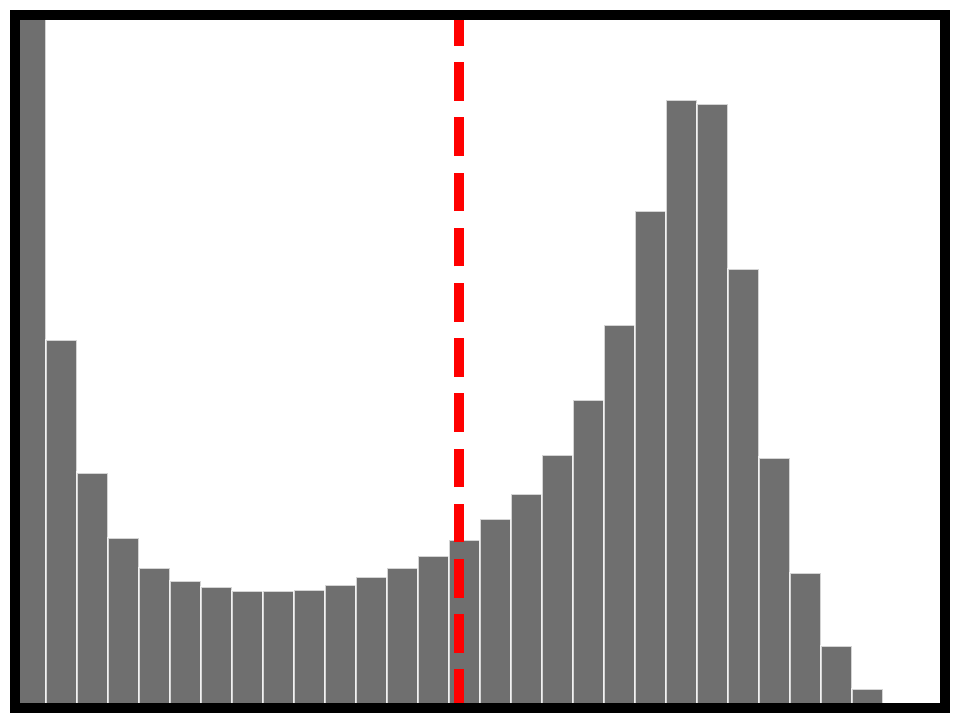}


    \myrowSix{\textsc{Playroom}}{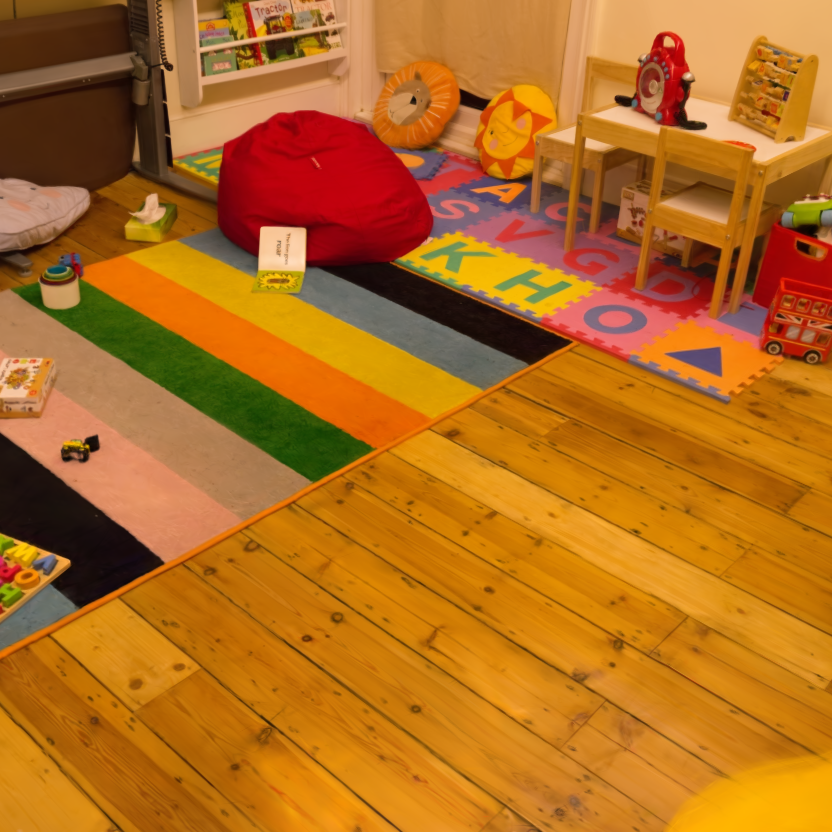}{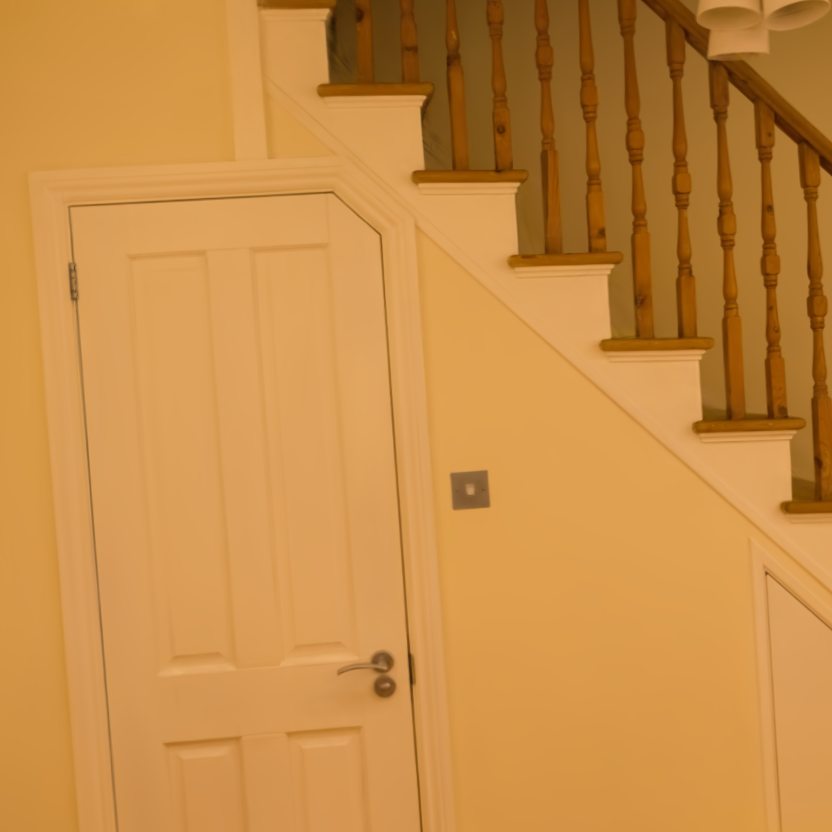}{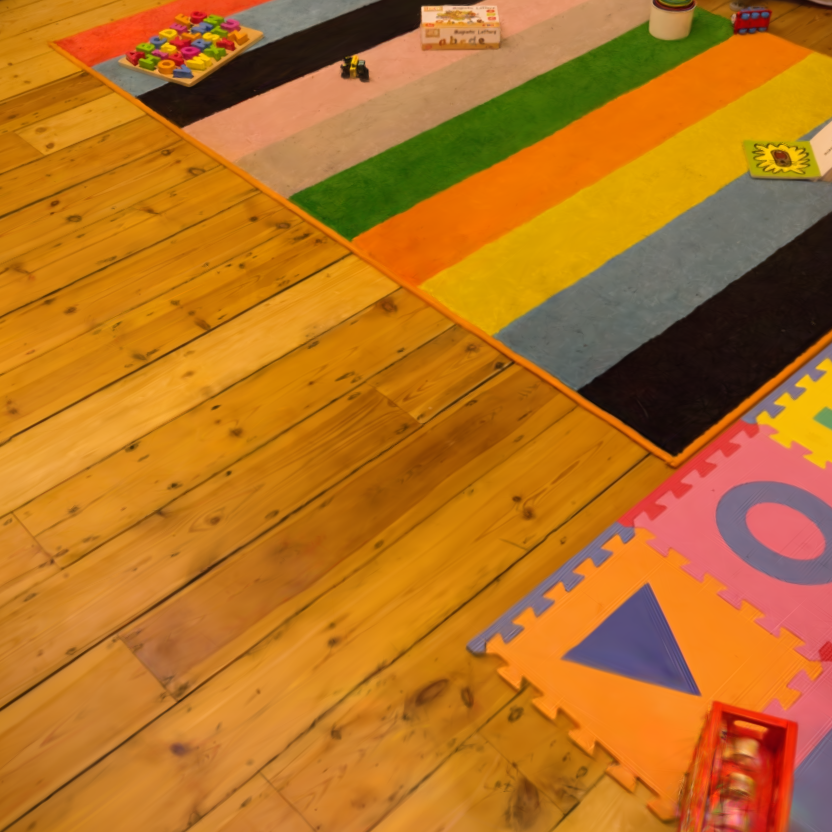}{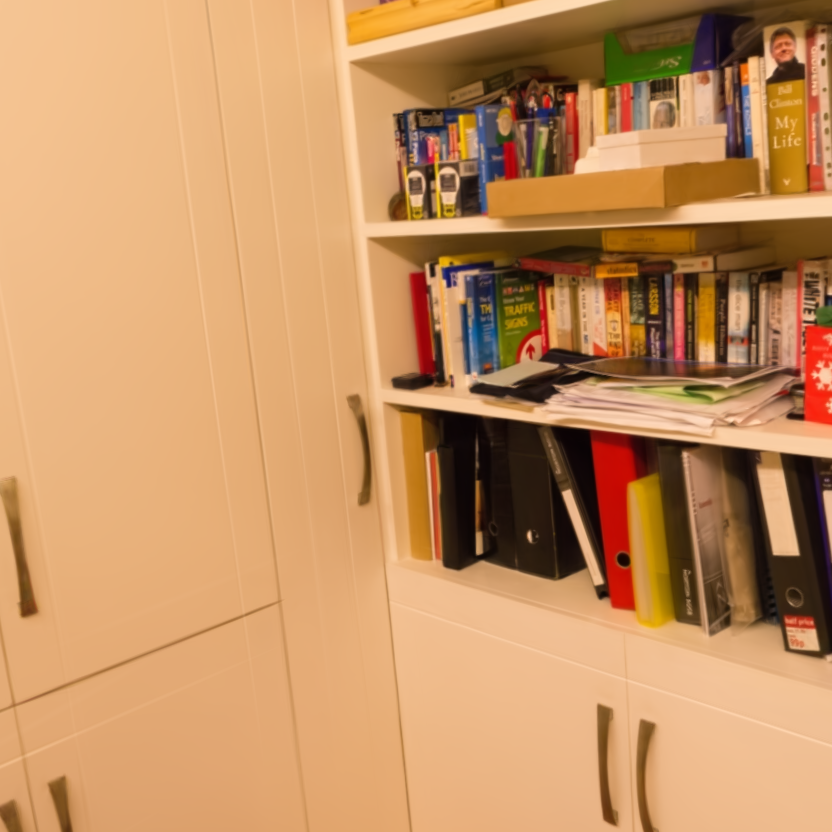}{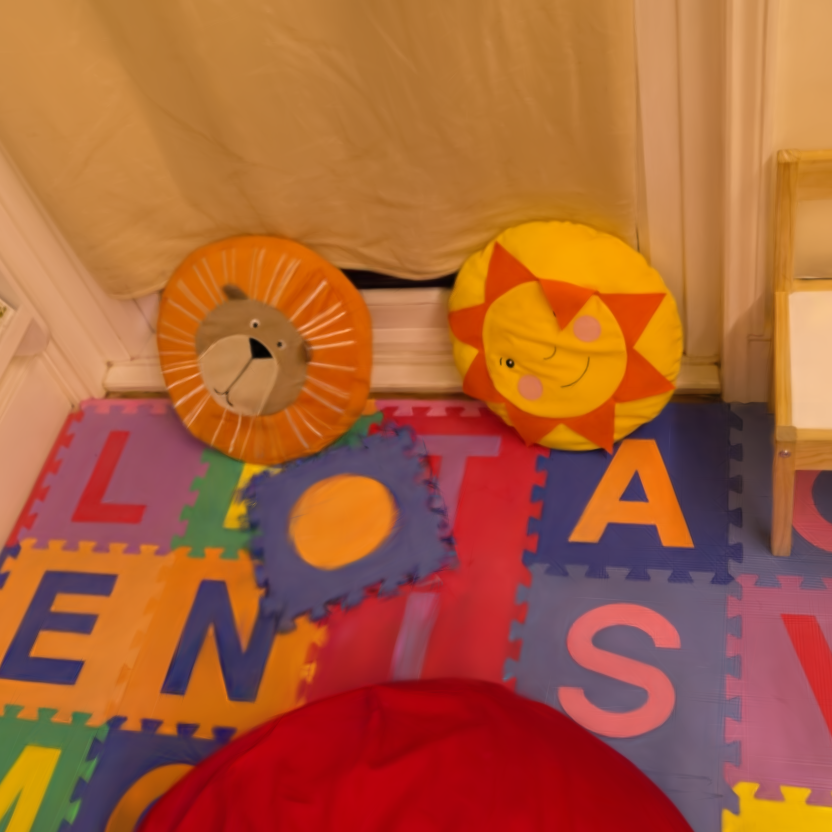}{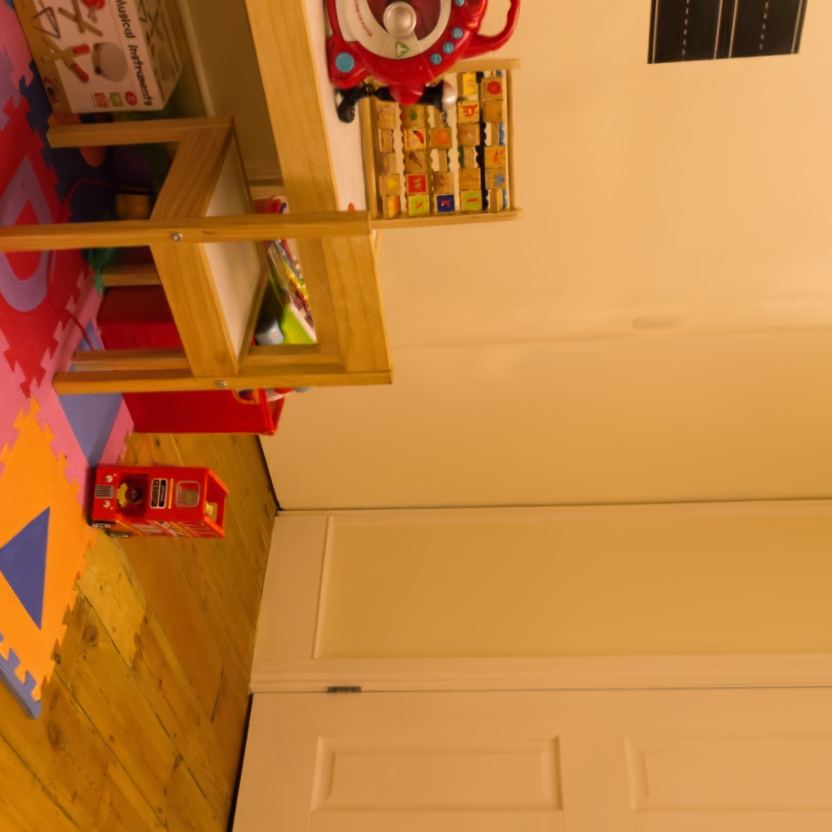}
    
    \myrowSix{Histogram}{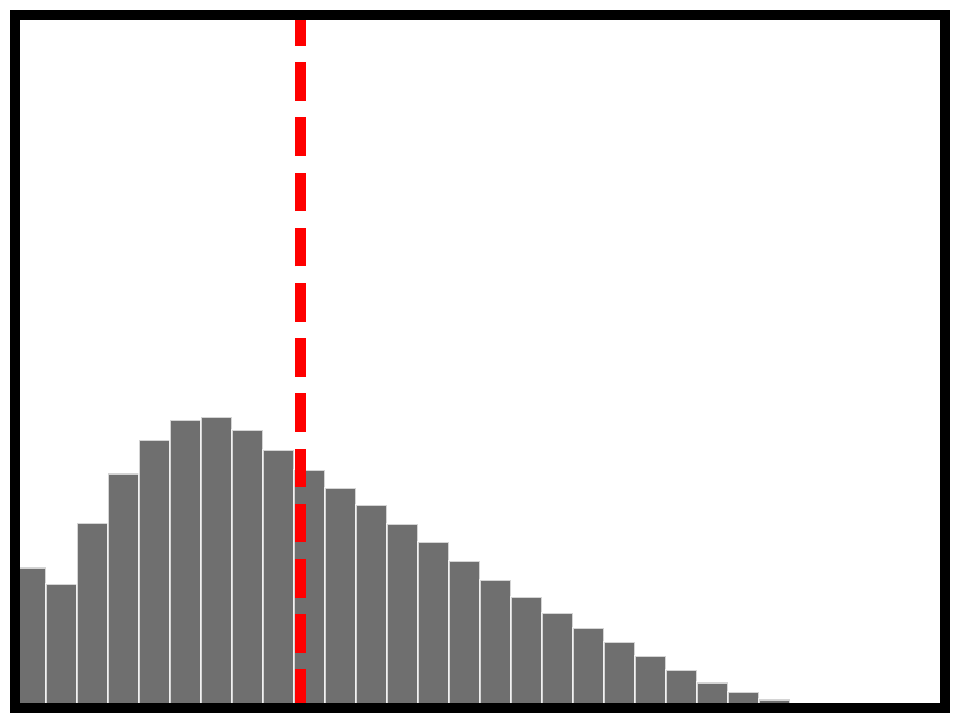}{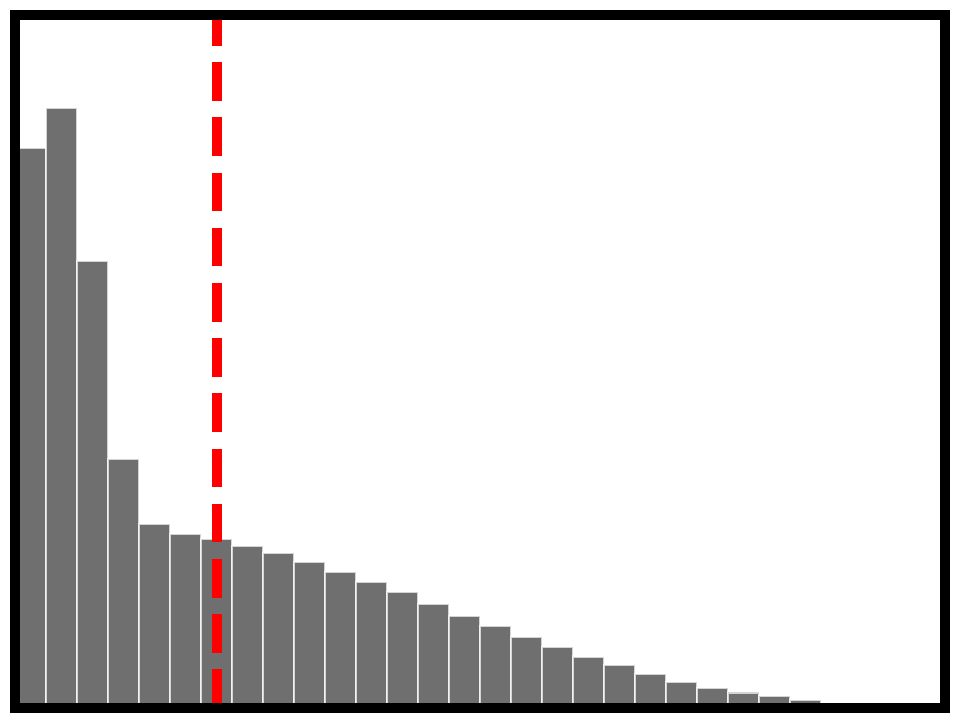}{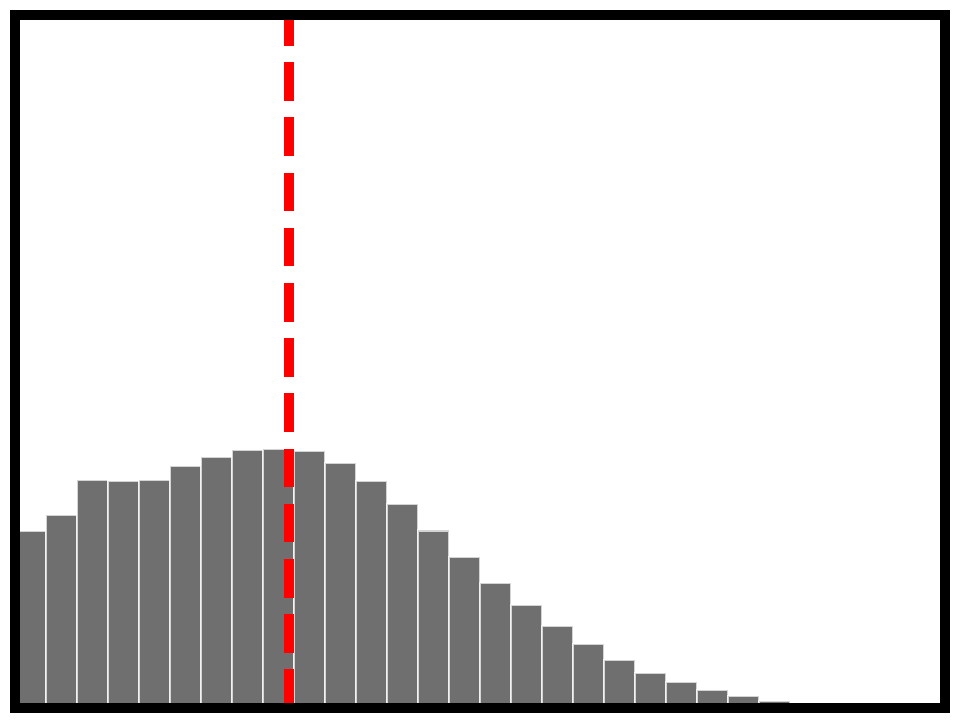}{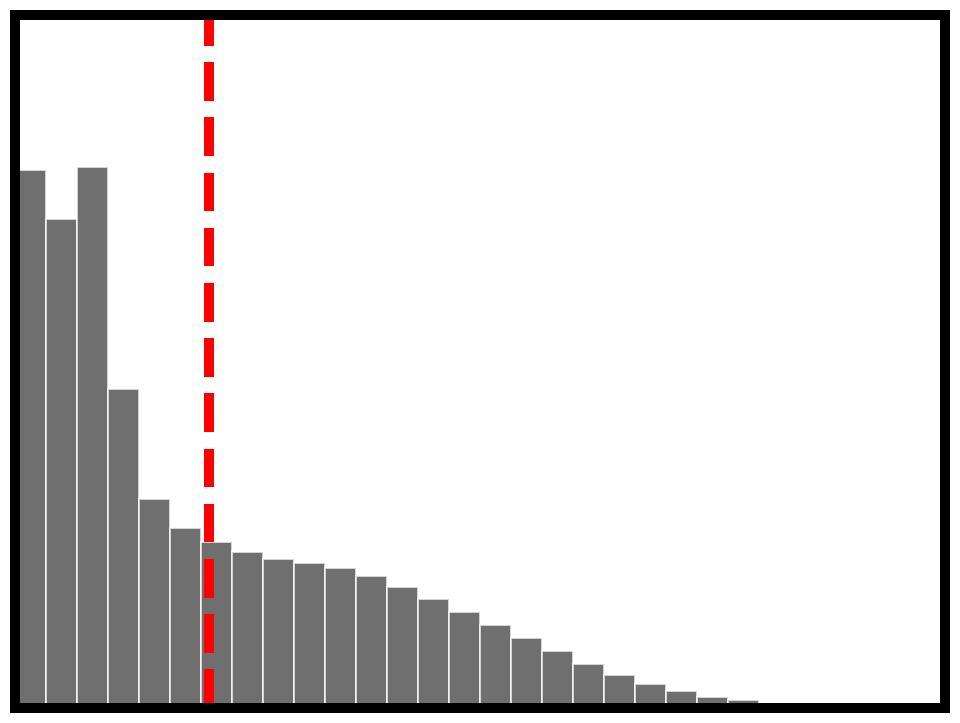}{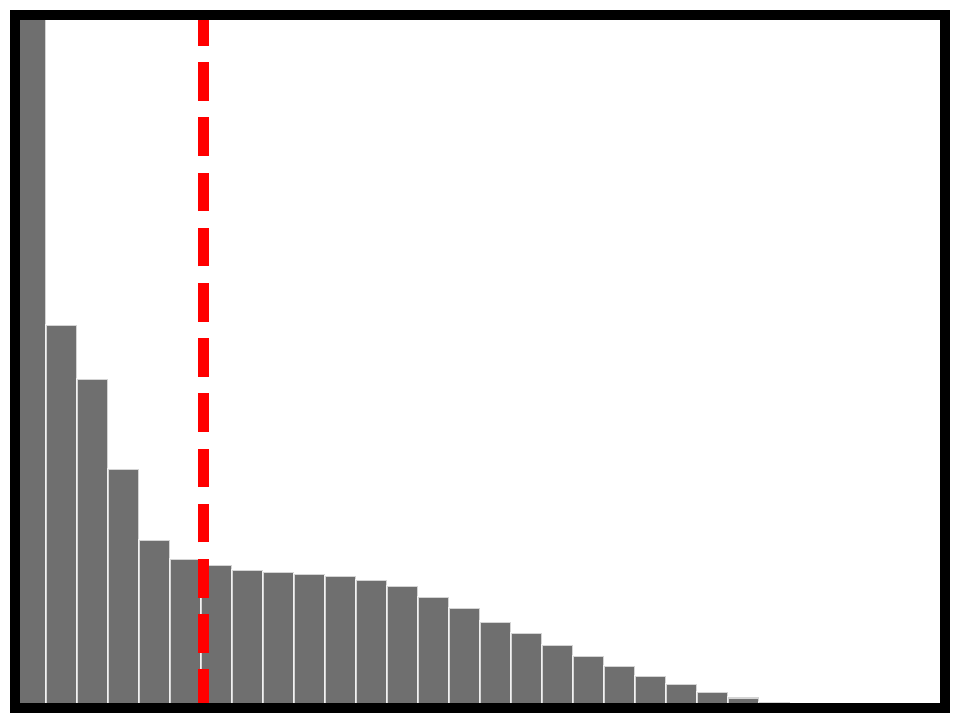}{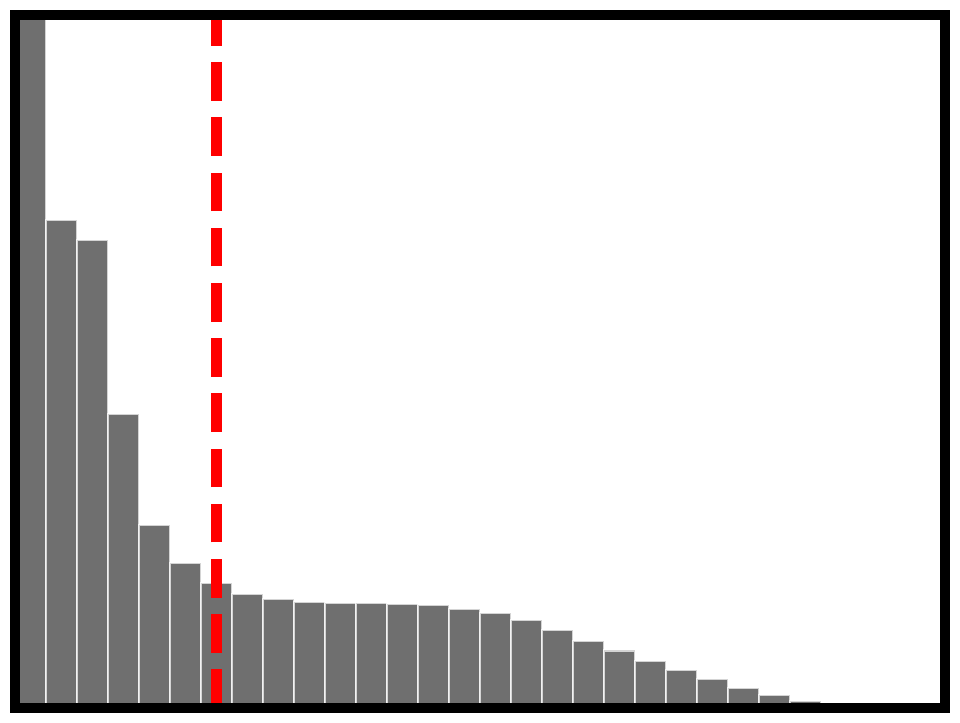}


    \myrowSix{\textsc{Room}}{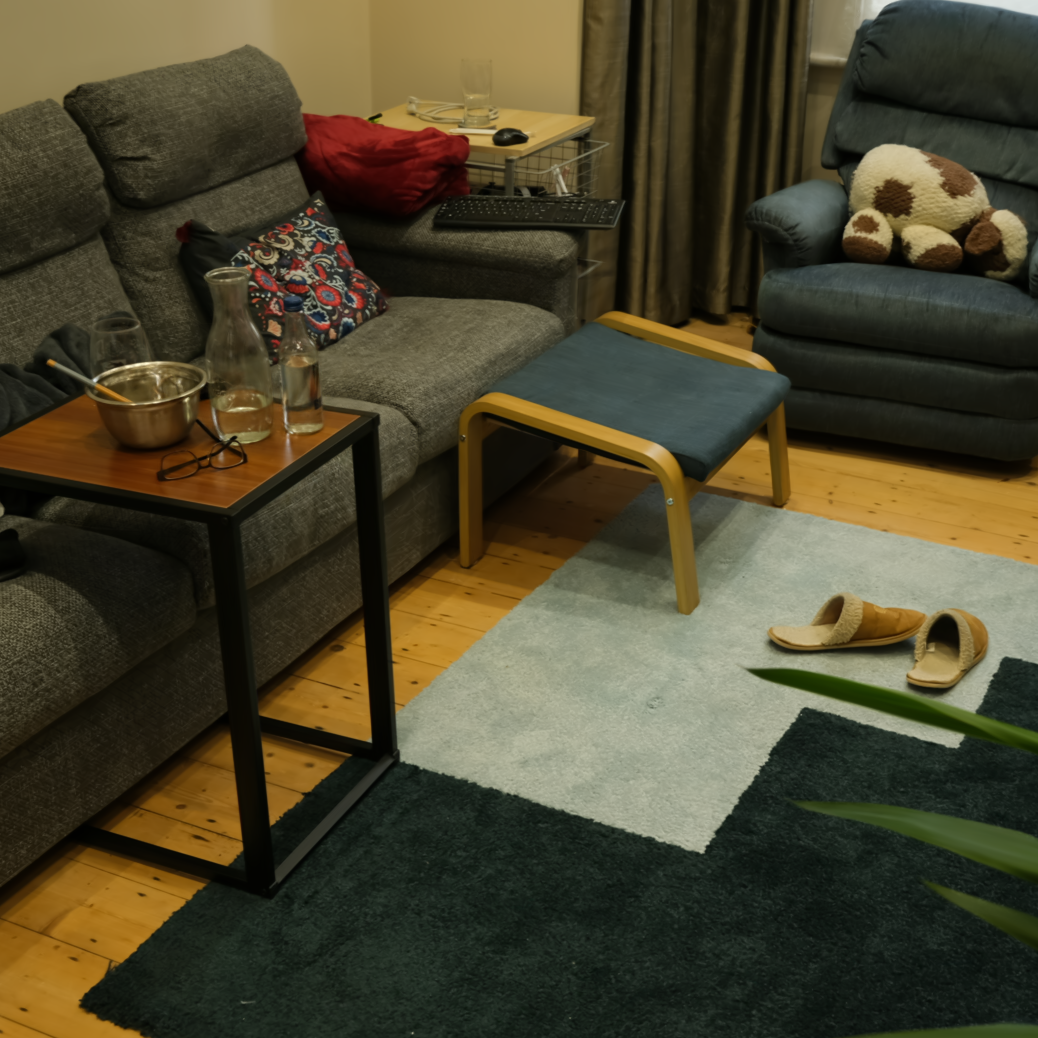}{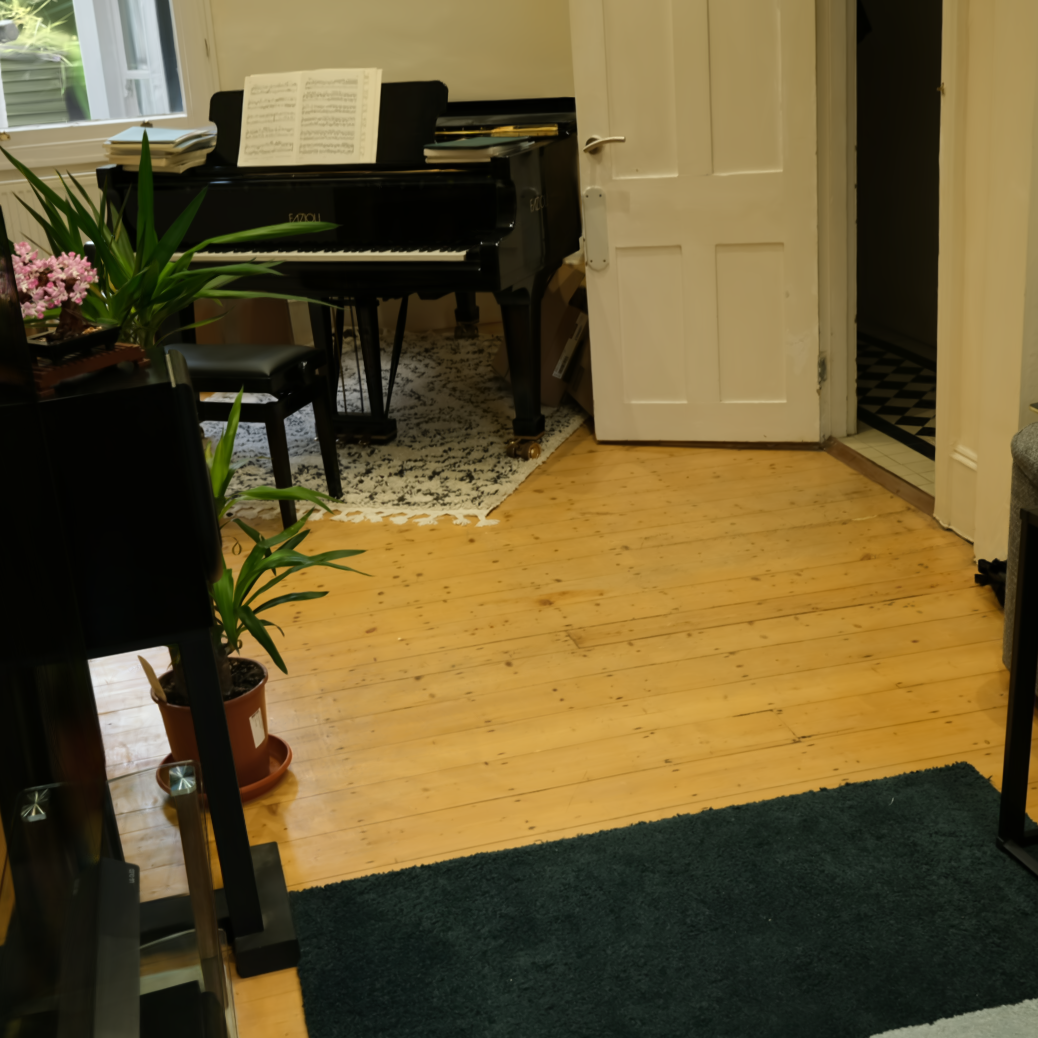}{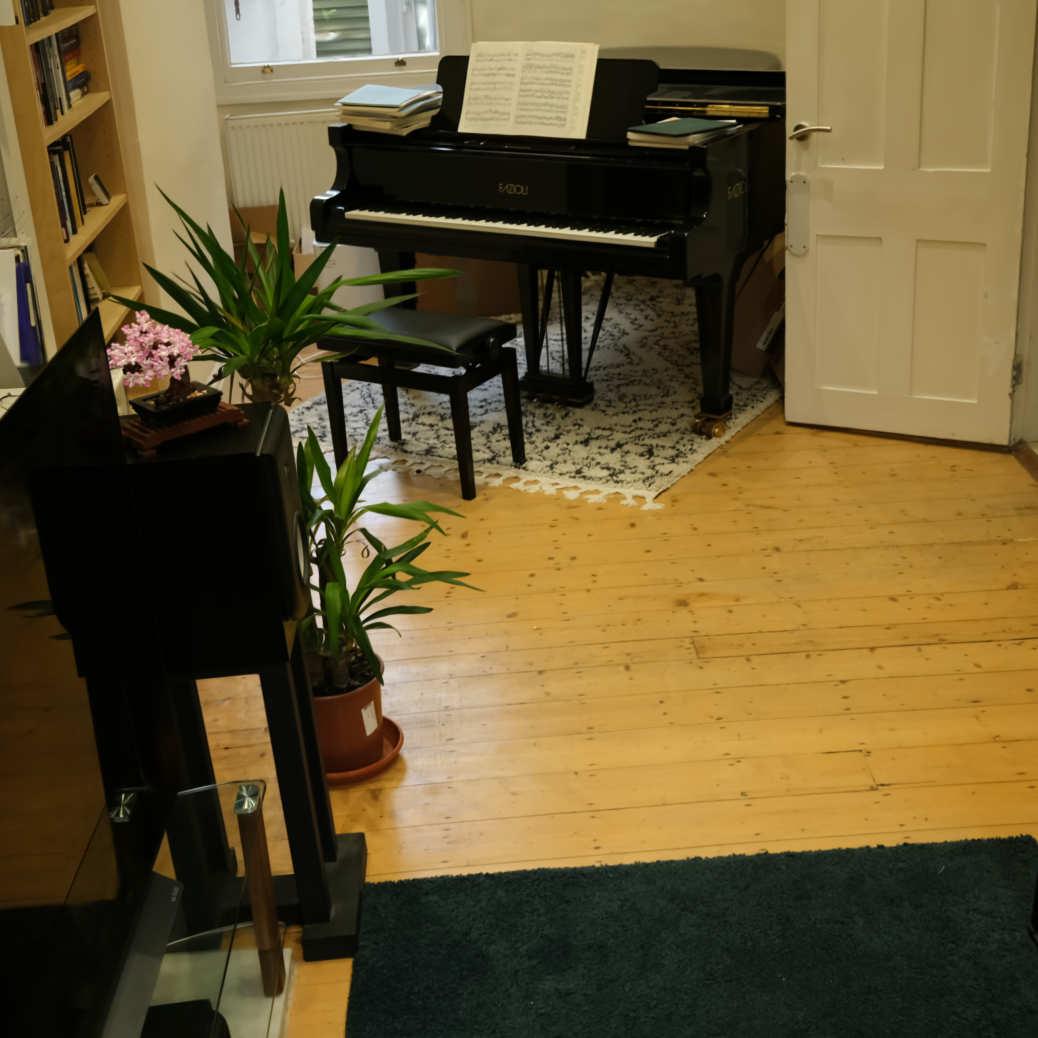}{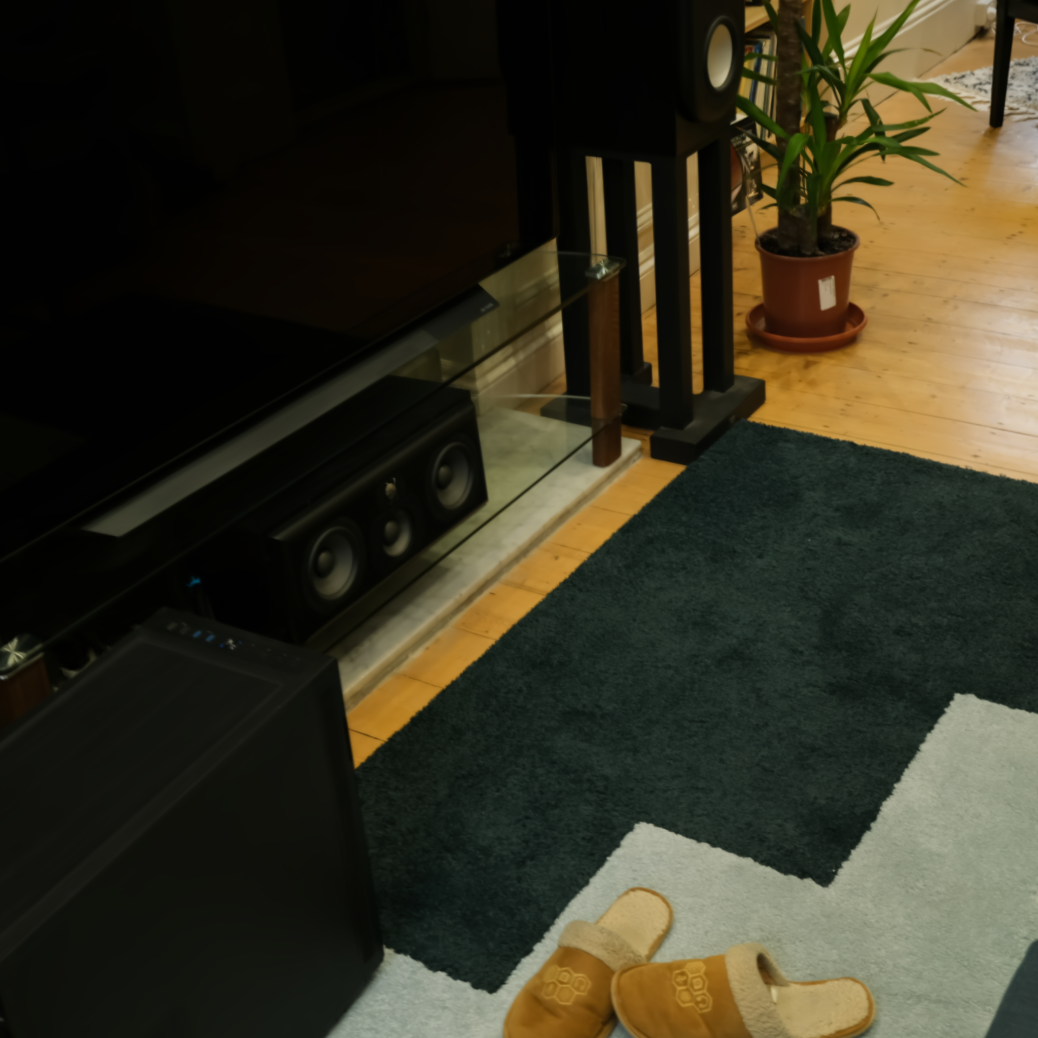}{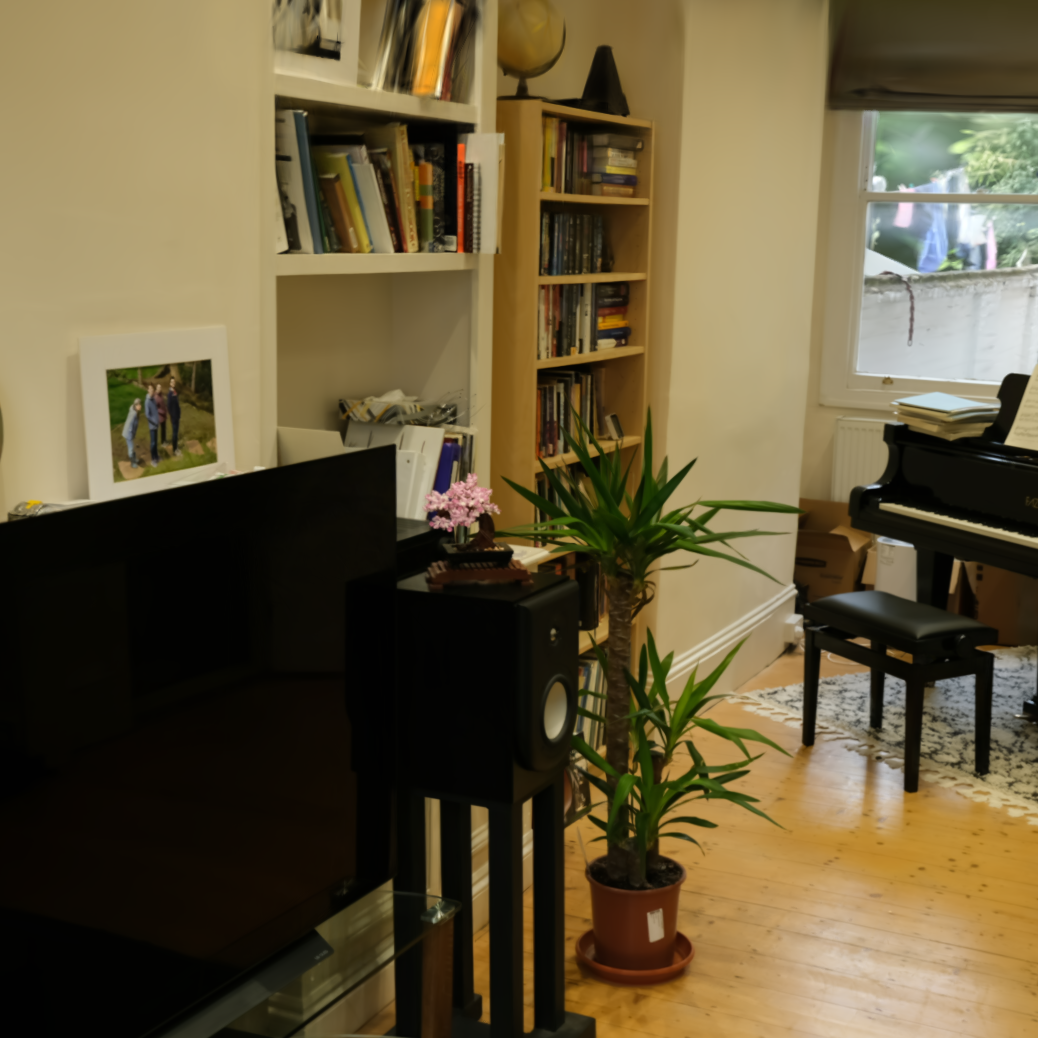}{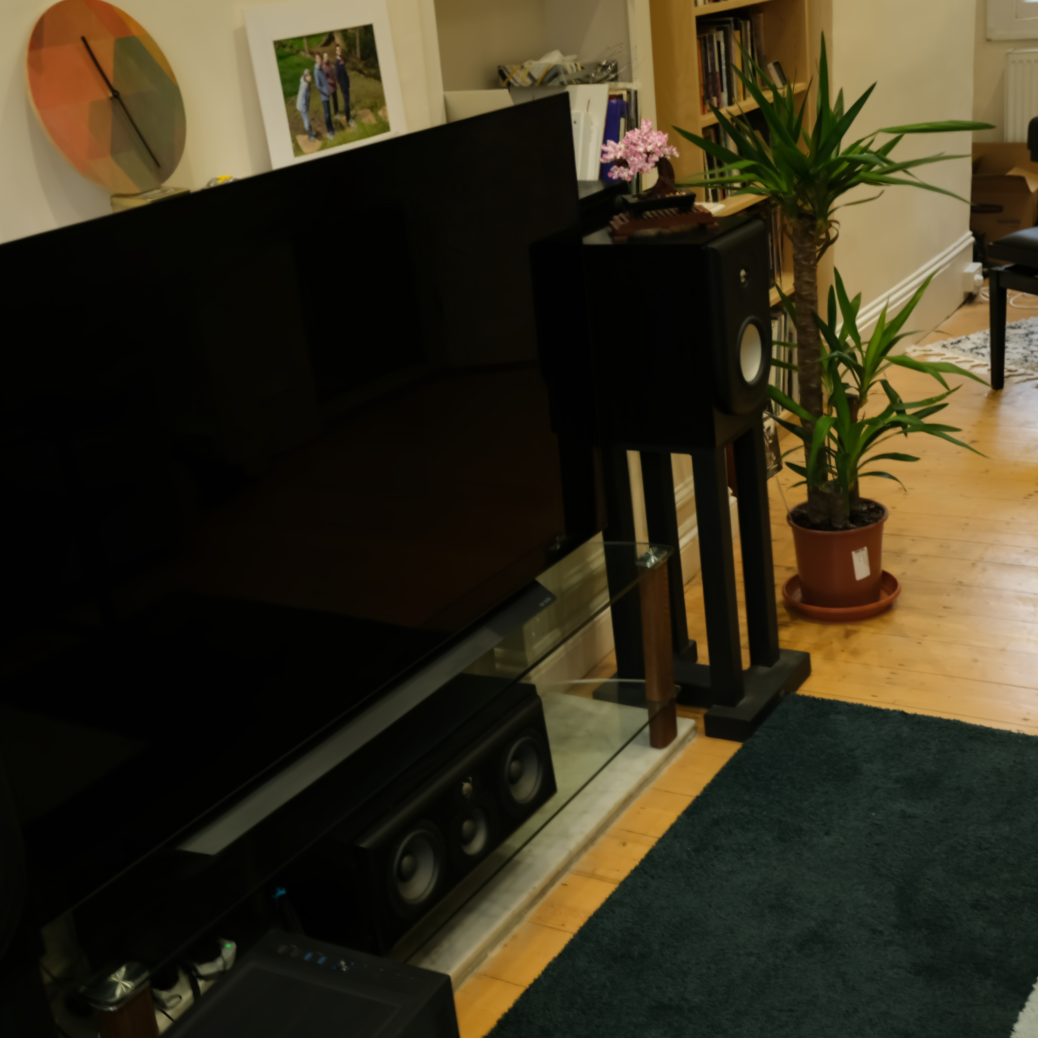}
    
    \myrowSix{Histogram}{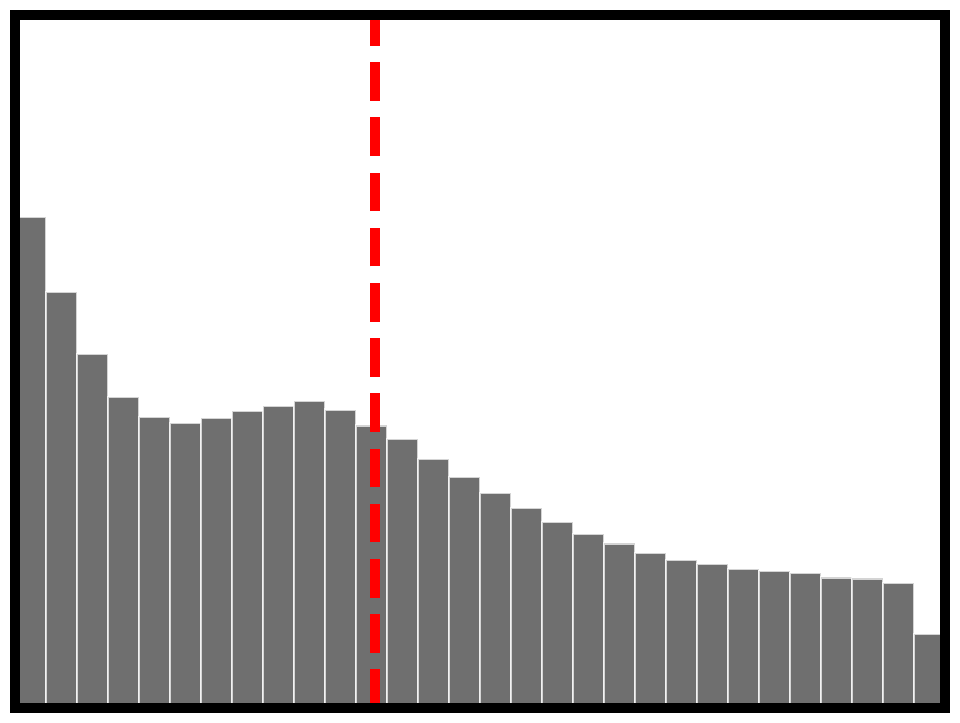}{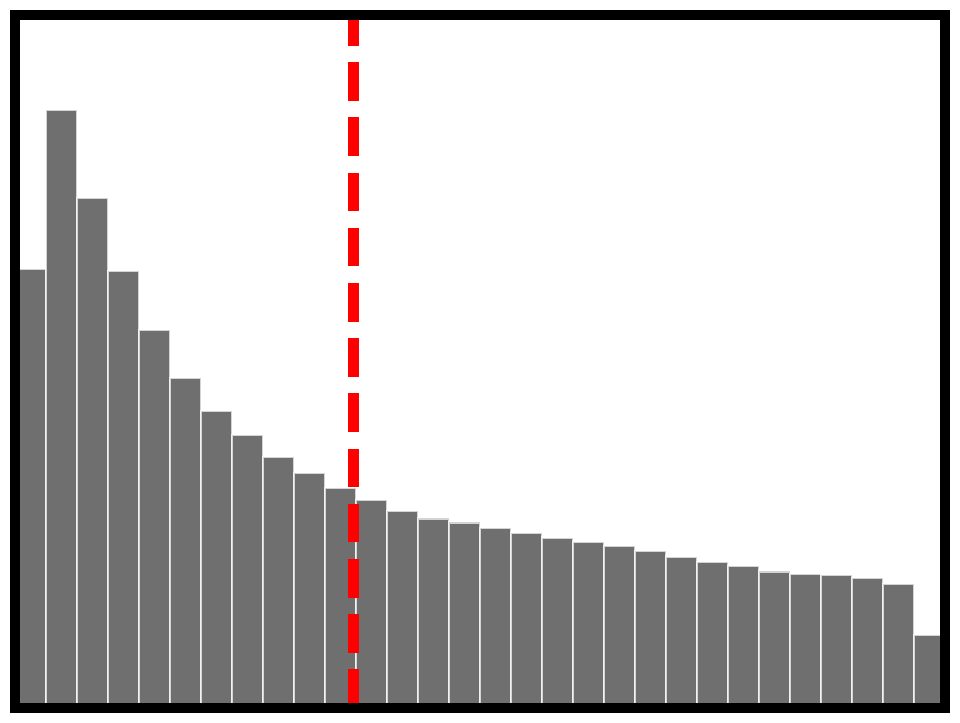}{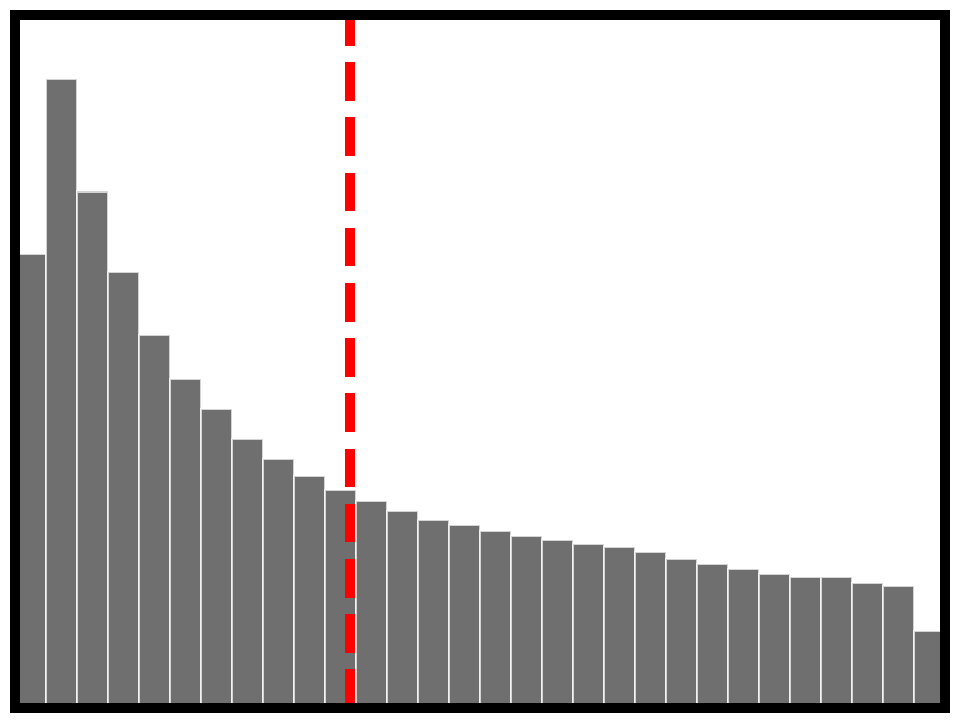}{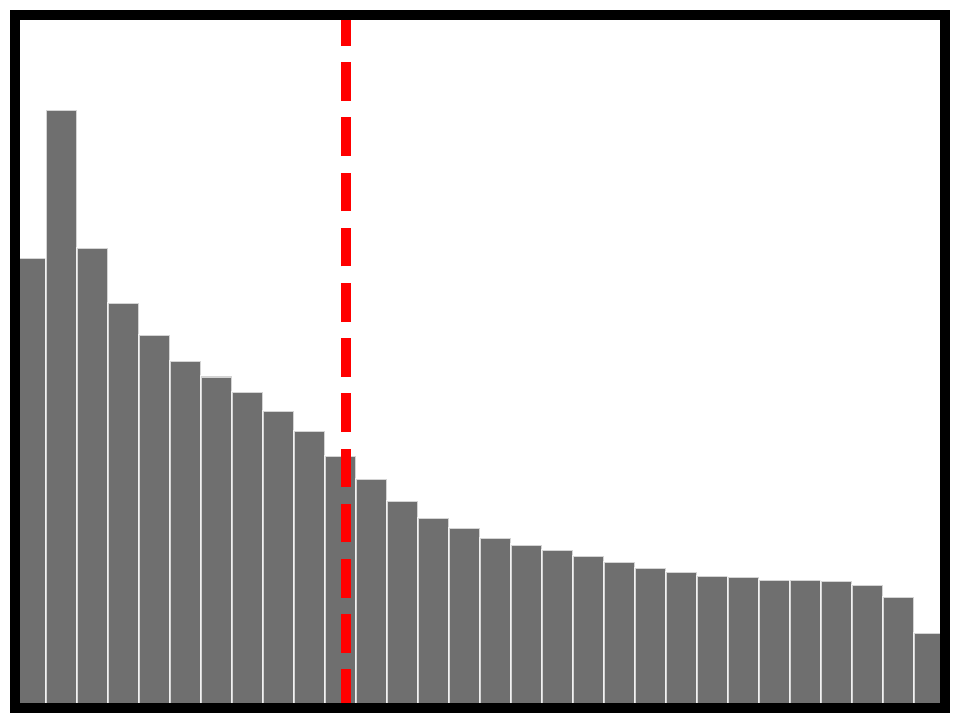}{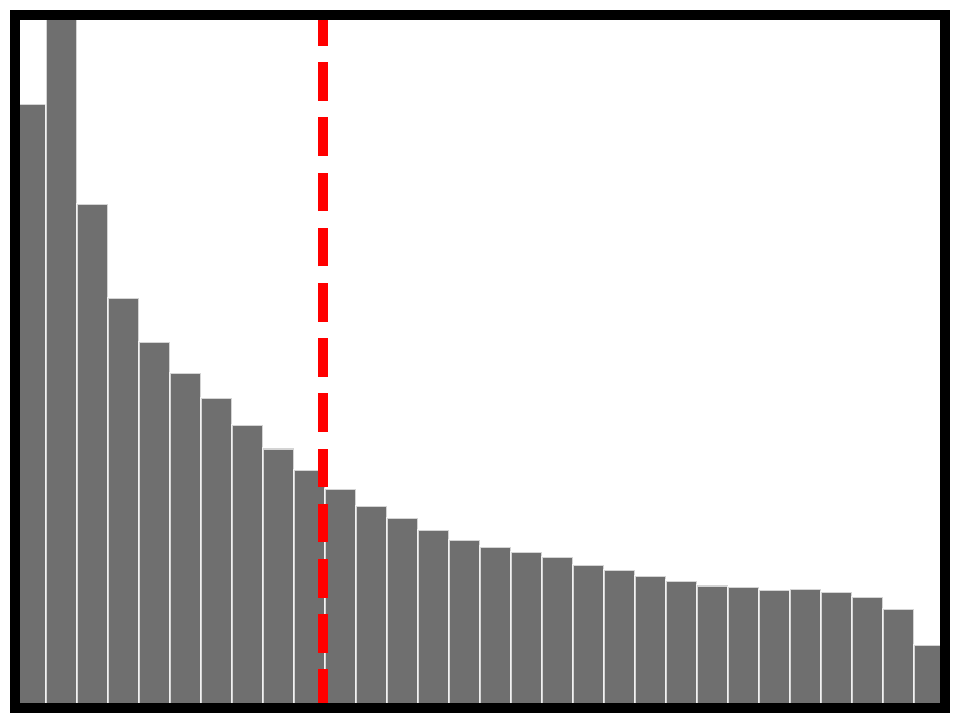}{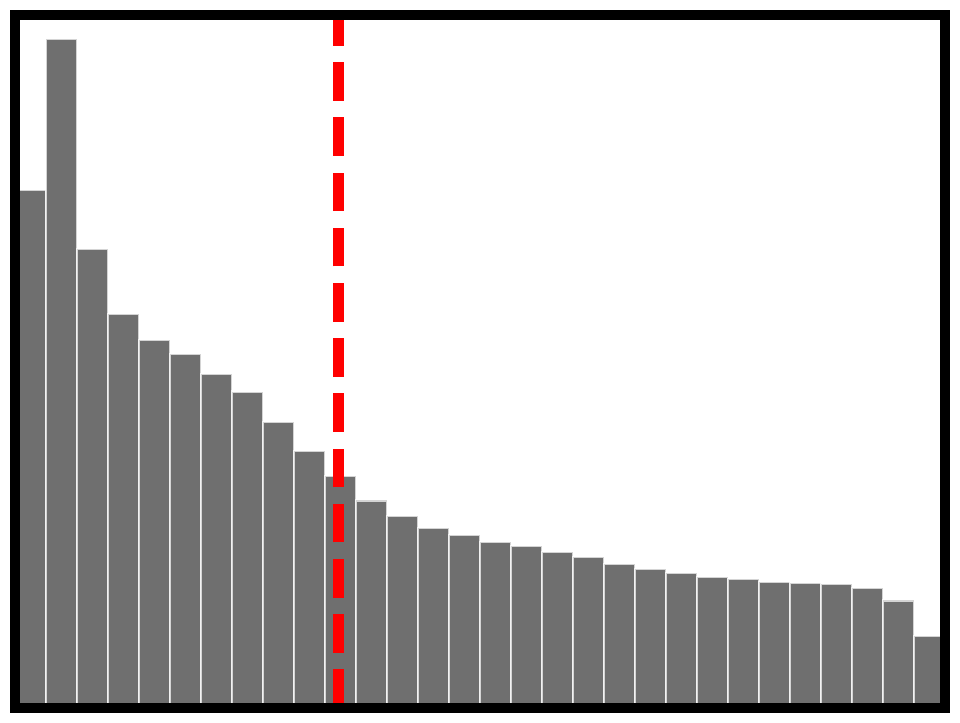}

    \caption{Distributions of our kernels across different scenes and views. Each column shows a different view. We show a histogram of $d_1$ (\cref{{eq:sample_d}}) of our learned kernels corresponding to the image above it, with the red dotted line indicating the mean; the range of the horizontal axis is [0,1].}
    \label{fig:dist_analysis}
\end{figure}

In~\cref{fig:dist_analysis}, we visualize the statistics of our learned kernels across views and scenes. The histogram of $d_1 = \Phi_{\operatorname{dec}}(0, \mathbf{z}_{2D})$ of each kernel, as defined in~\cref{eq:sample_d}, not only changes with the view due to our learned view dependency, but it also varies across different scenes, demonstrating our ability to \emph{adapt} the geometry of kernels to input data.

\subsubsection{Representation Efficiency.}
\begin{table}[h]
\centering
\caption{Comparisons on representation efficiency and runtime performance. We compare our approach with 3DGS~\cite{kerbl3Dgaussians}, 3DGS-MCMC~\cite{kheradmand20243dgsmcmc}, SSS~\cite{zhu2025sss} and DBS~\cite{liu2025deformablebetasplatting} on Tanks~\&~Temples dataset, with 3 different memory footprint for primitives. Quantitative measures, including reconstruction quality(PSNR), primitive count and rendering speed(FPS) are reported. Note that our memory footprint does not change with $k$. \#Prim. = number of primitives.}
\resizebox{\columnwidth}{!}{%
\begin{tabular}{l|ccc|ccc|ccc}
\toprule
\multicolumn{1}{c|}{Memory} & \multicolumn{3}{c|}{$\sim$25MB } & \multicolumn{3}{c|}{$\sim$50MB} & \multicolumn{3}{c}{$\sim$100MB} \\

\cmidrule(r){1-1} \cmidrule(lr){2-4} \cmidrule(lr){5-7} \cmidrule(l){8-10}

 & PSNR$\uparrow$ & \#Prim. & FPS$\uparrow$ & PSNR$\uparrow$ & \#Prim. & FPS$\uparrow$ & PSNR$\uparrow$ & \#Prim. & FPS$\uparrow$ \\
\midrule

3DGS & 23.12 & 446K & 295 & 23.36 & 893K & 232 & 23.38 & 1.79M & 198 \\
3DGS-MCMC & 22.83 & 446K & 301 & 23.40 & 893K & 235 & 23.68 & 1.79M & 165 \\
SSS & 23.90 & 391K & 130 & 24.20 & 781K & 104 & 24.38 & 1.56M & 80 \\
DBS & 23.57 & 417K & 319 & 23.79 & 833K & 249 & 24.24 & 1.67M & 178 \\
\midrule
Ours~($k$=2) & 22.80 & 328K & 123 & 24.18 & 657K & 80 & 25.22 & 1.32M & 42 \\
Ours~($k$=4) & 24.08 & 328K & 102 & 24.66 & 657K & 67 & 25.15 & 1.32M & 36 \\
Ours~($k$=8) & 24.19 & 328K & 96 & 24.70 & 657K & 55 & 25.04 & 1.32M & 28 \\
\bottomrule
\end{tabular}
}
\label{tab:efficiency}
\end{table}

\begin{figure*}
    \centering
    \newlength{\labelw}
    \setlength{\labelw}{0.12in} 
    
    \newcommand{\colw}{0.1415\linewidth}
    
    \newcommand{\headerfont}{\scriptsize} 

    \begin{minipage}{\labelw}
        \hfill 
    \end{minipage}%
    \begin{minipage}{\dimexpr\textwidth-\labelw\relax}
        \centering
        \setlength{\lineskip}{0pt}
        \begin{minipage}{\colw} \centering \subcaption*{\headerfont Ground-Truth} \end{minipage}\hfill
        \begin{minipage}{\colw} \centering \subcaption*{\headerfont Ours~($k=2$)} \end{minipage}\hfill
        \begin{minipage}{\colw} \centering \subcaption*{\headerfont Ours~($k=4$)} \end{minipage}\hfill
        \begin{minipage}{\colw} \centering \subcaption*{\headerfont Ours~($k=8$)} \end{minipage}\hfill
        \begin{minipage}{\colw} \centering \subcaption*{\headerfont 3DGS-MCMC} \end{minipage}\hfill
        \begin{minipage}{\colw} \centering \subcaption*{\headerfont DBS} \end{minipage}\hfill
        \begin{minipage}{\colw} \centering \subcaption*{\headerfont SSS} \end{minipage}
    \end{minipage}


    \begin{minipage}{\labelw}
        \centering
        \rotatebox{90}{\scriptsize $\sim$25MB}
    \end{minipage}%
    \begin{minipage}{\dimexpr\textwidth-\labelw\relax}
        \centering
        \begin{minipage}{\colw}
            \includegraphics[width=\linewidth, height=\linewidth]{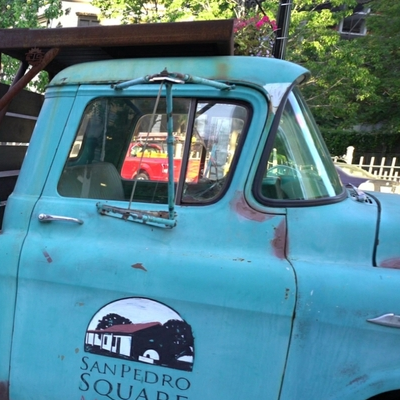}
                \put(-60,3){\alphabox{PSNR|SSIM|LPIPS}}
        \end{minipage}\hfill
        \begin{minipage}{\colw}
            \includegraphics[width=\linewidth, height=\linewidth]{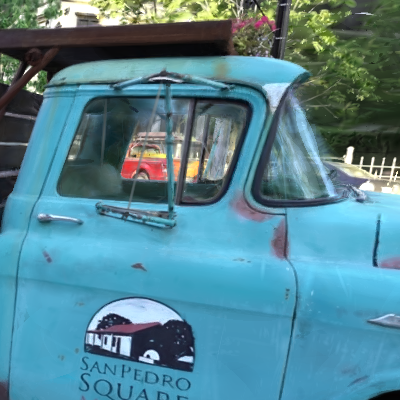}
            \put(-60,3){\alphabox{\color{white} 24.57|0.851|0.169}}
        \end{minipage}\hfill
        \begin{minipage}{\colw}
            \includegraphics[width=\linewidth, height=\linewidth]{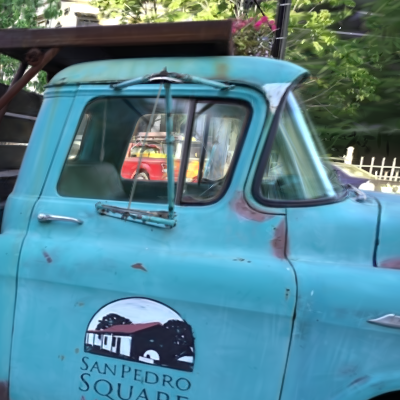}
            \put(-60,3){\alphabox{\color{white} 25.67|0.878|0.150}}
        \end{minipage}\hfill
        \begin{minipage}{\colw}
            \includegraphics[width=\linewidth, height=\linewidth]{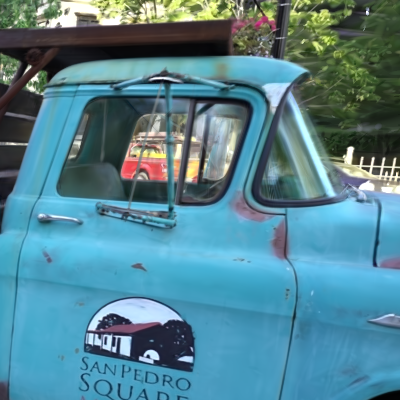}
            \put(-60,3){\alphabox{\color{white} 25.72|0.878|0.151}}
        \end{minipage}\hfill
        \begin{minipage}{\colw}
            \includegraphics[width=\linewidth, height=\linewidth]{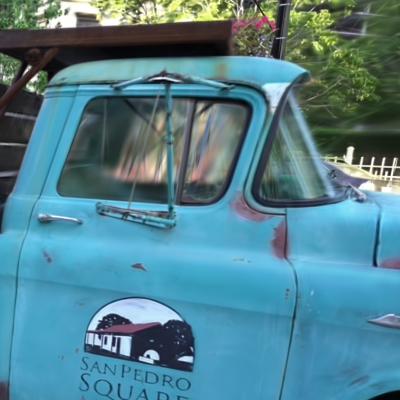}
            \put(-60,3){\alphabox{\color{white} 24.05|0.853|0.182}}
        \end{minipage}\hfill
        \begin{minipage}{\colw}
            \includegraphics[width=\linewidth, height=\linewidth]{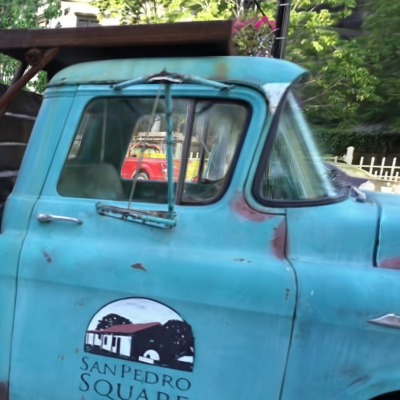}
            \put(-60,3){\alphabox{\color{white} 25.12|0.873|0.147}}
        \end{minipage}\hfill
        \begin{minipage}{\colw}
            \includegraphics[width=\linewidth, height=\linewidth]{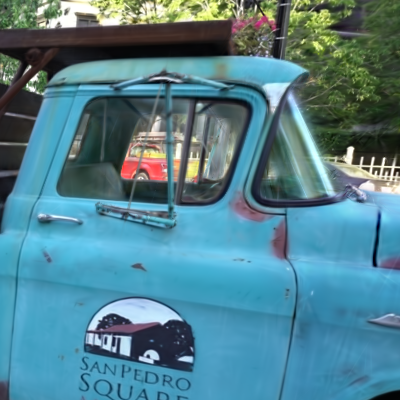}
            \put(-60,3){\alphabox{\color{white} 25.31|0.872|0.155}}
        \end{minipage}
    \end{minipage}


    \begin{minipage}{\labelw}
        \centering
        \rotatebox{90}{\scriptsize $\sim$50MB}
    \end{minipage}%
    \begin{minipage}{\dimexpr\textwidth-\labelw\relax}
        \centering
        \begin{minipage}{\colw}
            \includegraphics[width=\linewidth, height=\linewidth]{imgs/same_mem/2/crop_GT_00021.png}
                \put(-60,3){\alphabox{PSNR|SSIM|LPIPS}}
        \end{minipage}\hfill
        \begin{minipage}{\colw}
            \includegraphics[width=\linewidth, height=\linewidth]{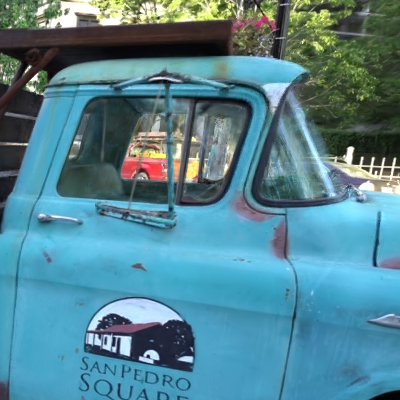}
            \put(-60,3){\alphabox{\color{white} 25.18|0.870|0.137}}
        \end{minipage}\hfill
        \begin{minipage}{\colw}
            \includegraphics[width=\linewidth, height=\linewidth]{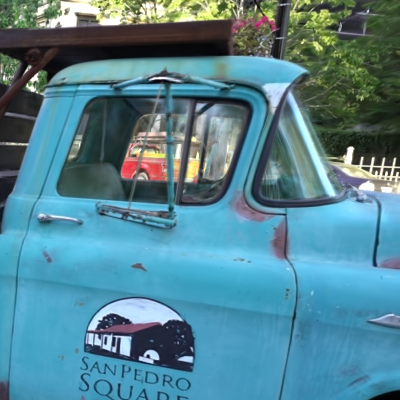}
            \put(-60,3){\alphabox{\color{white} 25.78|0.884|0.125}}
        \end{minipage}\hfill
        \begin{minipage}{\colw}
            \includegraphics[width=\linewidth, height=\linewidth]{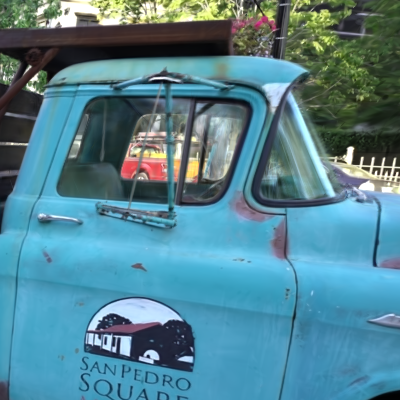}
            \put(-60,3){\alphabox{\color{white} 25.84|0.884|0.127}}
        \end{minipage}\hfill
        \begin{minipage}{\colw}
            \includegraphics[width=\linewidth, height=\linewidth]{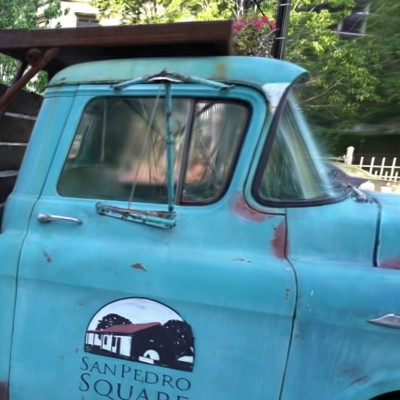}
            \put(-60,3){\alphabox{\color{white} 24.23|0.863|0.152}}
        \end{minipage}\hfill
        \begin{minipage}{\colw}
            \includegraphics[width=\linewidth, height=\linewidth]{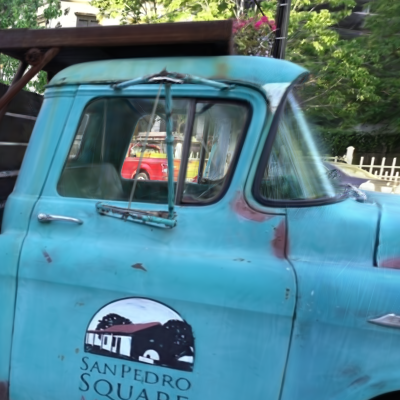}
            \put(-60,3){\alphabox{\color{white} 25.41|0.885|0.128}}
        \end{minipage}\hfill
        \begin{minipage}{\colw}
            \includegraphics[width=\linewidth, height=\linewidth]{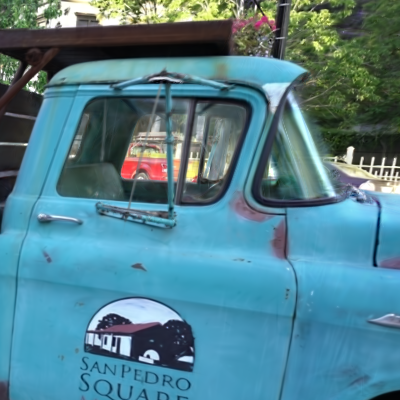}
            \put(-60,3){\alphabox{\color{white} 25.33|0.878|0.135}}
        \end{minipage}
    \end{minipage}


    \begin{minipage}{\labelw}
        \centering
        \rotatebox{90}{\scriptsize Memory = $\sim$100MB}
    \end{minipage}%
    \begin{minipage}{\dimexpr\textwidth-\labelw\relax}
        \centering
        \begin{minipage}{\colw}
            \includegraphics[width=\linewidth, height=\linewidth]{imgs/same_mem/2/crop_GT_00021.png}{
                \put(-60,3){\alphabox{PSNR|SSIM|LPIPS}}}
        \end{minipage}\hfill
        \begin{minipage}{\colw}
            \includegraphics[width=\linewidth, height=\linewidth]{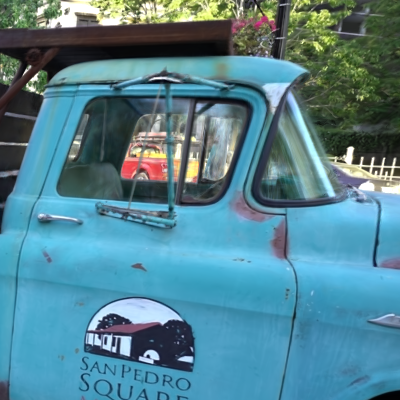}
            \put(-60,3){\alphabox{\color{white} 25.85|0.885|0.112}}
        \end{minipage}\hfill
        \begin{minipage}{\colw}
            \includegraphics[width=\linewidth, height=\linewidth]{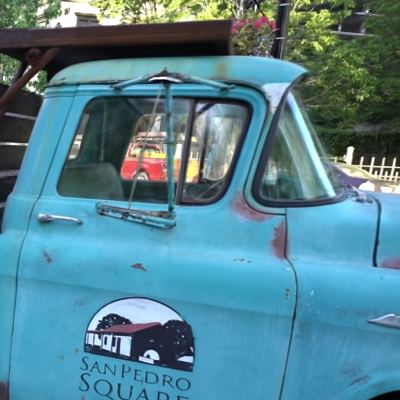}
            \put(-60,3){\alphabox{\color{white} 25.93|0.887|0.110}}
        \end{minipage}\hfill
        \begin{minipage}{\colw}
            \includegraphics[width=\linewidth, height=\linewidth]{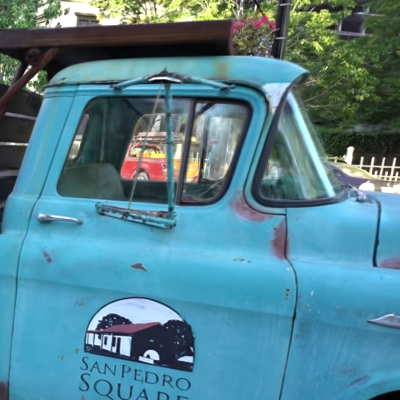}
            \put(-60,3){\alphabox{\color{white} 25.99|0.887|0.111}}
        \end{minipage}\hfill
        \begin{minipage}{\colw}
            \includegraphics[width=\linewidth, height=\linewidth]{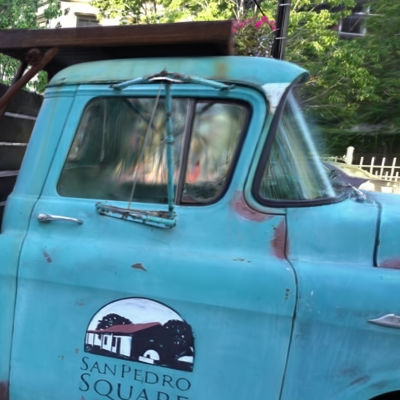}
            \put(-60,3){\alphabox{\color{white} 24.44|0.870|0.132}}
        \end{minipage}\hfill
        \begin{minipage}{\colw}
            \includegraphics[width=\linewidth, height=\linewidth]{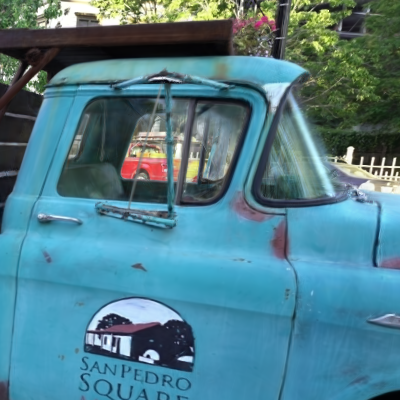}
            \put(-60,3){\alphabox{\color{white} 25.12|0.879|0.121}}
        \end{minipage}\hfill
        \begin{minipage}{\colw}
            \includegraphics[width=\linewidth, height=\linewidth]{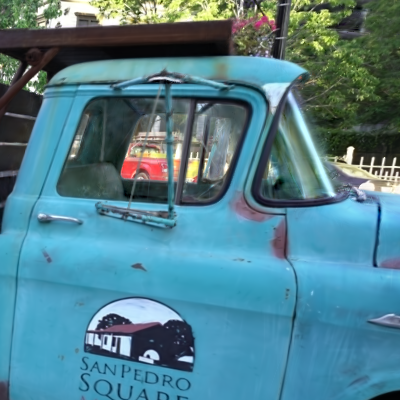}
            \put(-60,3){\alphabox{\color{white} 25.37|0.879|0.123}}
        \end{minipage}
    \end{minipage}

    \caption{Qualitative comparisons with state-of-the-art techniques under different memory footprint for primitives. We compare our approach ($k$=2, 4, 8) with 3DGS-MCMC~\cite{kheradmand20243dgsmcmc}, SSS~\cite{zhu2025sss} and DBS~\cite{liu2025deformablebetasplatting}. Quantitative errors in PSNR, SSIM and LPIPS are reported at the bottom of each related image.}
    \label{fig:primitive_number}
\end{figure*}

We compare the representation efficiency between our approach and state-of-the-art work of 3DGS, 3DGS-MCMC, SSS and DBS in~\cref{tab:efficiency}. Specifically, following 
the protocol detailed in the supplemental material of~\cite{liu2025deformablebetasplatting}, we disable all view-dependent color components to focus on the impact of the geometry of kernels. Next, we perform comparisons with a similar memory footprint for primitives from different methods. Our approach results in higher quality compared to the baselines, demonstrating improved representation efficiency. Moreover, our representation is scalable with respect to the size of available memory footprint. Please also refer to~\cref{fig:primitive_number} for qualitative comparisons.

\subsubsection{Extension to Image Representation.} 
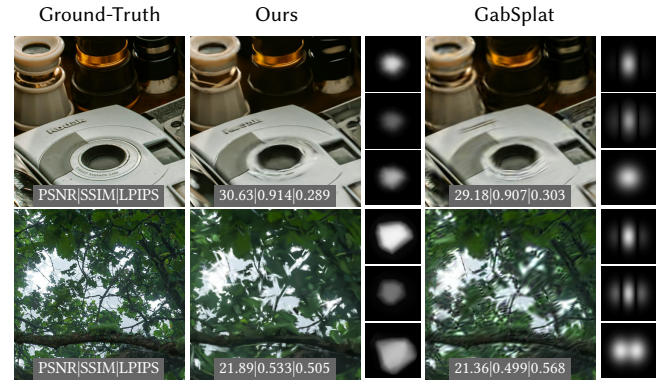
\begin{figure}[htbp]
    \centering
    \newcommand{\metricbox}[1]{%
        \begingroup
        \setlength{\fboxsep}{0.1pt}
        \colorbox{black!60}{%
            \makebox[50pt][c]{
                \vphantom{Hp}
                \scalebox{.7}{\color{white}#1}
            }%
        }%
        \endgroup
    }

    \begin{minipage}{\linewidth}
        \centering
        \begin{minipage}{.263\linewidth} \centering \subcaption*{\small Ground-Truth} \end{minipage} \hfill
        \begin{minipage}{.263\linewidth} \centering \subcaption*{\small Ours} \end{minipage} \hfill
        \begin{minipage}{.084\linewidth} \centering \subcaption*{} \end{minipage} \hfill
        \begin{minipage}{.263\linewidth} \centering \subcaption*{\small GabSplat} \end{minipage} \hfill
        \begin{minipage}{.084\linewidth} \centering \subcaption*{} \end{minipage}
    \end{minipage}

    \smallskip

    \begin{minipage}{\linewidth}
        \centering
        \begin{minipage}[t]{.263\linewidth}
            \hrule height 0pt
            \includegraphics[width=\linewidth]{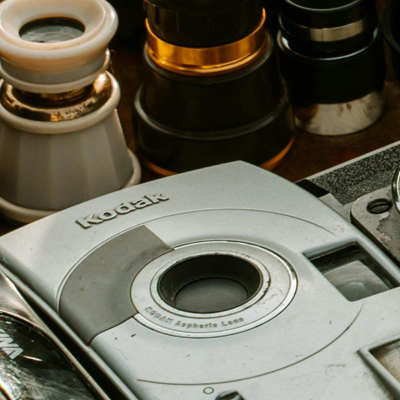}%
            \begin{picture}(0,0)\put(-57,2.5){\alphabox{PSNR|SSIM|LPIPS}}\end{picture}
        \end{minipage} \hfill
        \begin{minipage}[t]{.263\linewidth}
            \hrule height 0pt
            \includegraphics[width=\linewidth]{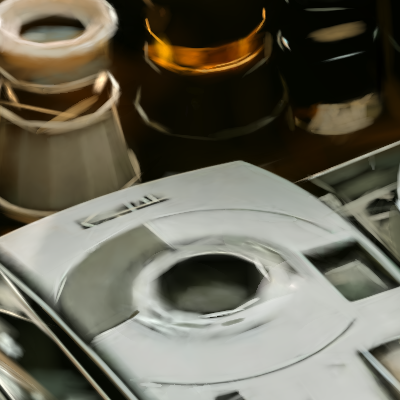}%
            \begin{picture}(0,0)\put(-55,2.5){\alphabox{30.63|0.914|0.289}}\end{picture}
        \end{minipage} \hfill
        \begin{minipage}[t]{.084\linewidth}
            \hrule height 0pt
            \includegraphics[width=\linewidth]{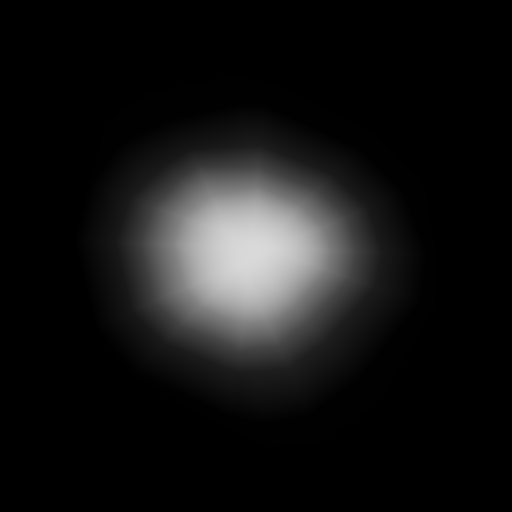} \\ 
            \includegraphics[width=\linewidth]{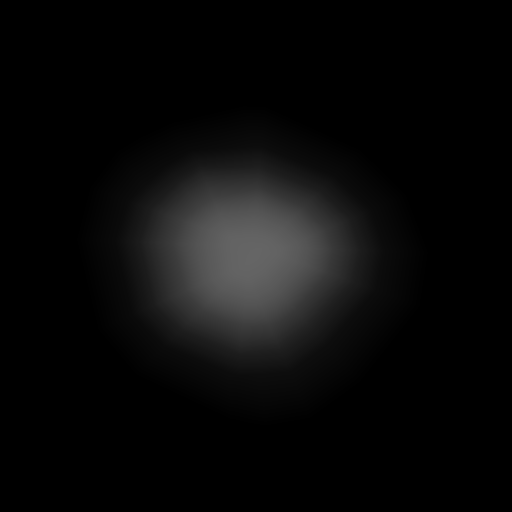} \\ 
            \includegraphics[width=\linewidth]{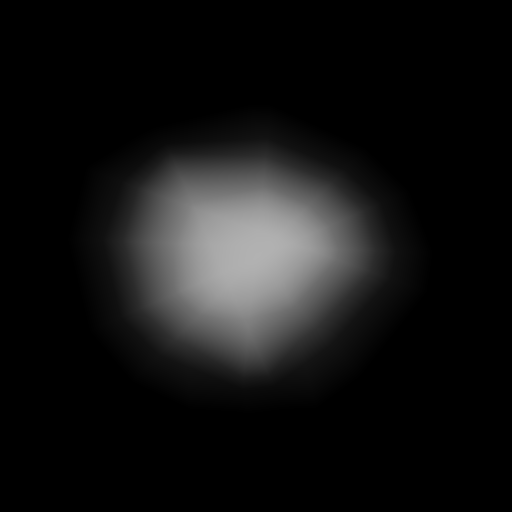}
        \end{minipage} \hfill
        \begin{minipage}[t]{.263\linewidth}
            \hrule height 0pt
            \includegraphics[width=\linewidth]{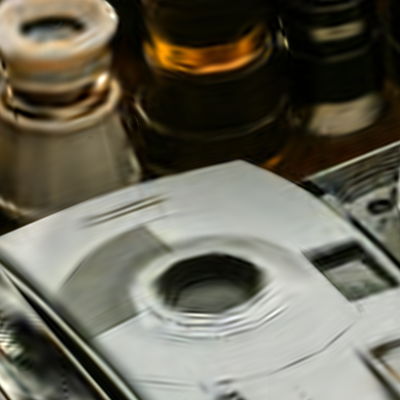}%
            \begin{picture}(0,0)\put(-55,2.5){\alphabox{29.18|0.907|0.303}}\end{picture}
        \end{minipage}
        \begin{minipage}[t]{.084\linewidth}
            \hrule height 0pt
            \includegraphics[width=\linewidth]{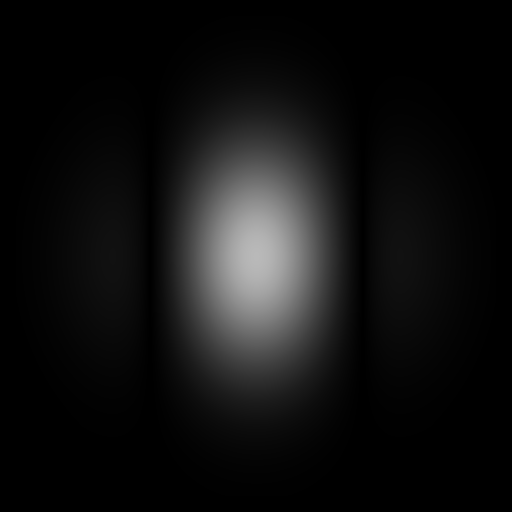} \\ 
            \includegraphics[width=\linewidth]{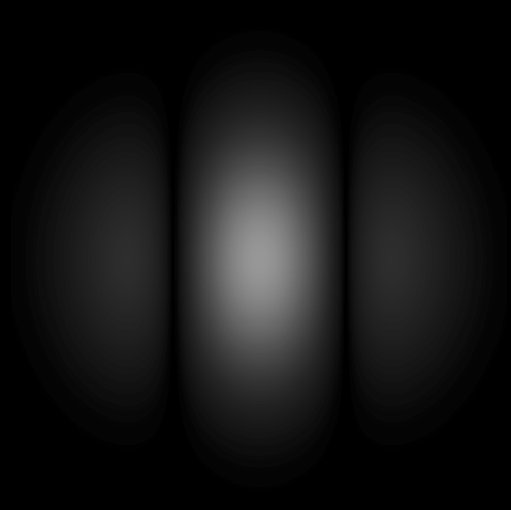} \\ 
            \includegraphics[width=\linewidth]{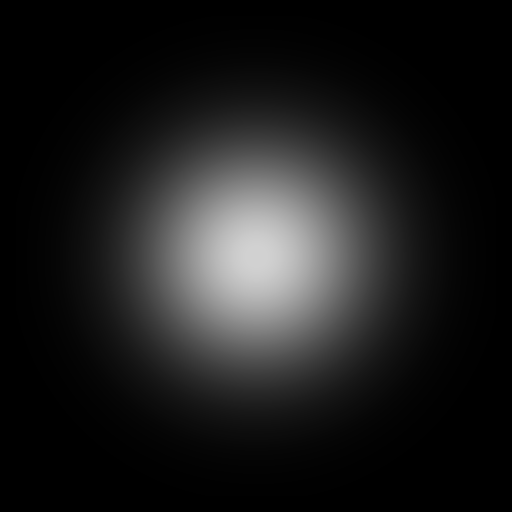}
        \end{minipage} \hfill
    \end{minipage}
    

    \begin{minipage}{\linewidth}
        \centering
        \begin{minipage}[t]{.263\linewidth}
            \hrule height 0pt
            \includegraphics[width=\linewidth]{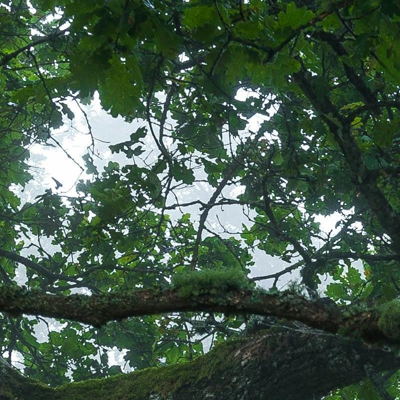}%
            \begin{picture}(0,0)\put(-57,2.5){\alphabox{PSNR|SSIM|LPIPS}}\end{picture}
        \end{minipage} \hfill
        \begin{minipage}[t]{.263\linewidth}
            \hrule height 0pt
            \includegraphics[width=\linewidth]{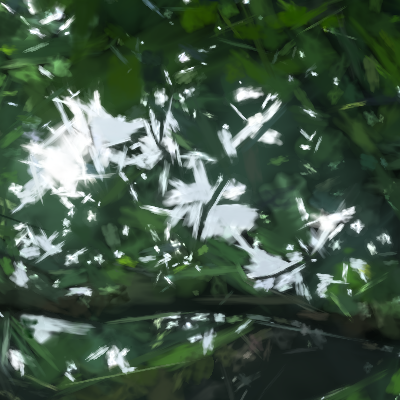}%
            \begin{picture}(0,0)\put(-55,2.5){\alphabox{21.89|0.533|0.505}}\end{picture}
        \end{minipage} \hfill
        \begin{minipage}[t]{.084\linewidth}
            \hrule height 0pt
            \includegraphics[width=\linewidth]{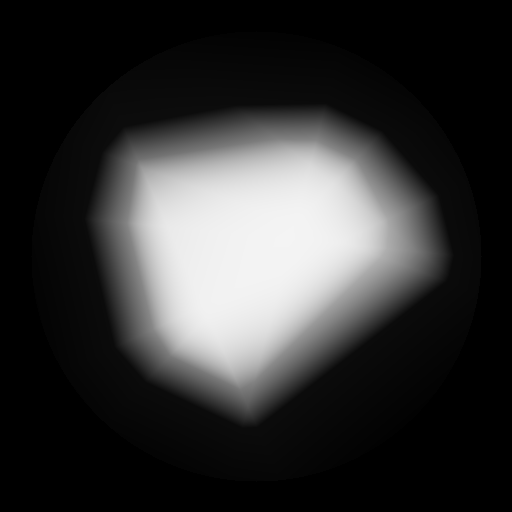} \\ 
            \includegraphics[width=\linewidth]{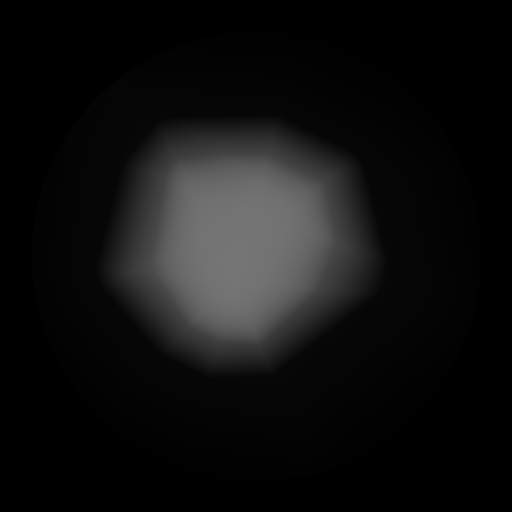} \\ 
            \includegraphics[width=\linewidth]{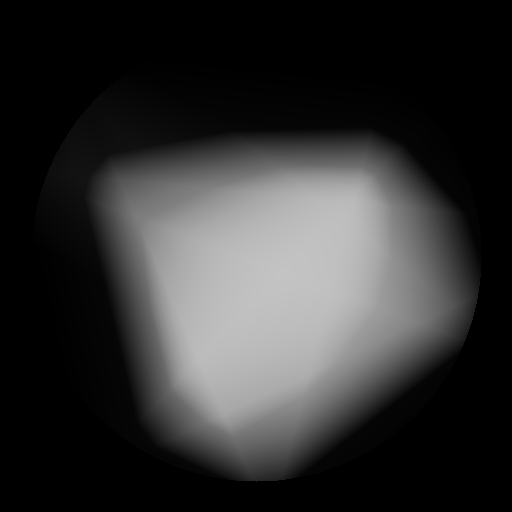}
        \end{minipage} \hfill
        \begin{minipage}[t]{.263\linewidth}
            \hrule height 0pt
            \includegraphics[width=\linewidth]{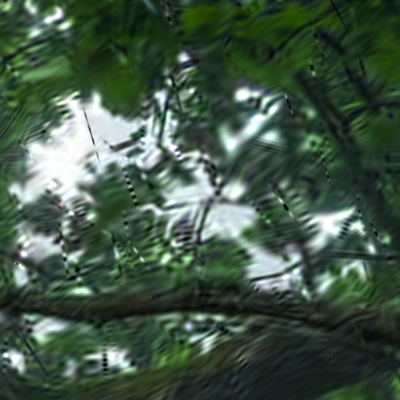}%
            \begin{picture}(0,0)\put(-55,2.5){\alphabox{21.36|0.499|0.568}}\end{picture}
        \end{minipage}
        \begin{minipage}[t]{.084\linewidth}
            \hrule height 0pt
            \includegraphics[width=\linewidth]{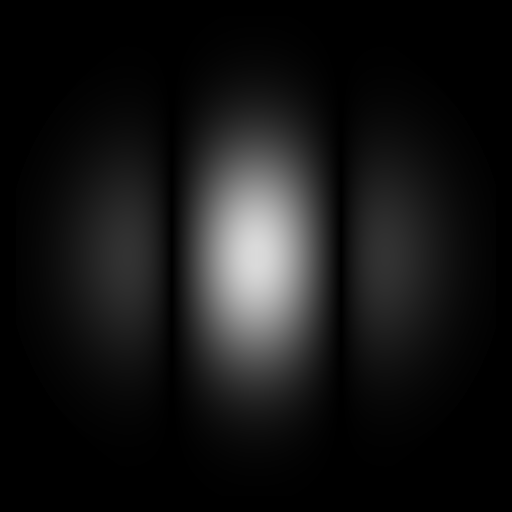} \\ 
            \includegraphics[width=\linewidth]{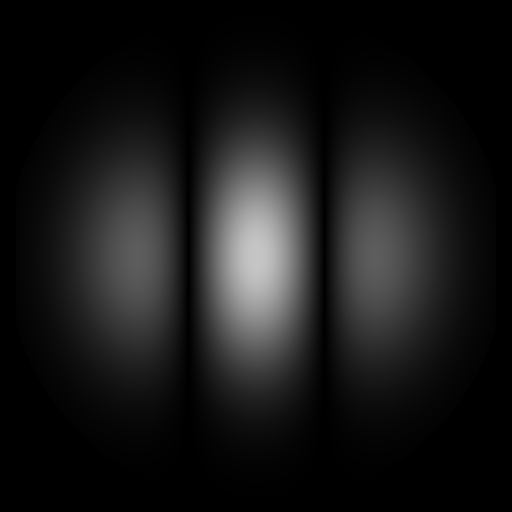} \\ 
            \includegraphics[width=\linewidth]{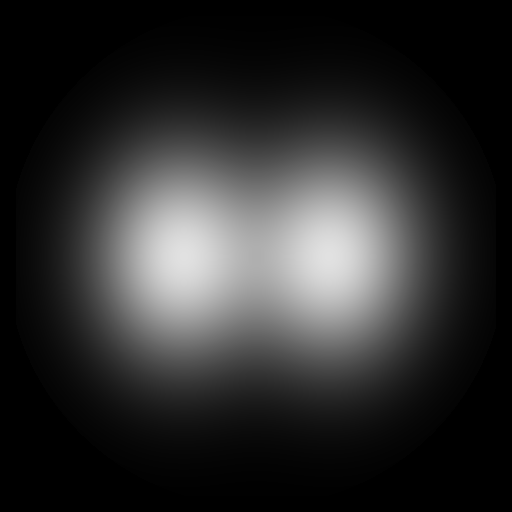}
        \end{minipage} \hfill
    \end{minipage}

    \caption{Comparisons between our approach and GabSplat~\cite{gabor2d} on efficiently representing 2D images, using a similar memory footprint for primitives. We show close-up views alongside with visualization of 2D kernel samples for each approach. Quantitative errors in PSNR, SSIM and LPIPS are reported at the bottom of each related image. Note that unlike in 3D splatting, our learned kernels here are general 2D ones and not restricted to being radially symmetric.}
    \label{fig:splat2d}
\end{figure}

We further extend our learned 2D kernels to efficient representation of generic 2D images in~\cref{fig:splat2d}. Our approach compares favorably against a state-of-the-art method that uses hand-crafted Gabor kernels~\cite{gabor2d}, with a similar memory footprint for primitives. Qualitative and quantitative results, along with visualization of primitives are shown in the figure.

\subsection{Ablations}
\label{sec:ablation}

We first evaluate the impact of different sets of input to the projection MLP $\Phi_{\operatorname{proj}}$ in~\cref{tab:ablation_inputs}. Our current choice of input achieves the best overall quality, by making the MLP spatial-, scale- and view-aware.
\begin{table}[h]
\centering
\caption{Ablation on the input to $\Phi_{\text{proj}}$ on two datasets. Quantitative errors in PSNR, SSIM and LPIPS are reported. Note that the last row corresponds to our current choice of the input.}
\footnotesize
\begin{tabular}{l|ccc|ccc}
\toprule
\multicolumn{1}{c|}{\multirow{2}{*}{Input to $\Phi_{\text{proj}}$}} & \multicolumn{3}{c|}{Tanks~\&~Temples} & \multicolumn{3}{c}{Deep Blending} \\
\multicolumn{1}{c|}{} & PSNR$\uparrow$ & SSIM$\uparrow$ & LPIPS$\downarrow$ & PSNR$\uparrow$ & SSIM$\uparrow$ & LPIPS$\downarrow$ \\
\midrule

$\mathbf{z}_{3D}$
& 24.92 & 0.873 & 0.144 
& 29.96 & 0.906 & 0.241 \\

$\mathbf{z}_{3D}$, $\boldsymbol{\omega}_o$ 
& 24.97 & 0.874 & 0.144 
& 30.09 & 0.911 & 0.235 \\

$\boldsymbol{\mu}^{\operatorname{cam}}_{3D}$, $\mathbf{s}$, $\mathbf{R}^{\operatorname{cam}}$ 
 & 24.82 & 0.871 & 0.144 & 30.49 & 0.913 & 0.235 \\ 
$\mathbf{z}_{3D}$, $\mathbf{s}$, $\mathbf{R}^{\operatorname{cam}}$ 
& 25.37 & 0.876 & 0.143 
& 30.06 & 0.914 & 0.239 \\
$\mathbf{z}_{3D}$, $\boldsymbol{\mu}^{\operatorname{cam}}_{3D}$, $\mathbf{R}^{\operatorname{cam}}$ 
& 25.45 & 0.876 & 0.143 
& 30.18 & 0.911 & 0.239 \\
$\mathbf{z}_{3D}$, $\boldsymbol{\mu}^{\operatorname{cam}}_{3D}$,
$\mathbf{s}$
& 24.53 & 0.832 & 0.175 
& 30.25 & 0.914 & 0.245 \\
\midrule
$\mathbf{z}_{3D}$, $\boldsymbol{\mu}^{\operatorname{cam}}_{3D}$,
$\mathbf{s}$, $\mathbf{R}^{\operatorname{cam}}$ & 25.42 & 0.877 & 0.141 & 30.52 & 0.913 & 0.233 \\
\bottomrule
\end{tabular}


\label{tab:ablation_inputs}
\end{table}

Next, \cref{tab:ablation_table} evaluates the impact of various factors (the dimension of $\mathbf{z}_{3D}$/$\mathbf{z}_{2D}$, and the width of $\Phi_{\operatorname{proj}}$/$\Phi_{\operatorname{enc}}$) over the reconstruction quality and memory footprint for primitives. Our default choice of these factors strikes a good balance between quality and memory footprint.
\begin{table}[h]
\centering
\caption{Ablations on the dimension of $\mathbf{z}_{3D}$/$\mathbf{z}_{2D}$ and the width of $\Phi_{\operatorname{proj}}$/$\Phi_{\operatorname{enc}}$ over the reconstruction quality and memory footprint on two datasets. Quantitative measures in PSNR and MB are listed. Mem. = memory footprint.}
\footnotesize
\setlength{\tabcolsep}{5pt} 

\begin{tabular}{l|cc|cc}
\toprule
\multirow{2}{*}{Variants} & \multicolumn{2}{c|}{Tanks~\&~Temples} & \multicolumn{2}{c}{Deep Blending} \\
& PSNR$\uparrow$ & Mem.(MB)$\downarrow$ & PSNR$\uparrow$ & Mem.(MB)$\downarrow$ \\
\midrule
$\operatorname{dim}(z_\text{3D})$=1          & 25.10 & 360.04 & 30.19 & 660.04 \\
$\operatorname{dim}(z_\text{3D})$=10         & 25.52 & 414.04 & 30.12 & 759.04 \\
$\operatorname{dim}(z_\text{2D})$=1          & 24.94 & 384.04 & 29.37 & 704.04 \\
$\operatorname{dim}(z_\text{2D})$=10         & 25.46 & 384.04 & 29.97 & 704.04 \\
$\operatorname{width}(\Phi_{\text{proj}})$=32  & 25.14 & 384.01 & 30.13 & 704.01 \\
$\operatorname{width}(\Phi_{\text{proj}})$=128 & 25.50 & 384.15 & 30.33 & 704.15 \\
$\operatorname{width}(\Phi_{\text{dec}})$=8    & 25.47 & 384.04 & 30.48 & 704.04 \\
$\operatorname{width}(\Phi_{\text{dec}})$=2    & 24.75 & 384.04 & 30.41 & 704.04 \\
\midrule 
Ours                    & 25.42 & 384.04 & 30.47 & 704.04 \\
\bottomrule
\end{tabular}

\label{tab:ablation_table}
\end{table}

Moreover, in the last 3 rows of~\cref{tab:efficiency}, we evaluate the impact of $k$ (Sec.~\ref{sec:acc}) over reconstruction quality with different memory footprints for primitives. With a small footprint, increasing $k$ results in higher reconstruction quality, as the expressiveness of learned kernels is increased. With a large footprint, this trend is not observed. We believe the reason is that the average screen-space size for each splat reduces with the increase in primitive count (i.e., memory footprint); simple conical-hat shaped kernels ($k=2$) suffice to produce satisfactory results, while achieving faster rendering speed compared with using a larger $k$. Note that our experiments are consistent with the main conclusion in~\cite{celarek2025does3dgaussiansplatting}.

We also investigate the impact of different target shapes for pre-training our kernel decoder $\Phi_{\operatorname{dec}}$ in~\cref{tab:ablation_init}. According to the table, pre-training outperforms no pre-training; we select the cosine shape for our pipeline (Sec.~\ref{sec:imp}), as it leads to the best results. Note that the learning rate is multiplied by 10 in no pre-training experiments.
\begin{table}[h]
\caption{Impact of different target shapes for pre-training $\Phi_{\text{dec}}$. Reconstruction errors on the Tanks~\&~Temples dataset in PSNR are listed in the table.}
\centering
\footnotesize
\begin{tabular}{lccccc} 
\toprule
 & no pre-train & Gaussian & polynomial & linear & cosine \\
\midrule
$k=2$ & 23.18 & 25.24 & 25.37 & 25.22 & 25.42 \\ 
$k=4$ & 23.72 & 25.04 & 25.37 & 25.26 & 25.34 \\ 
$k=8$ & 18.54 & 25.11 & 25.20 & 25.15 & 25.42 \\ 
\bottomrule
\end{tabular}
\label{tab:ablation_init}
\end{table}


In~\cref{tab:quantitative_eval}, we replace the view-dependent SH-based color model with a single constant color for each primitive, denoted as Ours~(3D,w/o SH), leading to lower-quality results. Color view-dependency and geometric one should not be viewed as equivalent as they work in different channels (i.e., RGB vs. alpha channel).

\begin{figure}[t]
    \centering
    \newcommand{\cellW}{0.325\linewidth}
    
    \newcommand{\InsetCell}[3]{%
        \begin{minipage}{\cellW}
            \centering
            \includegraphics[width=\linewidth, height=\linewidth]{#1}%
            \llap{%
                    \setlength{\fboxsep}{0pt}%
                    \setlength{\fboxrule}{0.5pt}%
                    \fcolorbox{white}{white}{%
                        \includegraphics[width=0.3\linewidth]{#2}%
                    }%
            }
        \end{minipage}%
    }

    \InsetCell{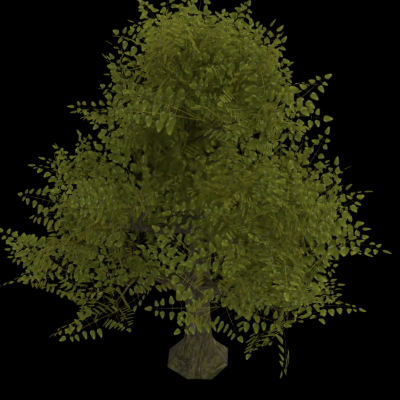}%
              {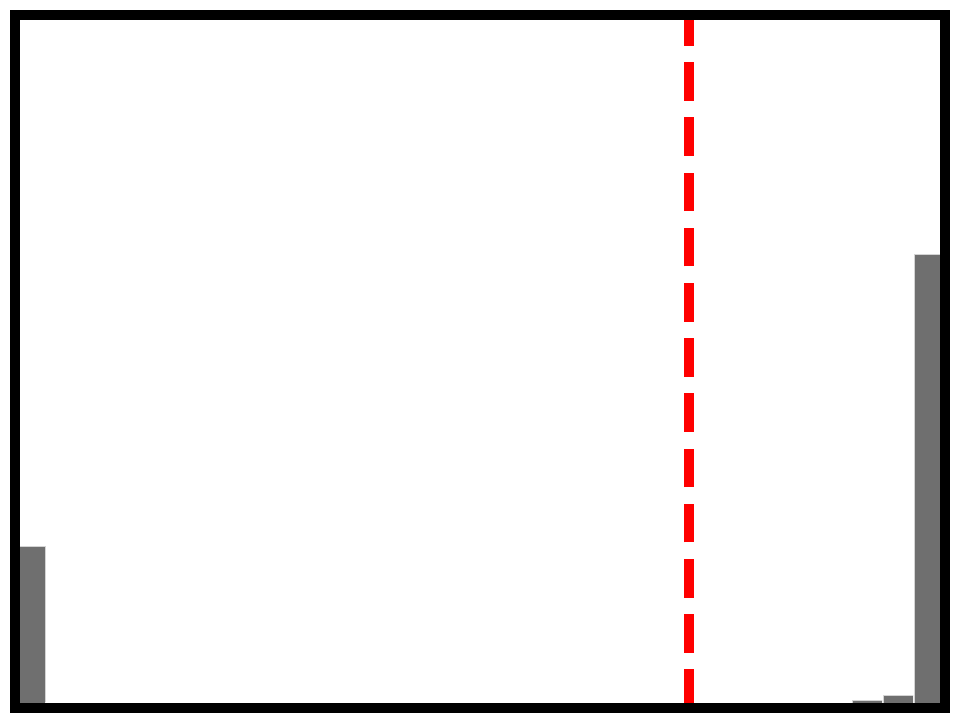}%
    \hfill
    \InsetCell{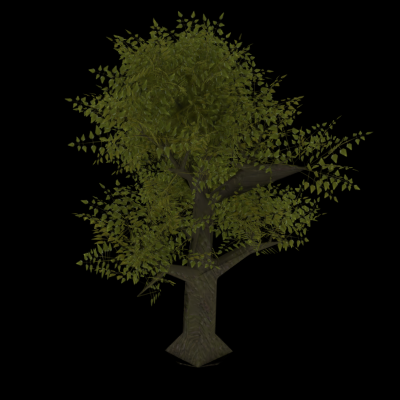}%
              {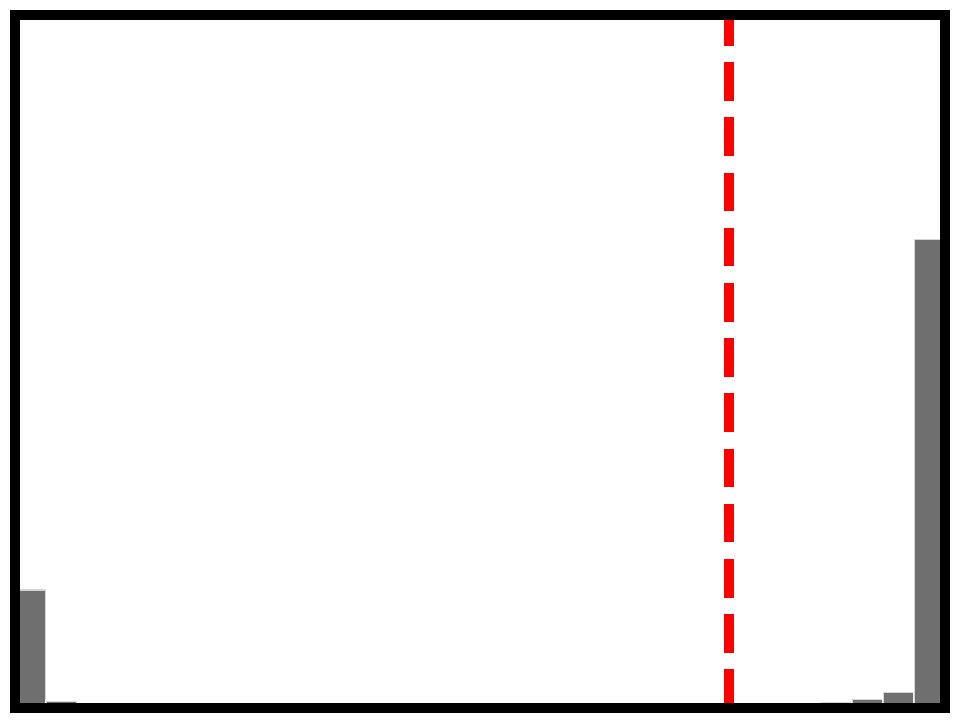}%
    \hfill
    \InsetCell{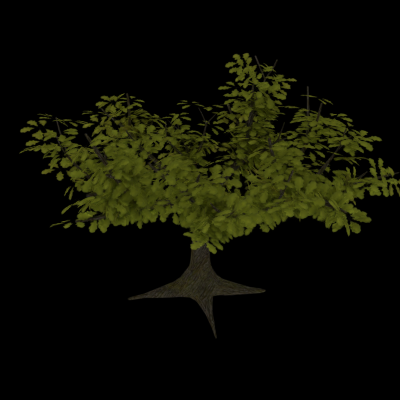}%
              {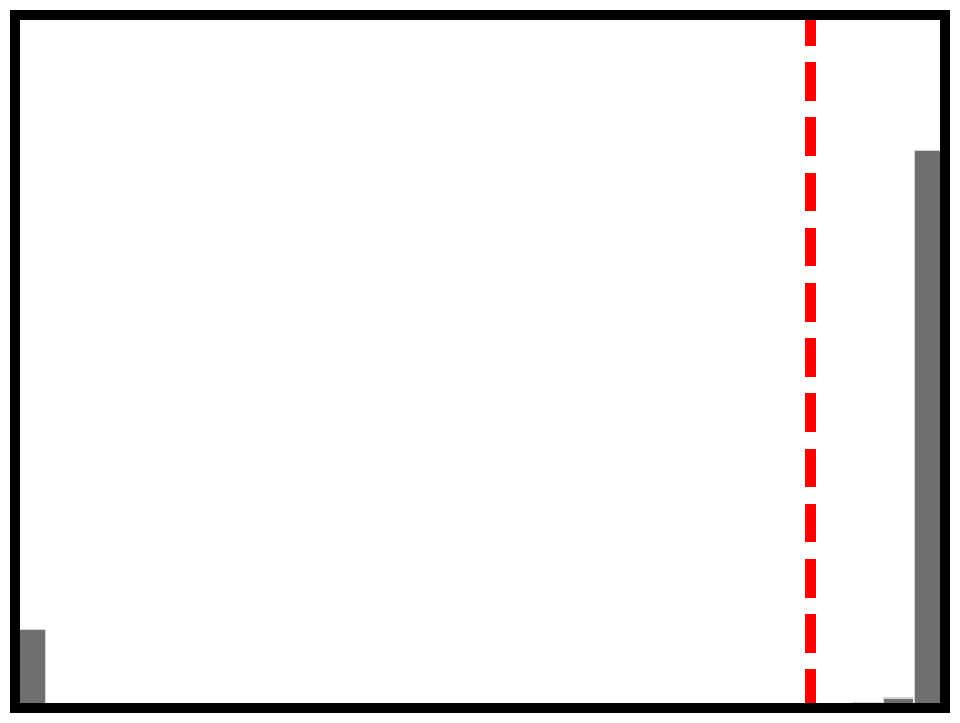}%
    

    \InsetCell{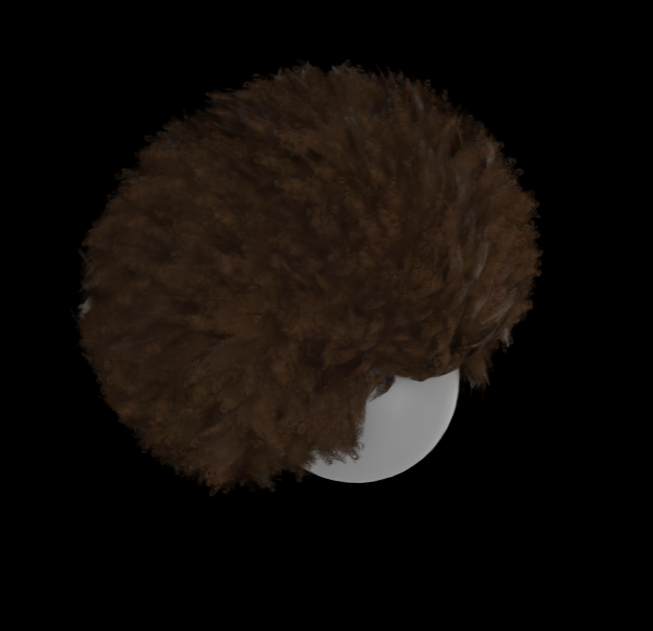}%
              {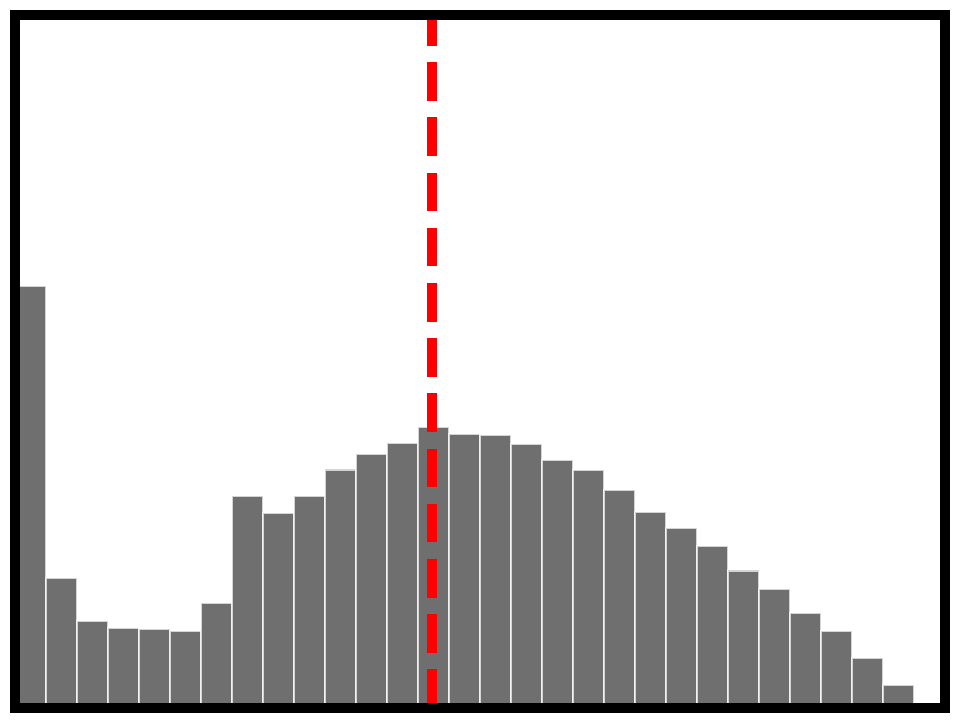}%
    \hfill
    \InsetCell{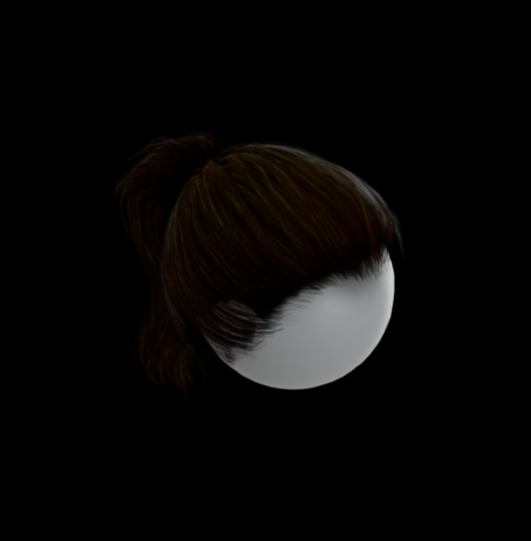}%
              {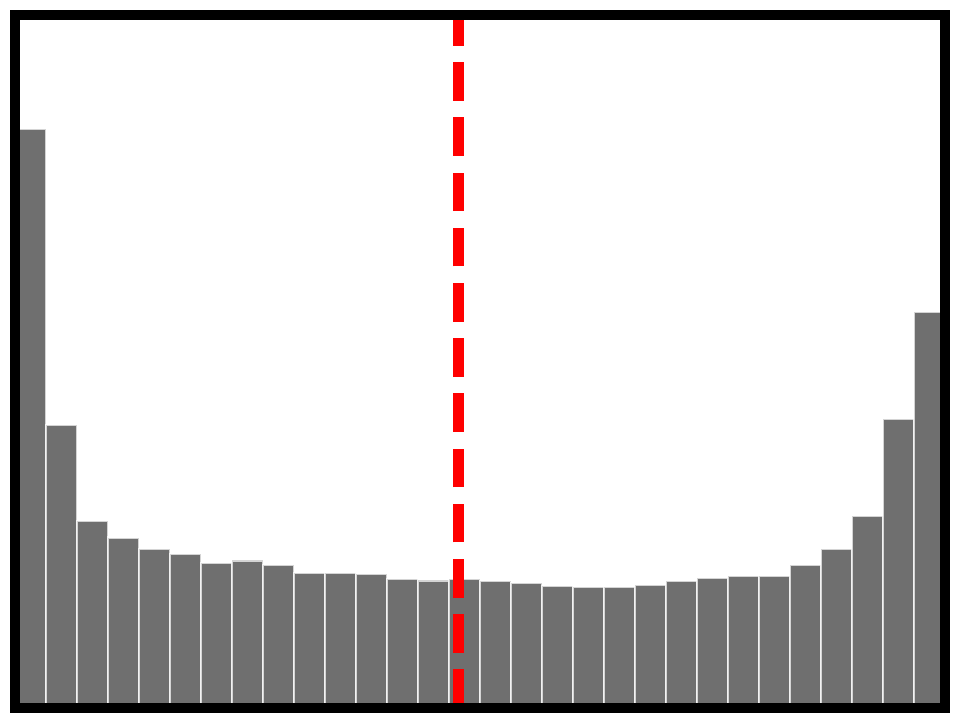}%
    \hfill
    \InsetCell{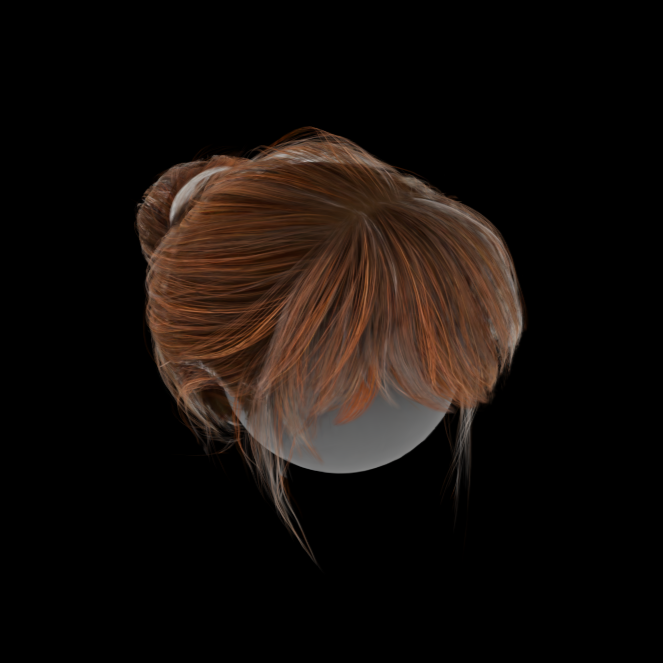}%
              {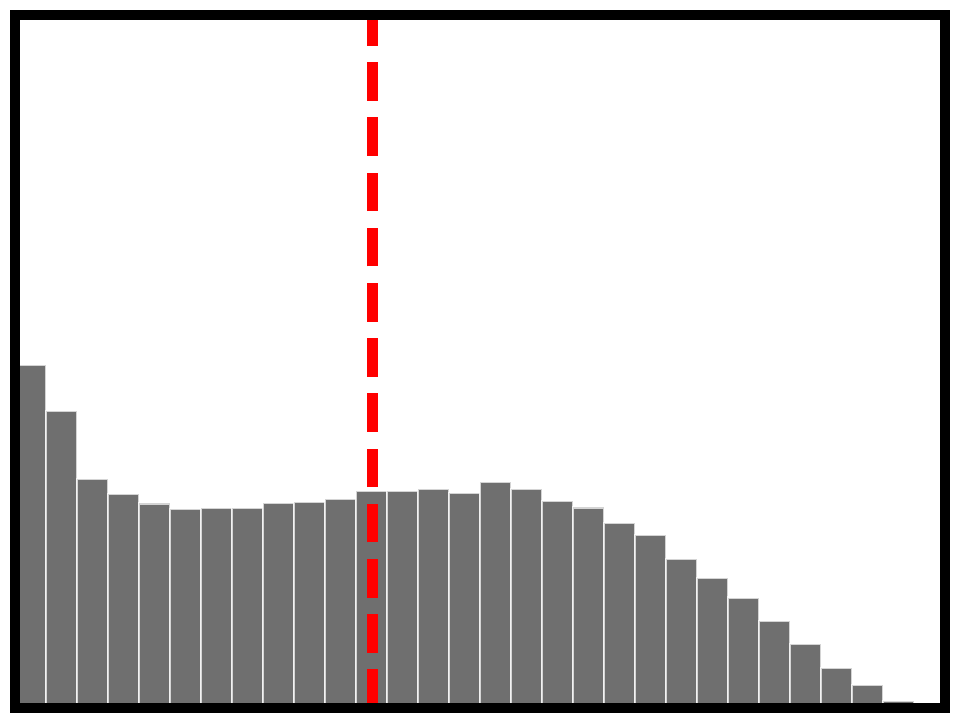}%

    \caption{Distributions of our kernels on different types of scenes. From the top row to bottom, vegetation and hair scenes. Each inset shows the corresponding histogram of $d_1$ (\cref{{eq:sample_d}}) of our learned kernels, with the red dotted line indicating the mean; the range of the horizontal axis is [0,1].}
    \label{fig:kernel_dist_inset_2x3}
\end{figure}

Finally, we evaluate in~\cref{fig:kernel_dist_inset_2x3} the distributions of learned volumetric primitives for two types of scenes, hair and vegetation. For each scene, we generate 100 training images, by randomly sampling a view the upper hemisphere, which points toward the center of the scene, and rendering the corresponding 3D scene with Blender. Strong correlations between primitive distributions and the scene type are shown in the figure. For each type of scene, the overall distributions of learned primitives look quite similar. This suggests interesting future work to further compress these similar primitives for more efficient representations. In other words, our framework could serve as a tool to \emph{automatically} generate novel, specialized splatting primitives (instead of general ones) that are tailored to efficiently represent certain types of scenes.

\begin{figure*}[t]
    \centering
    \setlength{\tabcolsep}{0pt}
    
    \newcommand{\wideW}{0.223\linewidth} 
    \newcommand{\sqW}{0.1485\linewidth}
    
    \begin{minipage}{\textwidth}
        \begin{minipage}{0.02\textwidth}\hspace{0pt}\end{minipage}%
        \hfill
        \begin{minipage}{0.975\textwidth}
            \centering
            \begin{minipage}{\wideW}\centering \subcaption*{\small }\end{minipage}\hfill
            \begin{minipage}{\sqW}\centering \subcaption*{\small Ground-Truth}\end{minipage}\hfill
            \begin{minipage}{\sqW}\centering \subcaption*{\small Ours}\end{minipage}\hfill
            \begin{minipage}{\sqW}\centering \subcaption*{\small 3DGS-MCMC}\end{minipage}\hfill
            \begin{minipage}{\sqW}\centering \subcaption*{\small DBS}\end{minipage}\hfill
            \begin{minipage}{\sqW}\centering \subcaption*{\small SSS}\end{minipage}
        \end{minipage}
    \end{minipage}
    

    \begin{minipage}{\textwidth}
        \begin{minipage}{0.02\textwidth}
            \centering
            \rotatebox{90}{\small \textsc{Room}}
        \end{minipage}%
        \hfill
        \begin{minipage}{0.975\textwidth}
            \centering
            \begin{minipage}{\wideW}
                \includegraphics[width=\linewidth, height=0.666\linewidth]{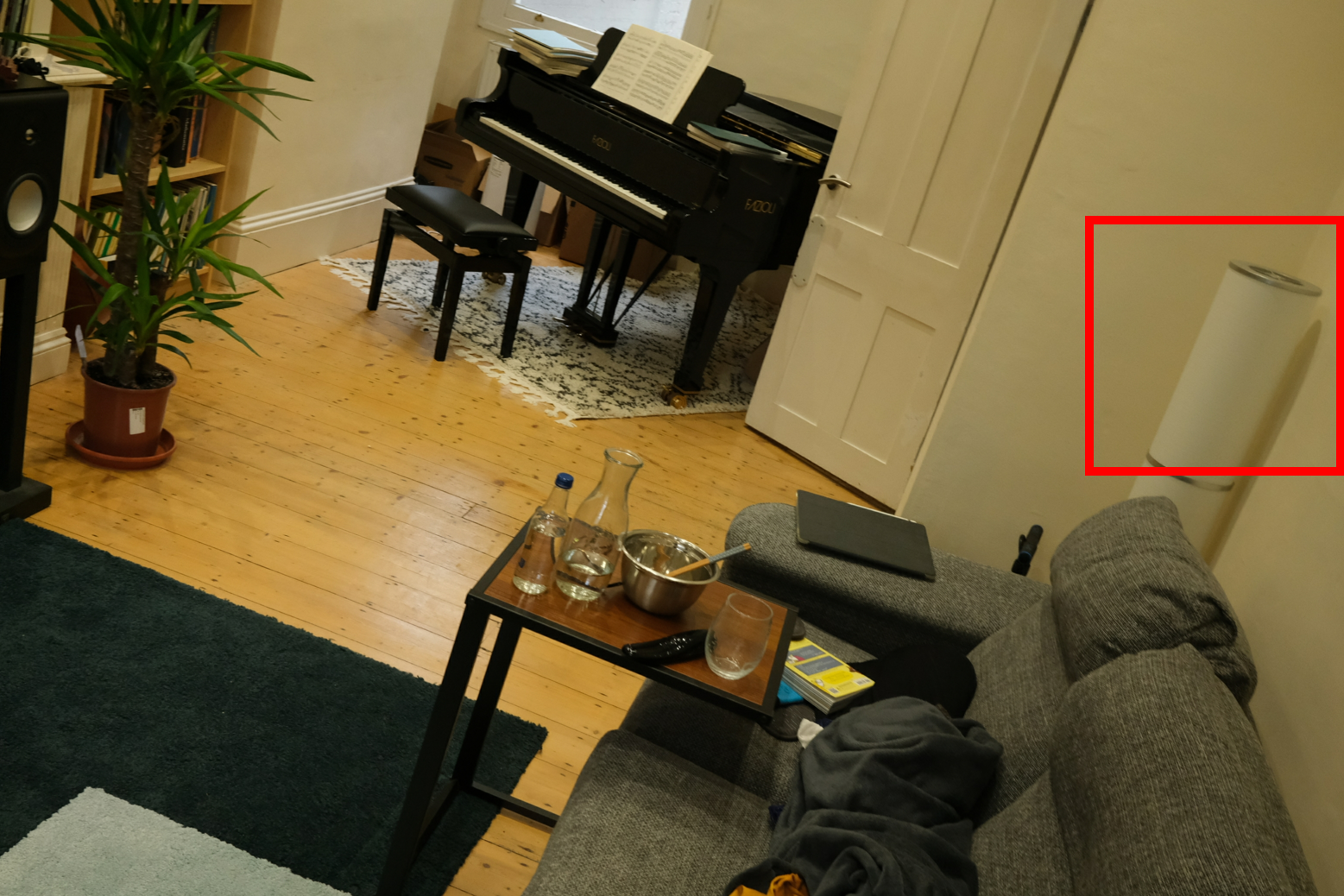}
            \end{minipage}\hfill
            \begin{minipage}{\sqW}
                \includegraphics[width=\linewidth, height=\linewidth]{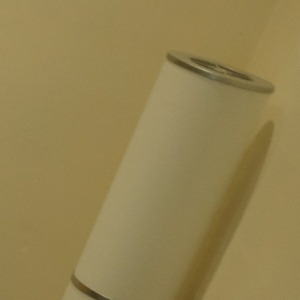}%
                \begin{picture}(0,0)\put(-60,3){\alphabox{PSNR|SSIM|LPIPS}}\end{picture}
            \end{minipage}\hfill
            \begin{minipage}{\sqW}
                \includegraphics[width=\linewidth, height=\linewidth]{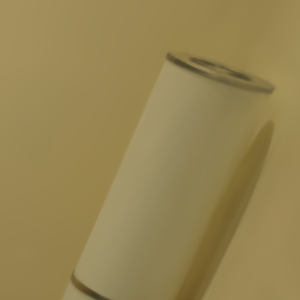}%
                \begin{picture}(0,0)\put(-60,3){\alphabox{35.27|0.941|0.167}}\end{picture}
            \end{minipage}\hfill
            \begin{minipage}{\sqW}
                \includegraphics[width=\linewidth, height=\linewidth]{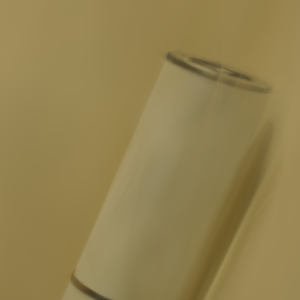}%
                \begin{picture}(0,0)\put(-60,3){\alphabox{33.78|0.938|0.174}}\end{picture}
            \end{minipage}\hfill
            \begin{minipage}{\sqW}
                \includegraphics[width=\linewidth, height=\linewidth]{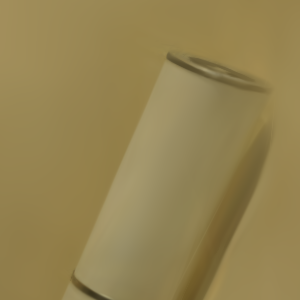}%
                \begin{picture}(0,0)\put(-60,3){\alphabox{33.73|0.943|0.161}}\end{picture}
            \end{minipage}\hfill
            \begin{minipage}{\sqW}
                \includegraphics[width=\linewidth, height=\linewidth]{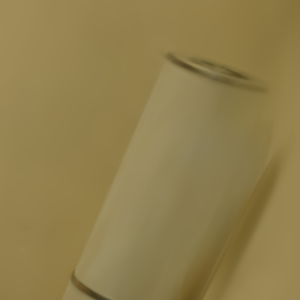}%
                \begin{picture}(0,0)\put(-60,3){\alphabox{33.25|0.940|0.166}}\end{picture}
            \end{minipage}
        \end{minipage}
    \end{minipage}


    \begin{minipage}{\textwidth}
        \begin{minipage}{0.02\textwidth}
            \centering
            \rotatebox{90}{\small \textsc{Bonsai}}
        \end{minipage}%
        \hfill
        \begin{minipage}{0.975\textwidth}
            \centering
            \begin{minipage}{\wideW}
                \includegraphics[width=\linewidth, height=0.666\linewidth]{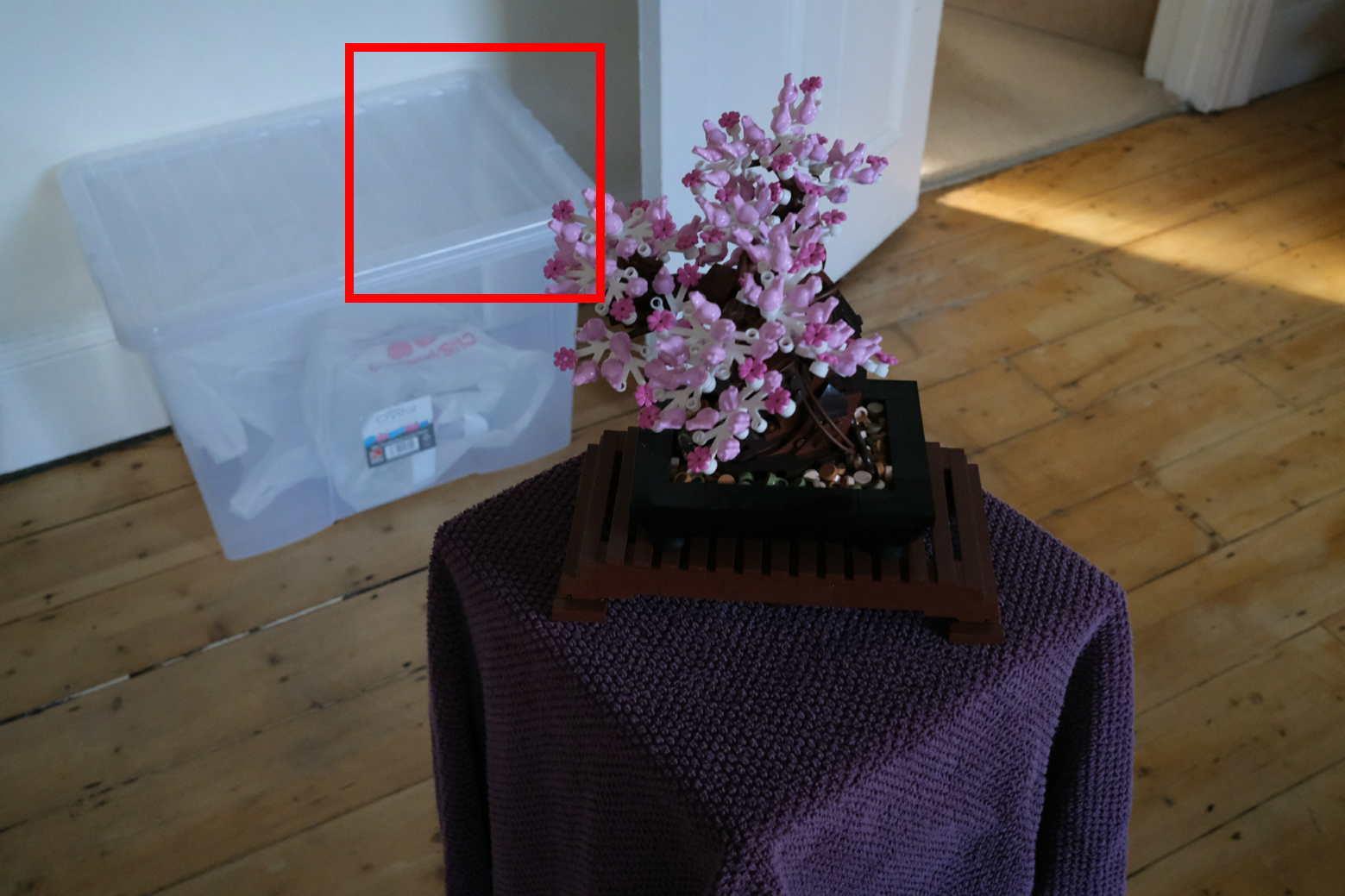}
            \end{minipage}\hfill
            \begin{minipage}{\sqW}
                \includegraphics[width=\linewidth, height=\linewidth]{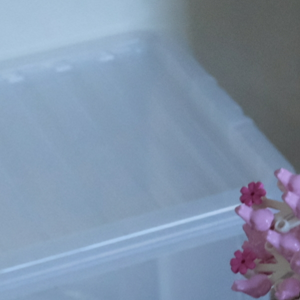}%
                \begin{picture}(0,0)\put(-60,3){\alphabox{PSNR|SSIM|LPIPS}}\end{picture}
            \end{minipage}\hfill
            \begin{minipage}{\sqW}
                \includegraphics[width=\linewidth, height=\linewidth]{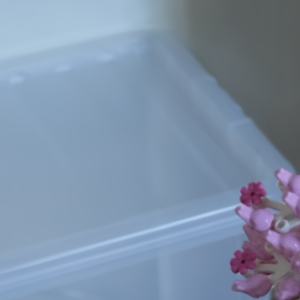}%
                \begin{picture}(0,0)\put(-60,3){\alphabox{38.16|0.966|0.209}}\end{picture}
            \end{minipage}\hfill
            \begin{minipage}{\sqW}
                \includegraphics[width=\linewidth, height=\linewidth]{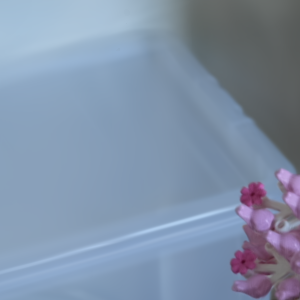}%
                \begin{picture}(0,0)\put(-60,3){\alphabox{35.65|0.961|0.228}}\end{picture}
            \end{minipage}\hfill
            \begin{minipage}{\sqW}
                \includegraphics[width=\linewidth, height=\linewidth]{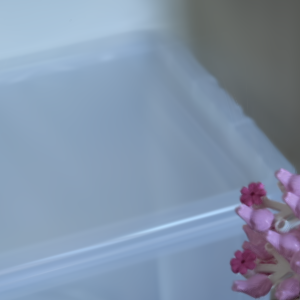}%
                \begin{picture}(0,0)\put(-60,3){\alphabox{36.01|0.961|0.215}}\end{picture}
            \end{minipage}\hfill
            \begin{minipage}{\sqW}
                \includegraphics[width=\linewidth, height=\linewidth]{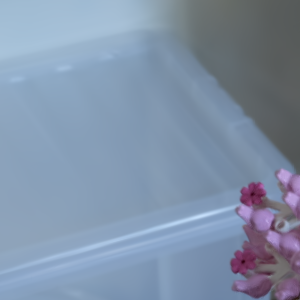}%
                \begin{picture}(0,0)\put(-60,3){\alphabox{36.92|0.964|0.212}}\end{picture}
            \end{minipage}
        \end{minipage}
    \end{minipage}


    \begin{minipage}{\textwidth}
        \begin{minipage}{0.02\textwidth}
            \centering
            \rotatebox{90}{\small \textsc{Train}}
        \end{minipage}%
        \hfill
        \begin{minipage}{0.975\textwidth}
            \centering
            \begin{minipage}{\wideW}
                \includegraphics[width=\linewidth, height=0.666\linewidth]{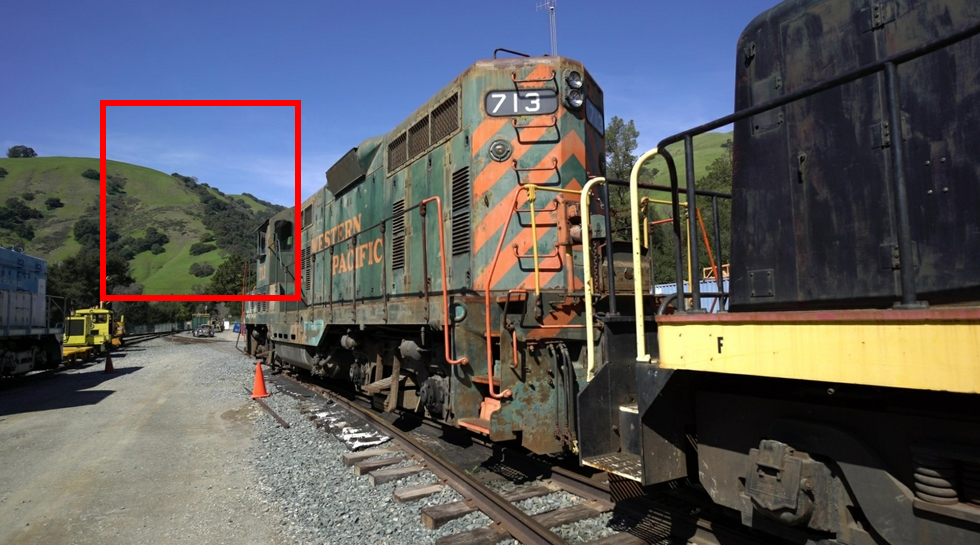}
            \end{minipage}\hfill
            \begin{minipage}{\sqW}
                \includegraphics[width=\linewidth, height=\linewidth]{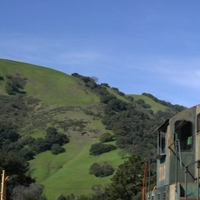}%
                \begin{picture}(0,0)\put(-60,3){\alphabox{PSNR|SSIM|LPIPS}}\end{picture}
            \end{minipage}\hfill
            \begin{minipage}{\sqW}
                \includegraphics[width=\linewidth, height=\linewidth]{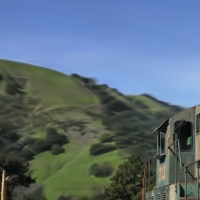}%
                \begin{picture}(0,0)\put(-60,3){\alphabox{27.50|0.884|0.174}}\end{picture}
            \end{minipage}\hfill
            \begin{minipage}{\sqW}
                \includegraphics[width=\linewidth, height=\linewidth]{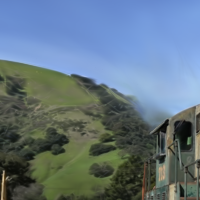}%
                \begin{picture}(0,0)\put(-60,3){\alphabox{24.90|0.865|0.190}}\end{picture}
            \end{minipage}\hfill
            \begin{minipage}{\sqW}
                \includegraphics[width=\linewidth, height=\linewidth]{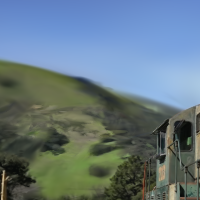}%
                \begin{picture}(0,0)\put(-60,3){\alphabox{24.39|0.854|0.190}}\end{picture}
            \end{minipage}\hfill
            \begin{minipage}{\sqW}
                \includegraphics[width=\linewidth, height=\linewidth]{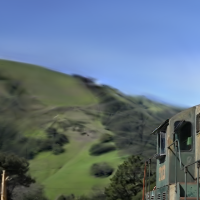}%
                \begin{picture}(0,0)\put(-60,3){\alphabox{24.67|0.873|0.164}}\end{picture}
            \end{minipage}
        \end{minipage}
    \end{minipage}


    \begin{minipage}{\textwidth}
        \begin{minipage}{0.02\textwidth}
            \centering
            \rotatebox{90}{\small \textsc{Playroom}}
        \end{minipage}%
        \hfill
        \begin{minipage}{0.975\textwidth}
            \centering
            \begin{minipage}{\wideW}
                \includegraphics[width=\linewidth, height=0.666\linewidth]{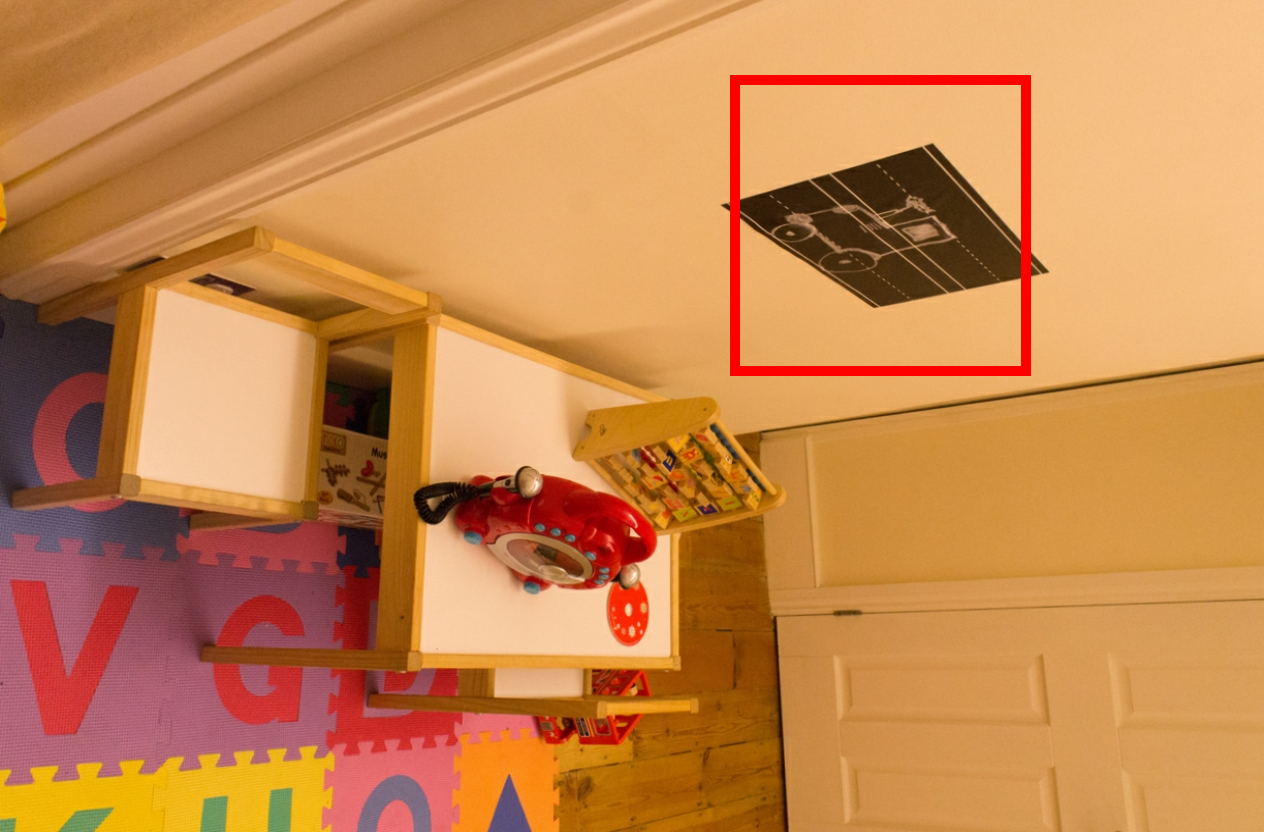}
            \end{minipage}\hfill
            \begin{minipage}{\sqW}
                \includegraphics[width=\linewidth, height=\linewidth]{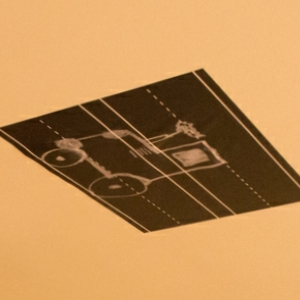}%
                \begin{picture}(0,0)\put(-60,3){\alphabox{PSNR/SSIM/LPIPS}}\end{picture}
            \end{minipage}\hfill
            \begin{minipage}{\sqW}
                \includegraphics[width=\linewidth, height=\linewidth]{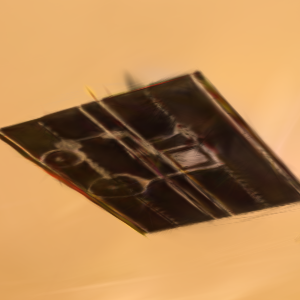}%
                \begin{picture}(0,0)\put(-60,3){\alphabox{28.04|0.897|0.240}}\end{picture}
            \end{minipage}\hfill
            \begin{minipage}{\sqW}
                \includegraphics[width=\linewidth, height=\linewidth]{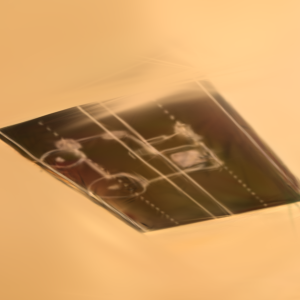}%
                \begin{picture}(0,0)\put(-60,3){\alphabox{25.72|0.896|0.242}}\end{picture}
            \end{minipage}\hfill
            \begin{minipage}{\sqW}
                \includegraphics[width=\linewidth, height=\linewidth]{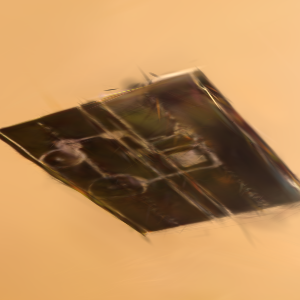}%
                \begin{picture}(0,0)\put(-60,3){\alphabox{27.45|0.901|0.239}}\end{picture}
            \end{minipage}\hfill
            \begin{minipage}{\sqW}
                \includegraphics[width=\linewidth, height=\linewidth]{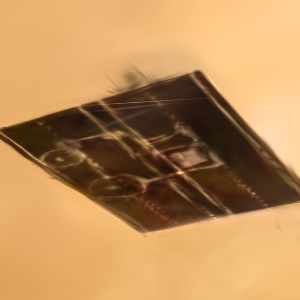}%
                \begin{picture}(0,0)\put(-60,3){\alphabox{26.83|0.896|0.246}}\end{picture}
            \end{minipage}
        \end{minipage}
    \end{minipage}
    
    \caption{Qualitative comparisons with state-of-the-art techniques. We compare our approach with 3DGS-MCMC~\cite{kheradmand20243dgsmcmc}, SSS~\cite{zhu2025sss} and DBS~\cite{liu2025deformablebetasplatting}. Quantitative errors in PSNR, SSIM and LPIPS are also reported at the bottom of each related image.}
    \label{fig:main_compact}
\end{figure*}
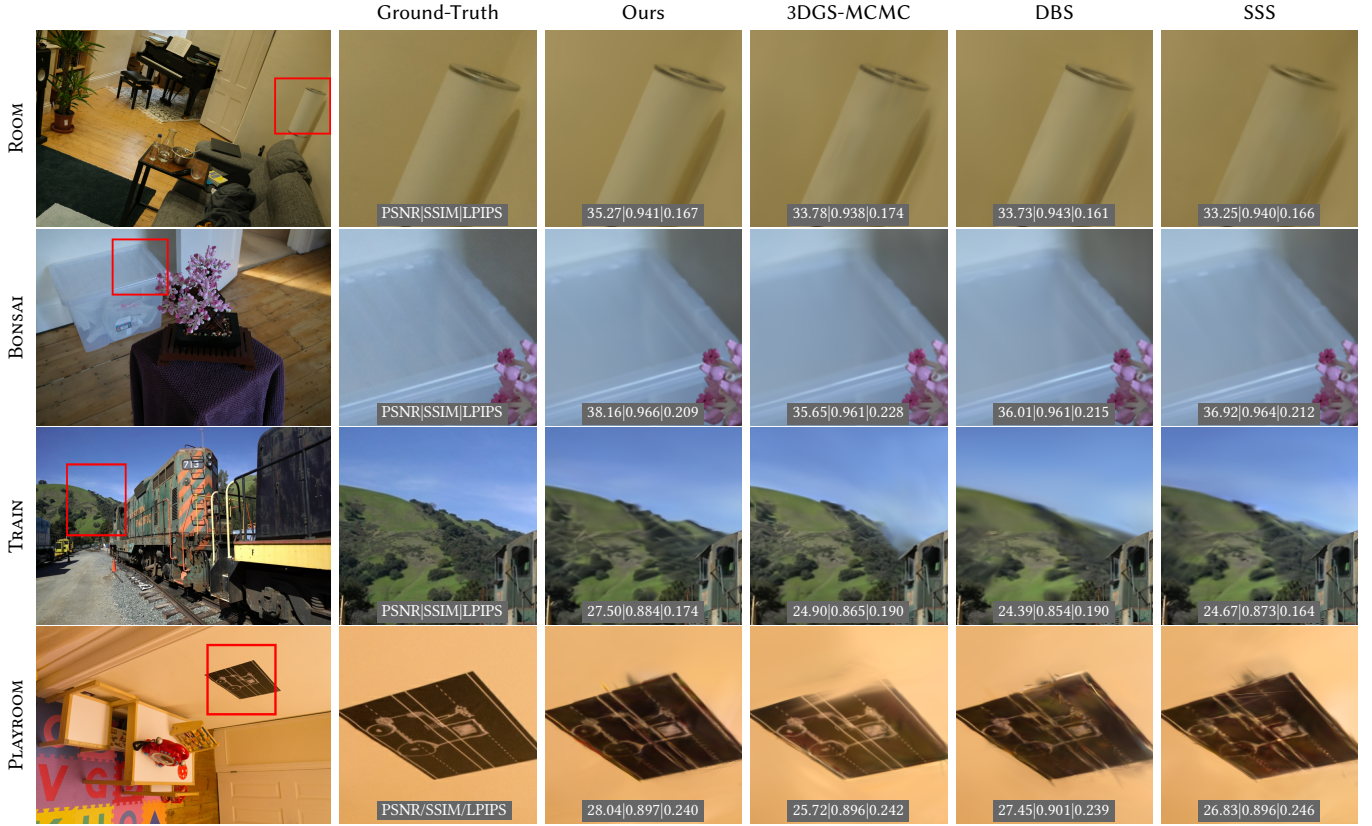

\section{Limitations and Future Work}


Our work is subject to a number of limitations. First, as shown in~\cref{tab:efficiency}, our rendering speed is lower than that of state-of-the-art techniques, even though in many cases we can achieve real-time performance (>50 fps). The reason is that $\Phi_{\operatorname{proj}}$ and $\Phi_{\operatorname{dec}}$ are primarily designed for reconstruction quality and representation efficiency; the current running time breakdown among $\Phi_{\operatorname{proj}}$, $\Phi_{\operatorname{dec}}$, and rasterization is roughly 9:2:1.
It will be promising to distill a high-performance version of the MLPs via various acceleration approaches (i.e., weight pruning, switching to FP16 precision, etc), or even derive equivalent analytical equations via, e.g., symbolic regression. Second, we do not take anti-aliasing into consideration. It seems possible to combine our approach with 
related techniques like~\cite{Yu2024MipSplatting}. In addition, no advanced encoding methods (e.g., positional/frequency) are adopted for input to $\Phi_{\operatorname{proj}}$/$\Phi_{\operatorname{dec}}$. We are interested in exploring different encodings to further improve the performance, especially for 2D splatting. 
 
We hope that our work could inspire more research into automatic design of graphics representations (e.g., developing efficient, specialized representations for particular types of scenes), or even the processing pipeline~\cite{zeng2025renderformer}. It is also intriguing to analyze what exactly is learned in the view dependency, and compare it with rigid Euclidean 3D consistency. Our learned consistency might be helpful in other tasks, such as regularizing generative 3D modeling. In addition, it is desirable to compute quantitative metrics over the temporal domain to systemically analyze the reconstruction quality in animation~\cite{liang2024perceptual}. Finally, we would like to combine with state-of-the-art work on relighting~\cite{bi2024rgs}, to support efficient image synthesis with novel view and lighting conditions.

\begin{acks}
Fig.~\ref{fig:teaser} \&~\ref{fig:kernel_dist_inset_2x3} use assets by khaloui with the extended use license, and by MagicCGIStudios with the standard license. This work is partially supported by NSF China (62332015, 62227806, \& 62421003), the XPLORER PRIZE, Information Technology Center, State Key Lab of CAD\&CG, Zhejiang University, and a gift from Adobe. 
\end{acks}
\bibliographystyle{ACM-Reference-Format}
\bibliography{OptimalPrimitive}

\clearpage








\end{document}